\documentclass[10pt,journal,compsoc]{IEEEtran}
\ifCLASSOPTIONcompsoc
  \usepackage[nocompress]{cite}
\else
  \usepackage{cite}
\fi
\ifCLASSINFOpdf
\else
\fi
\ifCLASSOPTIONcompsoc
\else
\fi
\ifCLASSINFOpdf
\else
\fi
\usepackage{float}
\usepackage[center]{caption2}
\usepackage{xspace}
\usepackage{balance}
\usepackage{algorithm}
\usepackage{algorithmic}
\usepackage{diagbox}
\usepackage{bbding}
\usepackage{amsthm}
\usepackage{amsmath,amssymb,amsfonts}
\usepackage{makecell}
\usepackage{diagbox}
\usepackage{multirow}
\usepackage{threeparttable}
\usepackage{wrapfig}
\usepackage{subfigure}
\usepackage{booktabs}
\usepackage{tikz}

\newtheorem{Theorem}{Theorem}[section]

\ifCLASSINFOpdf
\else
\fi

\begin{document}

\title{Efficiency Boosting of Secure Cross-platform Recommender Systems over Sparse Data}

\author{{Hao~Ren, Guowen~Xu, Tianwei~Zhang, Jianting~Ning,  Xinyi~Huang, Hongwei~Li, \\ Rongxing~Lu~\IEEEmembership{Fellow,~IEEE}}

\IEEEcompsocitemizethanks{\IEEEcompsocthanksitem Hao~Ren, Guowen~Xu (Corresponding author), and Tianwei~Zhang are with the School of Computer Science and Engineering, Nanyang Technological University. (e-mail: hao.ren@ntu.edu.sg; guowen.xu@ntu.edu.sg;  tianwei.zhang@ntu.edu.sg).

\IEEEcompsocthanksitem Jianting~Ning is with the College of Computer and Cyber Security, Fujian Normal University, Fuzhou, China (e-mail: jtning88@gmail.com).
\IEEEcompsocthanksitem Xinyi~Huang is with the Artificial Intelligence Thrust, Information Hub, Hong Kong University of Science and Technology (Guangzhou), Guangzhou, China, 511458 (e-mail: xinyi@ust.hk).

\IEEEcompsocthanksitem Hongwei~Li  is with the School of Computer Science and Engineering,  University of Electronic Science and Technology of China, Chengdu 611731, China. (e-mail: hongweili@uestc.edu.cn).

\IEEEcompsocthanksitem Rongxing~Lu  is with the School of  Computer Science, University of New Brunswick, Canada. (e-mail: rlu1@unb.ca).
}}



\IEEEcompsoctitleabstractindextext{
\begin{abstract}
\raggedright
Fueled by its successful commercialization, the recommender system (RS) has gained widespread attention. However, as the training data fed into the RS models are often highly sensitive, it ultimately leads to severe privacy concerns, especially when data are shared among different platforms. In this paper, we follow the tune of existing works to investigate the problem of secure sparse matrix multiplication for cross-platform RSs. Two fundamental while critical issues are addressed: preserving the training data privacy and breaking the data silo problem. Specifically, we propose two  concrete constructions with significantly boosted efficiency. They are designed for the sparse location insensitive case and location sensitive case, respectively. State-of-the-art cryptography building blocks including homomorphic encryption (HE) and private information retrieval (PIR) are fused into our protocols with non-trivial optimizations. %We eliminate the time-consuming rotation operations completely for both schemes, which is heavily invoked by the existing works. In addition, for the first times, our location sensitive scheme bridges the gap between PIR and popularly used message packing method for HE. In doing so,
As a result, our schemes can enjoy the HE acceleration technique without privacy trade-offs. We give formal security proofs for the proposed schemes and conduct extensive experiments on both real and large-scale simulated datasets. Compared with state-of-the-art works, our two schemes compress the running time roughly by $10\times$ and $2.8\times$. They also attain up to $15\times$ and $2.3\times$ communication reduction without accuracy loss.
\end{abstract}

}

\maketitle

\IEEEdisplaynontitleabstractindextext
\IEEEpeerreviewmaketitle

\IEEEraisesectionheading{\section{Introduction}
\label{sec:introduction}}
%1. The importance of RS. The evidence or the popularity of its widely application. 2. The very basic principles in RS, using the history data as the input to train a model, and the users history data to conduct prediciton.
\IEEEPARstart{T}{he} recommender system (RS) \cite{ma2011recommender} is now recognized as one of the most indispensable and powerful intelligent daily-life assistants, that can offer accurate and personalized recommendation services for large-scale users. Owing to the wide deployment of RSs, users are alleviated from making choices from the overwhelming amount of items. They can rely on RSs to explore interesting products and services according to their expenditure history.
%Evidence for proving the popularity and impacts of RSs is easy to find. On the one hand,
Technology giants such as Amazon, Google, and Byte Dance are applying RSs to dispatch advertisements to potential consumers. Indeed, RSs can not only provide huge commercial benefits to enterprises \cite{user_number_amazon, user_number_facebook} but also significantly promote the user experience for diverse applications \cite{tang2013social}. Especially when the social data are incorporated into the training datasets rather than solely using the rating data, much higher prediction accuracy can be achieved \cite{ma2011recommender}. Intuitively, the user's preference is likely to be similar to his close friends. In addition, existing works \cite{ma2011recommender, tang2013social} have already proved their effectiveness through real deployment. This paradigm is termed the cross-platform RS in this paper.

%Despite the importance of RS. The current RS is suffering two major problems. 1. Privcacy concerns. 2. The extreme data sparsity. 3. The con-current of these two issues make the designing of privacy-preserving RS more challenging. Therefore, it is urgent to...
While enjoying the benefits brought by the modern cross-platform RS, two inherent and intricate issues are emerging as the stone in the road towards its fast development. The first issue is the privacy concern introduced by the gathering and use of sensitive personal data especially when the data are transferred between two enterprises. The sharing of either social or rating data significantly raises the risk of information leakage, and breaches of user privacy are very likely. In some areas that have strong privacy cultures such as European, the use and transfer of personal data are strictly constrained by law (e.g., GDPR \cite {lu2021whois}). As a result, preserving data privacy in cross-platform RSs is paramount. The second issue is that the training data are extremely sparse, especially the social data. For instance, the social density in the commonly used testing dataset LiThing \cite{zhao2015improving} is roughly $0.02\%$. In the plaintext domain, we can trivially compress the sparse dataset, while the problem becomes challenging in the privacy-preserving context. Specifically, if the conventional secure multiparty computation (MPC) \cite{hastings2019sok} or homomorphic encryption (HE) \cite{acar2018survey} is applied,  we can train the RS model in a private way. However, this line of works \cite{chen2020secure,jumonji2021privacy} can hardly leverage the data sparsity as the datasets are either encrypted or shared. In consequence, prohibitive resource consumption becomes a longstanding unsolved problem. In this paper, we aim to conquer this dilemma by proposing schemes that fully exploit the data sparsity to boost efficiency, yet offer strong privacy preservation.

%%%%%%%%%%%%%%%%%%%%%%%%%%%%%%%%%%%%%%%%%%%%%%%%%%%%%%%%%%%%%%%%%%
\subsection{Related Works}
\label{subsec:Limitations of related works}
%Privacy-preserving RS has been actively investigated through diverse lens of privacy enhancing technologies for different RS models \cite{ge2022survey}. In term of RS model,
%逻辑整理：1.论文的基本技术背景，具体解决的问题在目前的相关工作中是什么情况。
In this paper, in terms of the RS model \cite{ge2022survey}, we focus on the popular collaborative filtering (CF) model \cite{ma2011recommender} that factorizes the rating matrix into two matrices and then conducts missing data prediction atop the factorized matrices. In this cross-platform setting, social data are incorporated into the training process. In specific, one party holds the rating data and the other holds the social data, and they collaboratively train a CF model. Technically, the used optimization algorithm can be boiled down to securely computing matrix multiplication for the two-party setting. Numerous arts \cite{demmler2015aby} have been proposed to solve this problem. In the following paragraphs, we briefly review the related works and analyze their advantages and limitations.

%删除的次要内容：\textit{Badsha et al.} \cite{badsha2017privacy} attempted to provide privacy-preserving recommendation service on the trained CF model. User's query data is encrypted by partial HE (PHE) scheme BGN \cite{boneh2005evaluating}. Constrained by BGN, the intermediate results have to be decrypted by a trusted third party that is difficult to seek in the real world.
%$\mathsf{SeSoRec}$ presented in \cite{chen2020secure} models this problem as conventional secure two party computation problem and solve it based on secret sharing \cite{demmler2015aby}. The sum of two rows/columns of input matrix is revealed. Such information may leads to serious privacy leakage (e.g., social relations) especially when the input social data is binary matrix.
%2.现有的基于HE的方案要么是没有使用不需要旋转的打包加速，矩阵分析等优化技术等技术，要么是没有考虑稀疏性，而且基本没有考虑夸平台的两方的方案。这三个问题普标存在于早期的方案。
Early work proposed by Jumonji \textit{et al.} \cite{jumonji2021privacy} turned to use fully HE (FHE) \cite{gentry2013homomorphic} to enable recommendation on the CF model without decryption during processing. To alleviate the heavy computational and communication loads brought by FHE, multiple messages are packed as one to compress the encryption/decryption costs. Huang \textit{et al.} proposed $uSCORE$ \cite{huang2021more}, an FHE based scheme for the data unbalanced scenario, that delegates most computational load to the service provider. In addition, a fast secure matrix multiplication algorithm is designed atop the secure sparse SVD optimization \cite{jiang2018secure}. Due to the use of packing methods \cite{viand2021sok}, the ciphertexts have to be rotated to obtain the encrypted results. Commonly, massive rotations are needed for FHE enabled matrix multiplication. Hence, this becomes the new performance bottleneck. It seems that we can not have the power of HE and high efficiency simultaneously.

%general MPC protocol \cite{demmler2015aby} is heavily used for calculating the input data intersection.
%3.在考虑两方的的方案中只有KDD21,CCS19和NIPS2三篇文章，但是依然受制于适用场景和效率问题。
However, the data sparsity is rarely utilized in schemes \cite{badsha2017privacy,jumonji2021privacy,huang2021more,chen2020secure} to promote the efficiency, not to mention specific customization for the corss-platform CF model. Thus, ROOM \cite{schoppmann2019make} introduces a novel cryptographic primitive, Read-Only-Oblivious Map, as a building block to achieve sparse matrix multiplication. Although data sparsity (only row/column sparsity) is somehow exploited, ROOM still suffers from large-volume communication and heavy computational load. Chen \textit{et al.} \cite{chen2021homomorphic} combines the FHE and secret sharing to enable multiplication for a sparse matrix (plaintext) and a dense matrix (encrypted). This method is custom designed for logistic regression where the client holds a small dense matrix and the server holds the model. Therefore, it only works well when one party's input is small and can hardly be extended for the large-scale dataset. The most related work to this paper is $\mathsf{S^{3}Rec}$ \cite{cui2021exploiting}. When the sparse locations are accessible, $\mathsf{S^{3}Rec}$ simply generates $O(\phi l\times m)$ Beaver’s triples \cite{hastings2019sok} to implement secure matrix multiplication, where $\phi$ is the density of the input matrix and $l, m$ are the dimensions. Such direct adoption of existing MPC scheme \cite{demmler2015aby} leads to unsatisfactory performance. When the sparse locations are agnostic, private information retrieval (PIR) \cite{angel2018pir} is used to fetch the non-spares values. To be compatible with PIR and preserve the confidentiality of the input dense matrix, each element has to be encrypted individually with PHE \cite{paillier1999public}, which results in massive additional computational cost. Therefore, a scheme that can enjoy the benefit of the packing method when working with PIR is desired.

%%%%%%%%%%%%%%%%%%%%%%%%%%%%%%%%%%%%%%%%%%%%%%%%%%%%%%%%%%%%%%%%%%%%%%%%%
\subsection{Technical Challenges}
\label{subsec:Technical challenges}
This paper aims to break the efficiency bottleneck of existing works and offer strong privacy preservation with provable security. We follow the tune of the state-of-the-art work \cite{cui2021exploiting}, which provides two schemes for sparse location insensitive and sensitive settings, respectively. However, it is non-trivial to conquer the current technical dilemma without seeking efficiency/privacy trade-offs. Through conducting a comprehensive analysis of recent advancements \cite{cui2021exploiting,schoppmann2019make}, we carefully condense out the following technical challenges.
\begin{itemize}
    %如何打包不旋转支持矩阵相乘即向量内积。
    \item \textit{How to enjoy the power of HE without impairing performance}? Existing works commonly use HE to implement matrix multiplication \cite{jiang2018secure}. Theoretically, arbitrary computation can be supported by HE over ciphertext. However, the powerful functionality is not free but costly. An effective method for computation/communication reduction is packing multiple messages into one message before encryption. As a side effect, existing works have to operate ciphertext rotations to obtain the encrypted vector inner product. Thus, massive rotations are needed when dealing with large matrices. Unfortunately, rotation is extremely expensive and consumes roughly $30\times$ more running time than the ciphertext multiplication \cite{huang2022cheetah}. This is a longstanding and challenging problem in related areas \cite{viand2021sok}. Significant performance gain will be achieved if we can design a rotation-free matrix multiplication scheme for cross-platform RSs.

    %如何在用PIR的时候数据库依然用HE打包来提速。如何压缩PIR本身的开销。
    \item \textit{How to compress the cost when PIR is applied}? In the sparse location sensitive setting, PIR is used for retrieving non-sparse elements without disclosing the queried location. To preserve the privacy of the queried matrix (dense matrix), each element has to be encrypted. Moreover, to compute the matrix multiplication, $\mathsf{S^{3}Rec}$ \cite{cui2021exploiting} chooses PHE to encrypt the dense matrix. As the elements have to be encrypted one by one due to the use of PIR, massive additional encryption costs are imposed on the participant. Straightforward adoption of existing packing methods can hardly support secure vector inner product not to mention matrix multiplication. Thus, how to bridge the gap between PIR and HE packing acceleration is vital and challenging. Furthermore, the underlying building block PIR is also constructed atop HE. It is non-trivial to compress the communication costs (upload and download volumes) on the basis of the current well-designed PIR protocol \cite{angel2018pir}.

    %在效率提高的同时，依然要可证明安全，然后使用了HE等技术，模型精度不该受影响。
    \item \textit{How to guarantee provable security and comparable accuracy}? We argue that this challenge indicates a vital and demanding requirement towards practicality. In spite of the charming performance promotion, the applied optimization methods should not undermine data privacy as well as model accuracy. In another word, we cannot adopt the approximate algorithm \cite{cheon2017homomorphic} for HE that will decrease the model accuracy. In terms of privacy, we cannot reveal additional information in exchange for better performance. Existing works \cite{chen2020secure,cui2021exploiting} suffer from either severe privacy risks or efficiency bottlenecks. Indeed, it is challenging to provide provable security and comparable accuracy beyond merely performance promotion.

\end{itemize}

%%%%%%%%%%%%%%%%%%%%%%%%%%%%%%%%%%%%%%%%%%%%%%%%%%%%%%%%%%%%%%%%%%%%%%%%%
\subsection{Our Contributions}
\label{subsec: our contributions}
In this paper, we propose two lean and fast sparse matrix multiplication schemes for RS model training with strong privacy preservation. In specific, $\mathsf{\Pi}_\mathrm{ins}$ stands for the scheme that can access the sparse locations in the input matrices, and $\mathsf{\Pi}_\mathrm{sen}$ denotes the scheme that sparse locations are agnostic. On addressing the above challenges, we make the following technical contributions.

\begin{itemize}
    \item We present $\mathsf{\Pi}_\mathrm{ins}$ that contributes two insights for efficiency boosting. First, we carefully analyze the computation task and convert it from standard matrix multiplication to Hadamard product \cite{horn2012matrix} between a dense matrix and an extremely spare matrix. This idea eliminates the costly rotation operations completely and can fully enjoy the high efficiency of the existing packing method. Second, to handle the case that we have to compute the vector inner product, a novel matrix packing method is adopted. In doing so, the ciphertext results can be extracted directly without rotation either.
    \item We present $\mathsf{\Pi}_\mathrm{sen}$ that conceals the sparse locations and enables efficient secure matrix multiplication simultaneously. We break through current performance bottlenecks by providing dual optimizations. The first new insight is using the packing based encryption acceleration method on the database (dense matrix) for PIR processing. To achieve this, we carefully design a new secure two-party sparse vector inner product protocol that for the first time bridges the gap between PIR and matrix packing. Second, the communication overheads brought by PIR including upload and download are further compressed by $2\times$ and $2.4\times$, respectively.
    \item Beyond boosting the efficiency, we provide formal security proofs for $\mathsf{\Pi}_\mathrm{ins}$ and $\mathsf{\Pi}_\mathrm{sen}$. In addition, extensive experiments are conducted on two popular testing datasets and two simulated large datasets. Compared with the existing effort, the proposed $\mathsf{\Pi}_\mathrm{ins}$ and $\mathsf{\Pi}_\mathrm{sen}$ compress the running time by at least $5\times$, and $2.8\times$, and achieve up to $15\times$ and $2.3\times$ in communication reduction, respectively.
\end{itemize}

%这个东西貌似不是很重要
%\textbf{Roadmap:} The remainder of this paper is organized as follows. In Section \ref{sec:technical background}, we review some building blocks used in this paper. In Section \ref{sec:problem statement}, we describe the system/threat models and work flow for the proposed schemes. Afterward, the technical details are given in Section \ref{sec:proposed scheme}. The security analysis and the performance evaluation are provided in Section \ref{sec:security analysis} and Section \ref{sec:performance evaluation}, respectively. Section \ref{sec:conlusion and future work} concludes the paper.

\section{Background}
\label{sec:technical background}
In this section, we first define the notations. Then, we give a brief introduction to the recommendation model and the related cryptographical tools, that serve as the building blocks of the proposed scheme.

\textbf{Notations.} We use the bold upper-case letters to denote the matrices (e.g., $\mathbf{M}$). The vectors are denoted as bold lower-case letters (e.g., $\mathbf{v}$). The element of $i$-th row and $j$-th column in matrix $\mathbf{M}$ is written as $\mathbf{M}[i,j]$. The $k$-th component of vector $\mathbf{v}$ is $\mathbf{v}[k]$. $[a]$ stands for the integer set $\{0,...,a-1\}$. We denote by lower-case letter with a circumflex symbol to represent a polynomial, such as $\widehat{m}$. The $i$-the coefficient of $\widehat{m}$ is written as $\widehat{m}[i]$. Given 2-power number $N$ and $q$ ($q>0$), let $R_{N,q}=\mathbb{Z}_{q}[X]/(X^{N}+1)$ to denote the integer polynomial set. Given two polynomials $\widehat{m}, \widehat{n} \in R_{N,q}$, the product $\widehat{s}=\widehat{m}\cdot \widehat{n} \in R_{N,q}$ is defined as
\begin{equation}
\label{product of polynomials}
\widehat{s}[i] = \sum_{0\le j \le i} \widehat{m} [j]\widehat{n}[i-j] - \sum_{i\le j \le N} \widehat{m}[j] \widehat{n}[N-j+i]\ \text{mod}\ q.
\end{equation}

\subsection{Recommendation Model}
\label{subsec:recommendation model}

%%推荐系统概念说明图。
\begin{figure}[H]
\centering
\includegraphics[width=0.2\textwidth]{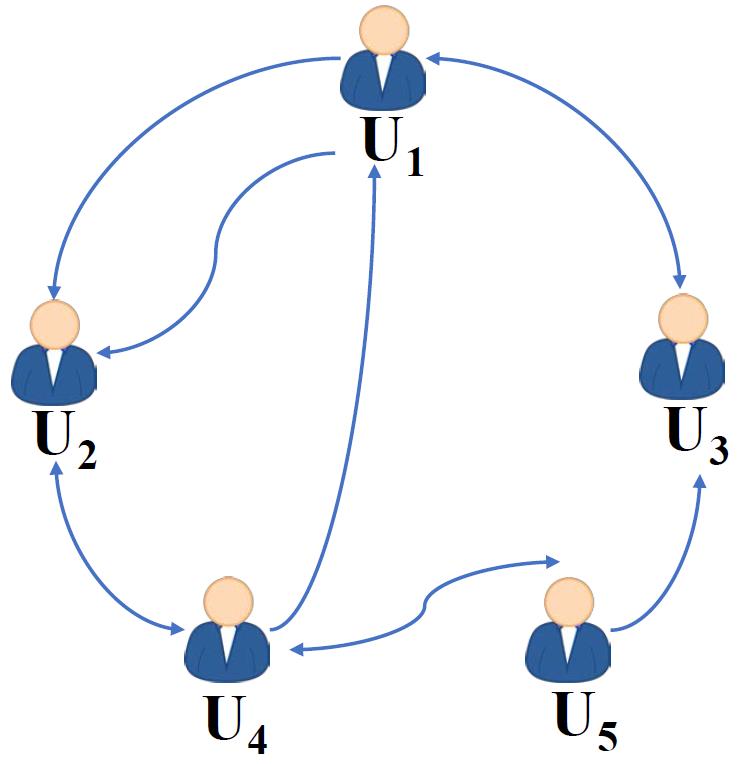}
\caption{A Toy Example of Social Network.}
\label{Fig.social network}
\end{figure}

%先将最基本的基于举证分解的推荐模型。然后，讲述社交网络信息融入推荐模型。最后讲优化目标的计算内容，加入两方的场景。

A classic and effective method \cite{ma2011recommender} \cite{tang2013social} to build a recommender system is to factorize the rating matrix $\mathbf{R}$ to obtain a user-specific matrix $\mathbf{U}$ and an item-specific matrix $\mathbf{V}$. The system then makes missing data prediction atop $\mathbf{U}$ and $\mathbf{V}$. To provide a more personalized and accurate prediction service, it is common to incorporate the data from social networks among users. This method is often termed as ``social recommender system" \cite{ma2011recommender}. The basic intuition of this method is easy to capture. The user's preference is likely to be similar to one's close friends. Thus, if the social data is embedded as the regularization constraint, the prediction results can be significantly improved \cite{bobadilla2013recommender}. Fig. \ref{Fig.social network} uses a directed graph to describe the topology of a social network, which can be characterized by adjacency matrix \cite{henry2007nodetrix}. The social matrix will be fed into the model training process.

In this paper, the main target is to boost the efficiency of privacy-preserving computation for the social recommender system, so we follow the state-of-the-art scheme \cite{cui2021exploiting} that uses the classic model \cite{ma2011recommender}. In specific, given the rating matrix $\mathbf{R}\in \mathbb{R}^{m\times n}$ and the social matrix $\mathbf{S}\in \mathbb{R}^{m\times m}$, the model's learning target is to obtain $\mathbf{U} \in \mathbb{R}^{l\times m}$ and $\mathbf{V} \in \mathbb{R}^{l\times n}$ through optimizing the objective function $\mathcal{L}$.
\begin{equation}
\label{equation:loss function}
\begin{split}
         \mathcal{L} &= \mathop{\mathbf{min}}\limits_{\mathbf{U},\mathbf{V}}\frac{1}{2} \sum_{i=1}^{m} \sum_{j=1}^{n} \mathbf{I}[i,j](\mathbf{R}[i,j]-\mathbf{U}[*,i]^{T}\mathbf{V}[*,j])^{2} \\
    &+\frac{\alpha}{2} \sum_{i=1}^{m} \sum_{f=1}^{m}\mathbf{S}[i,f]\|\mathbf{U}[*,i]- \mathbf{U}[*,f]\|_{F}^{2} \\
    &+\frac{\beta}{2}(\sum_{i=1}^{m}\|\mathbf{U}[*,i]\|_{F}^{2}+\sum_{j=1}^{n}\|\mathbf{V}[*,j]\|_{F}^{2}).
\end{split}
\end{equation}

In the function $\mathcal{L}$, the first term is the factorization of rating matrix $\mathbf{R}$, the second term indicates the social information, the last term is the regularizer. Matrix $\mathbf{I}[\cdot]$ records the rated items, $\alpha, \beta$ are hyper-parameters, and $\|\cdot\|_{F}^{2}$ is the Frobenius norm. Normally, we adopt gradient descent to solve $\mathcal{L}$ \cite{ma2011recommender}. Assume the diagonal matrix $\mathbf{A} \in \mathbb{R}$ with diagonal elements $a_i = \sum_{j=1}^{m}\mathbf{S}[i,j]$, the diagonal matrix $\mathbf{B} \in \mathbb{R}$ with diagonal elements $b_j=\sum_{i=1}^{m}\mathbf{S}[i,j]$. Let $\mathbf{D}=\mathbf{A}^{T}+\mathbf{B}^{T} $, then gradients of $\mathcal{L}$ can be written as:
\begin{equation}
\label{equation:partial U}
    \frac{\partial \mathcal{L}}{\partial \mathbf{U}} = \beta \mathbf{U}-\mathbf{V}((\mathbf{R}-\mathbf{U}^{T}\mathbf{V})^{T}\cdot \mathbf{I}) + (\frac{\alpha}{2}\mathbf{U}\mathbf{D}-\alpha\mathbf{U}\mathbf{S}^{T}),
\end{equation}
\begin{equation}
\label{equation:partial V}
    \frac{\partial \mathcal{L}}{\partial \mathbf{V}} = \beta \mathbf{V}-\mathbf{U}((\mathbf{R}-\mathbf{U}^{T}\mathbf{V})^{T}\cdot \mathbf{I}).
\end{equation}

Given the gradients of $\mathcal{L}$, the problem is boiled down to computing the matrix products and additions. Recall that, in this paper, the social matrix $\mathbf{S}$ and rating matrix $\mathbf{R}$ are held by two different platforms (i.e., party $\mathsf{P_0}$ has $\mathbf{R}$, party $\mathsf{P_1}$ has $\mathbf{S}$). $\mathsf{P_0}$ can compute first term of ${\partial \mathcal{L}}/{\partial \mathbf{U}}$ and ${\partial \mathcal{L}}/{\partial \mathbf{V}}$ locally. While $\mathsf{P_0}$ and $\mathsf{P_1}$ need to compute the second term of ${\partial \mathcal{L}}/{\partial \mathbf{U}}$ collaboratively in privacy-preserving way.

\subsection{Cryptographical Tools}
\label{subsec:cryptographical tools}
%1.加法秘密分享
%2.基于格密码的两种HE(RLWE, LWE)+半同态基于群的同态， paillier.
%3.single server PIR concept.
\noindent \textbf{Arithmetic Secret Sharing (SS).} SS \cite{demmler2015aby} is a fundamental technique used for secure multiparty computation (MPC). In this paper, we consider the two-party scenario. For example, $\mathsf{P_0}$ has a message $m$ in prime field $\mathbb{Z}_{p}$, and randomly samples $\left \langle m \right \rangle _{0} \in \mathbb{Z}_{p}$ as his share. Then, it computes $\left \langle m \right \rangle _1 = m- \left \langle m \right \rangle _0 \ \text{mod}\ p$ as $\mathsf{P_1}$'s share. To recover $m$, $\mathsf{P_0}$ and $\mathsf{P_1}$ computes $m = \left \langle m \right \rangle _{0} + \left \langle m \right \rangle _{1}\ \text{mod}\ p$. For simplicity, we omit the mod operation if the context is clear.

\noindent \textbf{Homomorphic Encryption (HE).} HE \cite{viand2021sok} generated ciphertexts enable versatile evaluations without decryption during the processing. According to the evaluation functionalities, HE schemes can be categorized into three types, that are partial HE (PHE), somewhat HE (SHE), and fully HE (FHE). In this paper, we use PHE \cite{paillier1999public} and lattice-based SHE \cite{viand2021sok} schemes to implement the proposed secure protocols. A typical addition PHE crypto-system, such as Paillier \cite{paillier1999public}, involves a pair of public and private keys $\{\mathsf{pk_P}, \mathsf{sk_P}\}$ and encryption/decryption algorithms $\{\mathsf{P.Enc}, \mathsf{P.Dec}\}$. Normally, $\mathsf{pk_P}$ is used to encrypt messages and $\mathsf{sk_P}$ is used for decryption. Given two messages $x,y$, Paillier encryption offers the following functions.

\begin{itemize}
\item Addition homomorphism ($\oplus$):\\
$\mathsf{P.Enc}(\mathsf{pk_P}, x+y) \triangleq \mathsf{P.Enc}(\mathsf{pk_P}, x)\oplus  \mathsf{P.Enc}(\mathsf{pk_P}, y)$.
\item Ciphertext-plaintext multiplication ($\otimes$):\\
$\mathsf{P.Enc}(\mathsf{pk_P}, x\cdot y) \triangleq \mathsf{P.Enc}(\mathsf{pk_P}, x)\otimes  y$.
\end{itemize}
The symbol $\triangleq$ indicates that two ciphertexts can be decrypted to the same plaintext, not numerically equal.

In this paper, we also apply lattice-based HE that is constructed atop the learning with errors (LWE) problem \cite{gentry2013homomorphic} or its ring variant (RLWE) \cite{fan2012somewhat}. These two types of HE schemes share the same public parameters $\textsf{HE.pp} = \{N,p,q,\sigma\}$, where $p,q \in \mathbb{Z}; q\gg p>0$, and $\sigma$ is the standard deviation of a discrete Gaussian distribution used for error sampling. In the RLWE scheme, the plaintext message is a polynomial in $R_{N,p}$. An RLWE scheme comprises three algorithms denoted by $\mathsf{\{R.KeyGen, R.Enc, R.Dec\}}$. In specific, $\mathsf{R.KeyGen}$ generates the secret and public keys $\{\mathsf{pk_R},\mathsf{sk_R}\} \in R_{N,q}$. We can invoke $\mathsf{R.Enc}$ to encrypt the message $\widehat{m}\in R_{N,p}$, and obtain its ciphertext $\mathsf{CT}\gets \mathsf{R.Enc}(\mathsf{pk_R}, \widehat{m})$, where $\mathsf{CT}\in R_{N,q}^{2}$. The decryption algorithm $\mathsf{R.Dec}$ takes the secret key $\mathsf{sk_R}$, the ciphertext $\mathsf{CT}$ as the input, and outputs the plaintext $\widehat{m}$. For LWE scheme, the plaintext space is $\mathbb{Z}_p$, and the ciphertext space is $\mathbb{Z}_{q}^{N+1}$. The syntax of LWE scheme is similar to RLWE, we write it as a tuple $\mathsf{\{L.KeyGen, L.Enc, L.Dec\}}$, which represents the key generation, encryption, and decryption algorithm respectively. The generated key pair is denoted as $\{\mathsf{pk_L},\mathsf{sk_L}\}\in R_{N,q}$. Interested readers can refer to the literatures \cite{fan2012somewhat, gentry2013homomorphic} for the technical details. Note that, existing LWE and RLWE based SHE schemes can be extended to FHE with bootstrapping module \cite{cheon2017homomorphic, viand2021sok}. In this paper, only linear homomorphic evaluations are applied, which are also termed SHE or linear HE \cite{acar2018survey}. For our purpose, we mainly focus on the following functions supported by RLWE scheme.

\begin{itemize}

\item Addition ($\boxplus$) and subtraction ($\boxminus$) homomorphism:\\
Given two plaintexts $\widehat{m}_1, \widehat{m}_2 \in R_{N, p}$, and their ciphertexts $\mathsf{CT_1}, \mathsf{CT_2}$, we have $\mathsf{R.Enc}(\mathsf{pk_R}, \widehat{m}_1+\widehat{m}_2) \triangleq \mathsf{CT_1}\boxplus  \mathsf{CT_2}$, and
$\mathsf{R.Enc}(\mathsf{pk_R}, \widehat{m}_1-\widehat{m}_2) \triangleq \mathsf{CT_1}\boxminus  \mathsf{CT_2}$.

\item Multiplication homomorphism ($\boxtimes$): \\
For message $\widehat{m}_1, \widehat{m}_2 \in R_{N, p}$, and the corresponding ciphertexts $\mathsf{CT_1}, \mathsf{CT_2}$, we have $\mathsf{R.Enc}(\mathsf{pk_R}, \widehat{m}_1\cdot \widehat{m}_2) \triangleq \widehat{m}_1\boxtimes \mathsf{CT_2}$, and $\mathsf{R.Enc}(\mathsf{pk_R}, \widehat{m}_1\cdot \widehat{m}_2) \triangleq \mathsf{CT_1}\boxtimes \mathsf{CT_2}$. Note that, the ciphertext-ciphertext and plaintext-ciphertext multiplication are different in the calculation. For simplicity, we use the same symbol $\boxtimes$ to represent them.

\item Extraction, $\mathsf{HE.Extract(\mathsf{CT}}, i)$: \\
For the message $\widehat{m}$ and its ciphertext $\mathsf{CT}$, this function can extract the $i$-th coefficient of $\widehat{m}$ from its ciphertext, and transfer it to a LWE format ciphertext. The corresponding LWE secrete (decryption) key is derived by a key switch algorithm. Only the specific required coefficient is revealed, which guarantees no extra information leakage incurred. Thus, this function is pretty elegant. Interested readers may refer to the literature \cite{chen2021efficient}.

%\item Substitution, $\mathsf{HE.Sub(\mathsf{CT}}, k)$: \\
%Given a message $\widehat{m}(x)$, its ciphertext $\mathsf{CT}$ and an odd integer $k$, this function can return an ciphertext of $\widehat{m}(x^{k})$. For instance, if $\widehat{m}(x) = 5+6x+7x^{3}$, then $\mathsf{HE.Sub(\mathsf{CT},} 3)$ will return an ciphertext of message $\widehat{m}(x^{3}) = 5+6x^{3}+7(x^{3})^{3} = 5+6x^{3}+7x^{9}$. Therefore, this function can move the encrypted coefficients to other locations.

\end{itemize}

%the Schematic diagram of PIR
\renewcommand\tablename{Fig}
\renewcommand \thetable{\arabic{table}}
\setcounter{table}{1}
\begin{table}
\centering
\begin{tabular}{|p{7cm}|}
\hline
 \textbf{Client} \ \ \ \ \ \  \ \ \ \ \ \ \ \ \  \ \ \ \ \ \ \ \ \ \ \ \ \ \ \textbf{Server} \\
 $q\leftarrow \mathsf{PIR.Query}(i)$ \ \ \ \ $\stackrel{q}{\longrightarrow}$   \\
 \ \ \ \ \ \ \ \ \ \ \ \ \ \ \ \ \ \ \ \ \ \  \ \ \ \ \ \  $\stackrel{r}{\longleftarrow}$ \ \ \ \ $r\leftarrow \mathsf{PIR.Response}(q, \mathsf{DB})$\\
  $d_i\leftarrow \mathsf{PIR.Extract}(r)$ \\
\hline
\end{tabular}
\vspace{2pt}
\caption{An overview of non-interactive PIR protocol.}
\label{figure:PIR protocol}
\end{table}

\noindent \textbf{Private Information Retrieval (PIR).} PIR \cite{melchor2016xpir} enables a client to send an encrypted query to the server, then the server returns the result without knowing the queried index. In this way, the query privacy is preserved. In this paper, we consider the single server setting \cite{angel2018pir}. Assume the server holds a database with $n$ elements denoted as $\mathsf{DB} = \{d_1,...,d_n\}$, and a client with the query index $i$. The classic PIR construction comprises the following three algorithms.

\begin{itemize}

\item $q\leftarrow \mathsf{PIR.Query}(i)$: the client runs this algorithm to obtain an encrypted query for the chosen index $i$, and send it to the server.

\item $r\leftarrow \mathsf{PIR.Response}(q, \mathsf{DB})$: upon receiving the encrypted query $q$, the server invokes this algorithm to compute the encrypted query response $r$ through the database $\mathsf{DB}$.

\item $d_i\leftarrow \mathsf{PIR.Extract}(r)$: this algorithm let the client extract the queried item ($i$-th item) from the returned response.

\end{itemize}

The non-interactive single server PIR protocol is illustrated in Fig. \ref{figure:PIR protocol}. Optimizations and trade-offs on this protocol are intensively investigated. In specific, existing schemes \cite{melchor2016xpir, angel2018pir,ali2021communication} mainly focus on compressing the upload and download communication costs, reducing and amortizing the server-side computational load. The applied optimization methods for this scheme will be introduced in Section \ref{sec:proposed scheme}.

\section{Problem Statement}
\label{sec:problem statement}
%这章主要内容：1. 系统模型以及主要要解决的问题。2. 威胁模型。

In this section, we elaborate on the system model of the proposed scheme. With the given model, we specify the work flow between the social platform and the rating platform. Afterward, we clarify the threat model and illustrate the potential risks of the system.

\subsection{System Model and Work Flow}
\label{subsec:system model and work flow}
\noindent \textbf{System Model.} The proposed scheme consists of two parties which are the rating platform and the social platform. Here, we use the same notations as the Section \ref{subsec:recommendation model}. $\mathsf{P_0}$ denotes the rating platform and $\mathsf{P_1}$ is the social platform. In the real world, $\mathsf{P_0}$ could be any E-commerce or advertising company. It holds the online shopping records and comments of users that can be represented as a rating matrix $\mathbf{R}$. $\mathsf{P_1}$ could be any social media such as Facebook, Wechat, etc. The relationship between users is characterized as a social matrix $\mathbf{S}$, which is highly sparse in nature. Since both parties are companies, they should be able to conduct computation intensive tasks like encryption/decryption and evaluation over the ciphertext. Note that the model can be extended to multi-party settings by incorporating MPC protocols.

\begin{table}
\centering
\begin{tabular}{|p{8cm}|}
\hline
\textbf{Globe Parameters:} Hyperparameters $\alpha$, $\beta$, learning rate $\eta$.   \\
\textbf{Input:} The rating matrix $\mathbf{R}$, the social matrix $\mathbf{S}$.  \\
\textbf{Output:} Return the user latent matrix $\mathbf{U}$, item latent matrix $\mathbf{V}$ to $\mathsf{P_0}$. \\
1. $\mathsf{P_0}$ initializes matrix $\mathbf{U}$ and $\mathbf{V}$.\\
2. \textbf{While} (not coverage), \\
3.\ \ \ $\mathsf{P_0}$ locally computes: \\
\ \ \ \ \ $\mathrm{T_1} \leftarrow \beta \mathbf{U}-\mathbf{V}((\mathbf{R}-\mathbf{U}^{T}\mathbf{V})^{T}\cdot \mathbf{I})$, \\
\ \ \ \ \ $\frac{\partial \mathcal{L}}{\partial \mathbf{V}} \leftarrow  \beta \mathbf{V}-\mathbf{U}((\mathbf{R}-\mathbf{U}^{T}\mathbf{V})^{T}\cdot \mathbf{I})$, \\
4.\ \ \ $\mathsf{P_0}$ AND $\mathsf{P_1}$ securely compute and share the result: \\
\ \ \ \ \ $\{\langle \mathrm{T_2} \rangle_0, \langle \mathrm{T_2} \rangle_1\} \leftarrow   \frac{\alpha}{2}\mathbf{U}\mathbf{D}-\alpha\mathbf{U}\mathbf{S}^{T}$, \\
\ \ \ \ \ \ \ \ \ // \textit{the cryptographical tools are applied}\\
5.\ \ \ $\mathsf{P_0}$ AND $\mathsf{P_1}$ computes: \\
\ \ \ \ \ $\mathbf{U}\leftarrow \mathbf{U}-\eta(\mathrm{T_1}+(\langle \mathrm{T_2} \rangle_0+\langle \mathrm{T_2} \rangle_1))$, \\
6.\ \ \ $\mathsf{P_0}$ locally computes: \\
\ \ \ \ \ $\mathbf{V}\leftarrow \mathbf{V}-\eta\frac{\partial \mathcal{L}}{\partial \mathbf{V}}$, \\
7. \textbf{Endwhile} \\
8. \textbf{return} $\mathbf{U}$ and $\mathbf{V}$ to party $\mathsf{P_0}$. \\
\hline
\end{tabular}
\vspace{2pt}
\caption{An overview of work flow.}
\label{figure:work flow}
\end{table}

\noindent \textbf{Work Flow.} As shown in Fig. \ref{figure:work flow}, we sketch the work flow step by step. The notations are exactly the same as Section \ref{sec:technical background}. The main task of both parties is to obtain the recommendation model in a privacy-preserving way. Specifically, $\mathsf{P_0}$ and $\mathsf{P_1}$ collaboratively calculate the factorization of the rating matrix $\mathbf{R}$ through optimizing the objective function $\mathcal{L}$. In another word, the optimization goal is to seek a pair of matrices $\{\mathbf{U}$, $\mathbf{V}\}$ whose production is an approximation of $\mathbf{R}$, i.e., $(\mathbf{R}\approx \mathbf{U}^{T}\cdot \mathbf{V})$. The used optimization method is gradient descent, then the problem is converted to calculating $\mathcal{L}$ in a privacy-preserving way. As discussed in Section \ref{subsec:recommendation model}, the first term of ${\partial \mathcal{L}}/{\partial \mathbf{U}}$, and ${\partial \mathcal{L}}/{\partial \mathbf{V}}$ can be computed locally by $\mathsf{P_0}$ without interacting with $\mathsf{P_1}$. However, the second term of ${\partial \mathcal{L}}/{\partial \mathbf{U}}$ contains both social and rating data. Therefore, to preserve data privacy, it needs to be collaboratively evaluated by $\mathsf{P_0}$ and $\mathsf{P_1}$ using the cryptographical tools introduced in Section \ref{subsec:cryptographical tools}. This corresponds to Step 4 in Fig. \ref{figure:work flow}. In this paper, two protocols with different information leakage settings are designed to fully explore data sparsity.

\subsection{Threat Model}
\label{subsec:threat model}
We argue that the threat model should match the real application scenario rather than pursuing an extremely high security level. In practice, heavy protection mechanisms often incur unacceptable efficiency degradation. On one hand, the essential motivation of this paper is to boost the efficiency of privacy-preserving recommender systems. On the other hand, the model accuracy directly affects the economical benefits of both social and rating platforms. Therefore, both parties have no interest in maliciously manipulating the data or deviating from the protocol. Considering this, we adopt the \textit{semi-honest} (i.e., honest-but-curious) threat model \cite{chen2021homomorphic}, which is the same as the-state-of-the-art work \cite{cui2021exploiting}. In specific, the probabilistic polynomial-time adversary can compromise one of the parties (non-conclusion) \cite{mohassel2017secureml} and observe the input/output view. The adversary aims to infer private information from the honest party by analyzing the corrupted party's view. This assumption is practical and widely applied to real-world scenarios \cite{chen2021homomorphic} that have privacy concerns.

%这部分貌似不是很有必要
%\subsection{Design Goals}
%\label{subsec:design goals}
%T The main design goals are two-fold: provable security and high efficiency. In this part, we break down the general goals into the following %several concrete targets.

%\textbf{High efficiency :} With comparable privacy protection strength, \cite{cui2021exploiting}, we aim to achieve high efficiency through in-depth technical analysis and modification. Besides, effective trade-offs between storage, computational and communication costs should be investigated. In specific, we follow the criterion \cite{ali2021communication} that additional overheads should bring about compelling improvement in overall performance. \textcolor[rgb]{0.00,0.00,1.00}{add the concrete performance gains when exp results comes out....}

%\textbf{Provable security :} The primary task is to conduct a comprehensive analysis of information leakage of the proposed protocols. Informally, in terms of privacy protection, the best-possible design should disclose no additional information except the allowed leakage. In this paper, we propose to offer different information leakage levels to meet the diverse privacy requirements. To support formal security proof, the ideal functions of protocols should be correctly defined. Then, we follow the simulator paradigm \cite{mohassel2017secureml} to prove the security by constructing an indistinguishable simulator, which is a non-trivial task.

%1. 概括整个章节的内容， 2. 给具体的两个不同信息信息泄露水平的方案一个overview. 3. 分别给出两个具体的技术设计内容  4.最后看篇幅是否需要给一个总结，或者是put together.

\section{Proposed Scheme}
\label{sec:proposed scheme}
In this section, we elaborate on the technical details of proposed protocols, that serve for two different leakage settings (i.e., $\mathsf{\Pi}_\mathrm{ins}$, $\mathsf{\Pi}_\mathrm{sen}$). As the key insight, we need to fully explore the sparsity of the social and rating data to promote performance. This operation may introduce mild additional information leakage. As discussed in Section \ref{subsec:threat model}, perfect privacy often imposes heavy computational burden to both parties \cite{cui2021exploiting}. Thus, in terms of concrete design, subtle trade-offs are made to mitigate such a dilemma. In the following subsections, we first give a scheme overview and then describe the privacy-preserving protocols. The optimization tricks and insights are presented appropriately in this section.

\subsection{Scheme Overview}
\label{subsec:scheme overview}
In this paper, two secure and highly efficient schemes are proposed. The first one is designed for the scenario that the data sparse location is insensitive. As discussed in work \cite{cui2021exploiting}, this information can be fully applied to promote efficiency. The second scheme aims to conceal the sparse locations from another party while supporting the same functionality as the insensitive case. For example, assume that party $\mathsf{P_1}$ holds a sparse matrix $\mathbf{Y}\in \mathbb{R}^{m\times m}$. The non-sparse locations can be denoted as a set (also can be written as a vector) $\mathbf{loc}\leftarrow \{(i,j)|\mathbf{Y}[i,j]\neq 0; i,j\in [m]\}$. Assume that party $\mathsf{P_0}$ has dense matrix $\mathbf{X}\in \mathbb{R}^{n\times m}$. As shown in Fig. \ref{figure:work flow} (Step 4), $\mathsf{P_0}$ and $\mathsf{P_1}$ need to conduct secure matrix multiplication $\mathbf{X}\cdot \mathbf{Y}$. In location insensitive scheme, $\mathsf{P_1}$ shares $\mathbf{loc}$ with $\mathsf{P_0}$. While in location sensitive scheme only vector size $|\mathbf{loc}|$ are revealed to $\mathsf{P_0}$. In practice, the general sparsity level (i.e., $|\mathbf{loc}|$) is often regarded as a public statistic \cite{cui2021exploiting}. In this view, sharing of $|\mathbf{loc}|$ will not brings about additional privacy leakage. To make the technical details easier to follow, we itemize the basic steps (i.e., Step 4 in Fig. \ref{figure:work flow}) for two schemes. Note that, we omit the operations that are conducted by $\mathsf{P_0}$ or $\mathsf{P_1}$ locally.

The \textit{location insensitive scheme} is dubbed as $\mathsf{\Pi}_\mathrm{ins}$. We achieve $\mathsf{\Pi}_\mathrm{ins}$ as follows.

%泄露位置方案的概括： 1. 用同态加密产生各种参数并公开PK。 此处的公钥无需分享。
%2.利用矩阵S产生D，并且将D明文下打包加密（SV）,SV是基础，是基于SV的SV打包,加密之前需要收到U的稀疏位置（此处是列稀疏），要充分利用双方的稀疏性来压缩计算开销。发送给P0后，P0在密文上计算UD并以秘密分享的方式把这步的结果发送给P1.
%3. P1将S压缩（如何充分利用稀疏性，减少密文数量需要单独设计个小trick）打包加密（cheetah）并发送给P0，可用矩阵整体打包（这个其实很trivial）。
%4. P0在密文上计算US。然后同样以秘密分享的方式返回结果。

\begin{enumerate}
  \item $\mathsf{P_1}$ invokes the RLWE based HE scheme to generate the public/private keys $\{\mathsf{pk_R},\mathsf{sk_R}\} \in R_{N,q}$. The model training public parameters (see Fig. \ref{figure:work flow}) are generated by $\mathsf{P_0}$. The cryptographic related public parameters (see Section \ref{subsec:cryptographical tools}) are generated by $\mathsf{P_1}$. In addition, $\mathsf{P_0}$ needs to share the \textit{non-sparse locations} (equivalent to sparse locations) of matrix $\mathbf{U}$, written as $\mathbf{loc}_{\mathbf{U}}$, with $\mathsf{P_1}$.

  \item $\mathsf{P_1}$ generates the diagonal matrix $\mathbf{D}$ atop the social matrix $\mathbf{S}$. By checking $\mathbf{loc}_{\mathbf{U}}$, $\mathsf{P_1}$ can directly delete the corresponding elements. For a simple example, if the $j$-th column of $\mathbf{U}$ is sparse, the element $\mathbf{D}[j,j]$ can be set as $0$ (i.e., deleted). Afterward, the SV packing method \cite{smart2014fully} (designed based on Chinese Remainder Theory) is applied to further compress the ciphertext size of $\mathbf{D}$. Then, $\mathsf{P_1}$ uses $\mathsf{pk_R}$ to encrypt the compressed and packed $\mathbf{D}$. At last, the ciphertext will be sent to $\mathsf{P_0}$. Note that, the packing size is shared as a public parameter.

  \item $\mathsf{P_0}$ deletes the sparse elements of $\mathbf{U}$, and computes $\mathbf{U}\cdot \mathbf{D}$, by utilizing the multiplication homomorphism property of RLWE-based HE. The result is then masked and split into two secret shares. $\mathsf{P_0}$ keeps one share and sends the other to party $\mathsf{P_1}$.

  \item $\mathsf{P_1}$ shares the non-sparse locations of $\mathbf{S}$ (written as $\mathbf{loc}_\mathbf{S}$) with $\mathsf{P_0}$. After deleting the sparse elements, $\mathsf{P_1}$ packs $\mathbf{S}^{T}$ by mapping its elements to the coefficients of ring polynomials. The packed matrix will be encrypted in exactly the same way as Step 2 of $\mathsf{\Pi}_\mathrm{ins}$. Similarly, ciphertext should be sent to $\mathsf{P_0}$.

  \item According to the sparse locations of $\mathbf{S}$, $\mathsf{P_0}$ deletes the sparse elements of $\mathbf{U}$, then computes $\mathbf{U}\cdot \mathbf{S}^{T}$. The result is also masked and split into two secret shares. $\mathsf{P_0}$ keeps one share and sends the other to party $\mathsf{P_1}$.

  \item At last, party $\mathsf{P_0}$ and $\mathsf{P_1} $ collaboratively reconstruct the final calculation result of $(\alpha \mathbf{U}\mathbf{D}/2-\alpha\mathbf{U}\mathbf{S}^{T})$.
\end{enumerate}

The \textit{location sensitive scheme} is dubbed as $\mathsf{\Pi}_\mathrm{sen}$. We achieve $\mathsf{\Pi}_\mathrm{sen}$ as follows.

%不泄露位置方案的概括： 1. 用同态加密产生各种参数并公开PK。
%2.利用矩阵S产生D，并且将D明文下打包加密（SV）,这一步和ins方案的唯一区别是无法利用U的稀疏性来压缩计算开销。发送给P0后，P0在密文上计算UD并以秘密分享的方式把这步的结果发送给P1.
%3. P1利用优化后的PIR查询U的(到底是哪一方查询是一个trade-off,还有稀疏性和打包规模也是需要平衡的。这个可以展开稍微叙述一下。)，获得查询结果（依然是个同态密文）。在结果上做内积，数据库可以打包加密。
%4. 另一方拿到查询结过后需要进行明文和密文的乘积并累加。最后以秘密分享的形式返回结果。

\begin{enumerate}
  \item $\mathsf{P_0}$ generates the PHE private and public key pair, $\mathsf{P_1}$ generates the RLWE HE private and public key pair. The public parameters are set and shared in the same way as the first step of $\mathsf{\Pi}_\mathrm{ins}$.

  \item $\mathsf{P_1}$ obtains the diagonal matrix $\mathbf{D}$. Then, $\mathsf{P_1}$ directly packs $\mathbf{D}$ (SV) and encrypts (RLWE) it using the same method as $\mathsf{\Pi}_\mathrm{ins}$. The ciphertext will be sent to $\mathsf{P_0}$.

  \item $\mathsf{P_0}$ computes $\mathbf{U}\cdot \mathbf{D}$, and forwards the encrypted result (secrete share) with $\mathsf{P_1}$. One of the shares is kept by $\mathsf{P_0}$.

  \item $\mathsf{P_1}$ leverages the optimized PIR methods to fetch the elements of $\mathbf{U}$ from $\mathsf{P_0}$ . To preserve the privacy of $\mathbf{U}$, $\mathsf{P_0}$ adopts the SV packing method and PHE to encrypt $\mathbf{U}$. $\mathsf{\Pi}_\mathrm{sen}$ proposes a packing-compatible secure vector inner product method for matrix multiplication.

  \item Upon receiving the query result, $\mathsf{P_1}$ calculates and remasks $\mathbf{U}\cdot \mathbf{S}^{T}$ by applying the homomorphic property of PHE. Afterward, $\mathsf{P_1}$ sends a secret share of the encrypted result to $\mathsf{P_0}$. Another share is kept by $\mathsf{P_1}$.

  \item  Same as $\mathsf{\Pi}_\mathrm{ins}$, $\mathsf{P_0}$ and $\mathsf{P_1} $ collaboratively reconstruct the plaintext result of $(\alpha \mathbf{U}\mathbf{D}/2-\alpha\mathbf{U}\mathbf{S}^{T})$.
\end{enumerate}

$\mathsf{\Pi}_\mathrm{ins}$ and $\mathsf{\Pi}_\mathrm{sen}$ are presented in detail in the following two subsections. In addition to the technical designs, we also show the merits of our proposed optimization tricks.

%按照协议流程把方案的具体设计写出来，然后点一下优化的部分，以及优化方案背后的思考。 按照implementation of 1 and 2的方式写。
%整个协议大体分为两个部分。第一个部分实现对角矩阵乘法UD。第二部分为实现矩阵乘法US。
\subsection{Sparse Location Insensitive Scheme $\mathsf{\Pi}_\mathrm{ins}$}
\label{subsec:Sparse Location Insensitive Scheme}
%需要单独描述的工具：普通矩阵和对角矩阵乘积的计算方法介绍，就是为啥可以用SV。SV的打包和解包。以及其和SIMD的关系； cheetah方案的打包方案，矩阵编码方案。

In this part, we illustrate the technical details of $\mathsf{\Pi}_\mathrm{ins}$. Except for the basic cryptographic tools introduced in Section \ref{subsec:cryptographical tools}, several advanced computing acceleration techniques are applied either. Besides, we also fully explore the sparsity and the linear algebra tricks to co-design the optimization methods.

The first task of $\mathsf{\Pi}_\mathrm{ins}$ is to securely compute $\mathbf{U}\mathbf{D}$. Since $\mathbf{D}$ is a diagonal matrix, it can be regarded as an extremely sparse matrix that only has one element in each row (column). To take advantage of this character, we can convert this problem to Hadamard product \cite{horn2012matrix} between $\mathbf{U}$ and $\mathbf{D}$ if the diagonal elements of $\mathbf{D}$ are noted as a vector. For instance, given two vectors $\mathbf{x}$ and $\mathbf{y}$ with $m$ elements, the Hadamard product can be written as $\mathbf{x}\star \mathbf{y}=(\mathbf{x}[0]\cdot \mathbf{y}[0],...,\mathbf{x}[m-1]\cdot \mathbf{y}[m-1])$. Then, let vector $\mathbf{d}[i]=\mathbf{D}[i, i], i\in [m]$ and $\mathbf{U} \in \mathbb{R}^{l\times m}$, $\mathbf{U}\mathbf{D}$ is computed as follows.
\begin{equation}\label{entrywise multiplication of UD}
\mathbf{U}\mathbf{D} =
\begin{bmatrix}
\mathbf{U}[0,*]\star \mathbf{d} \\
\mathbf{U}[1,*]\star \mathbf{d} \\
... \\
\mathbf{U}[l-1,*]\star \mathbf{d}
\end{bmatrix}
\end{equation}

The Equation \ref{entrywise multiplication of UD} indicates that the computation cost of $\mathbf{U}\mathbf{D}$ can be further reduced if we consider the sparsity of matrix $\mathbf{U}$. Upon receiving $\mathbf{loc}_{\mathbf{U}}$, party $\mathsf{P_1}$ can only encrypt the non-sparse elements. Accordingly, the computational load on the on party $\mathsf{P_0}$ becomes lighter. Another interesting benefit of computing $\mathbf{U}\mathbf{D}$ in this way is that the SV packing based HE acceleration method can be perfectly embedded while eliminating the time-consuming rotation operations \cite{acar2018survey}. We expand on this as follows.

\textbf{Why we choose SV packing.} The SV encoding method \cite{smart2014fully} is designed to pack multiple plaintexts into one message. In the context of ciphertext, the homomorphic evaluation cost can be amortized by a factor of $1/N$, if $N$ is the packing size. This useful property is often termed as single instructing multiple data (SIMD) \cite{acar2018survey}. Here, we give a brief description of SIMD. Assume that two vectors $\mathbf{x},\mathbf{y}$ with the same size $N$, and the SV encoding/decoding algorithms are denoted as $\mathsf{SV.En}(\cdot)$ and $\mathsf{SV.De}(\cdot)$. If $\mathbf{x}$ and $\mathbf{y}$ are encoded and encrypted using the SV packing and the same HE scheme, the addition, and subtraction homomorphism are perfectly preserved. In another word, the homomorphic operators $\boxplus$ and $\boxminus$ can be directly applied to obtain the ciphertext of $\mathbf{x}+\mathbf{y}$ and $\mathbf{x}-\mathbf{y}$. Similarly, the entrywise multiplication homomorphism also holds: $\{\mathbf{x}[0]\cdot\mathbf{y}[0],...,\mathbf{x}[N-1]\cdot\mathbf{y}[N-1]\} = \mathsf{SV.De}(\mathsf{R.Dec}(\mathsf{R.Enc}(\mathsf{pk_R},\mathsf{SV.En}(\mathbf{x})) \boxtimes \mathsf{R.Enc}(\mathsf{pk_R},\mathsf{SV.En}(\mathbf{y}))))$. Indeed, the SV based SIMD method is an ideal choice for boosting the efficiency of securely computing Hadamard product in Equation \ref{entrywise multiplication of UD}.

However, it is challenging to tackle the standard matrix multiplication (i.e., vector inner product) by solely applying SIMD. In specific, given a ciphertext that is the encryption of the Hadamard product of two vectors, written as $\mathsf{R.Enc}(\mathsf{pk_R},\mathsf{SV.En}(\mathbf{x})) \boxtimes \mathsf{R.Enc}(\mathsf{pk_R},\mathsf{SV.En}(\mathbf{y}))$, no straightforward method can be employed to obtain the ciphertext of $\mathbf{x}\cdot \mathbf{y}$. To address this, existing works \cite{juvekar2018gazelle} propose to homomorphically rotate the ciphertext by multiplying it with a rotation key. In doing so, the positions of the packed elements are changed. After each round of rotation, one needs to invoke operator $\boxplus$ to accumulate the ciphertexts. Through conducting certain rounds of rotation (i.e., $O(logN)$), the generated HE ciphertext implies the vector inner product $\mathbf{x}\mathbf{y}$. Note that the homomorphic rotation is extremely expensive in the realm of RLWE/LWE based HE. In specific, it is nearly $30\times$ more expensive than the multiplication operator \cite{huang2022cheetah}. To conclude, the massive heavy rotations become the major bottleneck of HE based secure matrix multiplication protocols, and ultimately lead to the inefficiency of the recommender system.

%从前面讲述了SV的SIMD 优势和缺陷，此处应该过度如何找到适合更适合内积的，无需旋转的编码方式。 在描述这个idea的时候要把其优势，以及背后的设计思想简述一下。然后，转折表示这还不够，应该设计支持矩阵乘向量的编码方式。首先用自然语言描述编码的方式。然后给个例子。然后把编码公式放上。最后给个有关这两个映射的分析。到这里就就把US计算的核心部分讲完了。

\textbf{Exploring new and fast packing method.} Restricted by the SV packing, when computing matrix multiplication (e.g., $\mathbf{U}\mathbf{S}^{T}$), most existing schemes \cite{juvekar2018gazelle} seek to adopt the particular prime technique \cite{gentry2012homomorphic} to mitigate the heavy computational load over homomorphic rotations, yet the security level is reduced as the side effect. To attain a certain security level, the lattice dimension has to be increased. As a result, all the consecutive homomorphic operations will be slower. After conducting a comprehensive investigation, we find that there exists a seesaw effect between security level and efficiency in rotation based schemes. To solve this dilemma, we propose to use a rotation-free packing method that fits for matrix multiplication to securely compute $\mathbf{U}\mathbf{S}^{T}$. Recall that the plaintext of RLWE HE scheme is a polynomial (see Equation \ref{product of polynomials}). Thus, in theory, a batch of messages can be packed as the polynomial coefficients so as to amortize the costs \cite{rathee2020cryptflow2, huang2022cheetah}. In specific, as shown in Equation \ref{product of polynomials}, the product of two polynomials $\widehat{m}\cdot \widehat{n}$ implies the inner product of these two coefficients vectors. Therefore, if the input vectors are arranged appropriately as the coefficients, we can obtain the inner product over the ciphertext without rotation. Accordingly, this idea can be extended for matrix multiplication by conducting multiple matrix-vector multiplications.

%矩阵相乘示例
%\renewcommand\tablename{Figure}
%\renewcommand \thetable{\arabic{table}}
%\setcounter{table}{2}
\begin{table}
\centering
\begin{tabular}{|p{8cm}|}
\hline
\vspace{-2pt}
A toy example over $\mathbb{Z}_{2^{5}}$ ($\text{mod}\ 2^{5}$). \\
\vspace{-10pt}
\begin{equation*}
\mathbf{X} =
\begin{bmatrix}
1 & 3 & 5 \\
7 & 9 & 11 \\
\end{bmatrix},
\mathbf{y} =
\begin{bmatrix}
2 \\
4 \\
6 \\
\end{bmatrix} \Rightarrow \mathbf{z} = \mathbf{X}\mathbf{y} \equiv
\begin{bmatrix}
\textcolor{blue}{14} & \textcolor{red}{20} \\
\end{bmatrix}^{T}
\end{equation*} \\
\hline
\vspace{-2pt}
Compute $\mathbf{z}$ using $\pi_1$ and $\pi_2$ ($\text{mod}\ (X^{8}+1, 2^{5})$). \\
\vspace{-2pt}
$\pi_1(\mathbf{X})\rightarrow \widehat{x} = 5X^{0}+3X^{1}+1X^{2}+11X^{3}+9X^{4}+7X^{5}$  \\
$\pi_2(\mathbf{y})\rightarrow \widehat{y} = 2X^{0}+4X^{1}+6X^{2}$  \\
\vspace{-2pt}
\ \ \ \ \ \ \ \ \ \ \ \ \ \ \ \ $\Downarrow \widehat{z}\leftarrow \widehat{x}\cdot \widehat{y}$ \\
\vspace{-2pt}
$\widehat{z} = a_{0}X^{0}+a_{1}X^{1}+\textcolor{blue}{14}X^{2}+a_{3}X^{3}+a_{4}X^{4}+\textcolor{red}{20}x^{5}+a_{6}X^{6}+a_{7}X^{7}$ \\
\vspace{-2pt}
\ \ \ \ \ \ \ \ \ \ \ \ \ \ \ \ $\Downarrow$ Extract the values in $\mathbf{z}$ from $\widehat{z}$.\\
\vspace{-2pt}
\textbf{If} the $i$-th coefficient in $\widehat{z}$ is colored, \textbf{Do} \\
\ \ \ \ Assume that the RLWE ciphertext of $\widehat{z}$ is $\mathsf{RCT}_{\widehat{z}}$;  \\
\ \ \ \ Compute LWE ciphertext: $\mathsf{LCT}_{\widehat{z}[i]} \leftarrow \mathsf{HE.Extract}(\mathsf{RCT}_{\widehat{z}}, i)$; \\
\ \ \ \ Arrange $\mathsf{LCT}_{\widehat{z}[i]}$ into vector $\mathbf{z}$ according to \textbf{Theorem} \ref{theorem:matrix multiplication using cheetah}; \\
\textbf{Return} the LWE ciphertext $\mathsf{LCT}_{\mathbf{z}}$ for vector $\mathbf{z}$. \\
\hline
\end{tabular}
\vspace{2pt}
\caption{A toy example for $\pi_1,\pi_2$ with $N=8$ and $p=2^{5}$.}
\label{figure:toy example of packing method}
\end{table}

Intuitively, the aforementioned packing method can be regarded as linear mappings from the original matrix/vector to the ring polynomial space. Formally, the mapping functions of the matrix and vector $\pi_{1}: \mathbb{Z}_{p}^{l\times m} \rightarrow R_{N,p}; \pi_{2}: \mathbb{Z}_{p}^{m} \rightarrow R_{N,p}$ are defined as follows:
\begin{equation}
\label{equation:packing method}
\begin{split}
  \widehat{x}  = \pi_{1}(\mathbf{X})\ &\text{where}\  \widehat{x}[i\cdot m+m-1-j] = \mathbf{X}[i,j], \\
  \widehat{y}  = \pi_{2}(\mathbf{y})\ &\text{where}\  \widehat{y}[j] = \mathbf{y}[j].
\end{split}
\end{equation}
For $\pi_1$ and $\pi_2$, s.t. $i\in [l], j\in [m]$. Note that all the rest coefficients of $\widehat{x}, \widehat{y}$ are set as $0$. Accordingly, the multiplication $\mathbf{z} = \mathbf{X}\mathbf{y}\ \text{mod}\ p$ is embedded in the coefficients of the polynomial $\widehat{z} = \widehat{x}\cdot \widehat{y}$. Since the number of the coefficients of a polynomial is limited to $N$ (i.e., $\widehat{x}, \widehat{y} \in R_{N,p}$), the  constraint condition $l\cdot m \leq N$ must hold to guarantee the correctness of Equation \ref{equation:packing method}. Formally, we give the following theorem to specify the mathematical relationship between $\mathbf{z}$ and $\widehat{z}$.
\begin{Theorem}[Matrix-vector multiplication]
\label{theorem:matrix multiplication using cheetah}
Given a matrix $\mathbf{X} \in \mathbb{Z}_{p}^{l\times m}$, a vector $\mathbf{y}\in \mathbb{Z}_{p}^{m}$, and two polynomials $\widehat{x} = \pi_{1}(\mathbf{X})$, $\widehat{y} = \pi_{1}(\mathbf{y})$; set $\widehat{z} \leftarrow \widehat{x}\cdot \widehat{y}$ and $\mathbf{z}\leftarrow \mathbf{X}\cdot \mathbf{y}$; for all $i \in [l], j\in [m]$, we have $\sum_{0\leq j< m}\widehat{x}[m-j]\cdot \widehat{y}[j] = \sum_{0\leq j< m} \mathbf{X}[i,j]\cdot \mathbf{y}[j]$, which indicates $\mathbf{z}[i] = \widehat{z}[i\cdot m+m-1]$.
\end{Theorem}
The correctness proof of \textbf{Theorem} \ref{theorem:matrix multiplication using cheetah} can be proved by expanding the multiplication result and then comparing the corresponding values of polynomial coefficients with the inner products. Note that the values of $\mathbf{z}$ can be extracted from the coefficients of $\widehat{z}$ by applying the function $\mathsf{HE.Extract(\cdot)}$ described in Section \ref{subsec:cryptographical tools}. The extracted ciphertexts are in decryptable LWE format. Given these ciphertexts, one can arrange them into a vector according to \textbf{Theorem} \ref{theorem:matrix multiplication using cheetah}. Finally, the LWE ciphertext of the matrix-vector multiplication $\mathsf{LCT}_{\mathbf{z}}$ is returned and will be fed into the next step of $\mathsf{\Pi}_\mathrm{ins}$. To facilitate the understanding, we provide a toy example of the whole processing in Fig. \ref{figure:toy example of packing method}.

%位置不敏感协议实现细节
%\renewcommand\tablename{Figure}
%\renewcommand \thetable{\arabic{table}}
%\setcounter{table}{3}
\begin{table*}
\centering
\begin{tabular}{|p{17cm}|}
\hline
\begin{center}
  Implementation of $\mathsf{\Pi}_\mathrm{ins}$
\end{center}
\textbf{Public Parameters:} $\mathsf{pp} = \{\alpha, \mathsf{HE.pp}, \mathsf{pk_R}, l, m, l_w, m_w\}$.   \\
$\bullet$ $\{l,m\}$ are the input matrix dimensions, and $\{l_w,m_w\}$ are the partition window size, where $0<l_w\leq l$, $0<m_w\leq m$, and $l_w {m_w}\leq N$ holds. \\
\textbf{Input:} $\mathsf{P_1}$ holds the social matrix $\mathbf{S} \in \mathbb{Z}_p^{m\times m}$, and the diagonal matrix $\mathbf{D}\in \mathbb{Z}_p^{m\times m}$, $\mathsf{P_0}$ holds the matrix $\mathbf{U}\in \mathbb{Z}_{p}^{l\times m}$. $\mathsf{P_0}, \mathsf{P_1}$ shares the sparse locations to each other in matrices $\mathbf{U}, \mathbf{S}$.  \\
\textbf{Output:} $\mathsf{P_0}$ and $\mathsf{P_1}$ obtain two shares $\langle \mathbf{Z}\rangle_0, \langle \mathbf{Z}\rangle_1\in \mathbb{Z}_{p}^{l\times m}$, respectively, where $\mathbf{Z} = \alpha\mathbf{U}\mathbf{D}/2-\alpha\mathbf{U}\mathbf{S}^{T}$.   \\
\vspace{-2pt} \\
\hline
\vspace{-2pt} \\
$\blacksquare$ \textbf{Securely compute $\mathbf{U}\mathbf{D}$:}
\begin{enumerate}[1:]

  \item $\mathsf{P_0}$ sends the non-sparse locations $\mathbf{loc}_{\mathbf{U}}$ of $\mathbf{U}$ to $\mathsf{P_1}$. Then $\mathsf{P_0}$ deletes the sparse columns on $\mathbf{U}$, and obtain the compressed matrix $\overline{\mathbf{U}}$. $\mathsf{P_0}$ partitions $\overline{\mathbf{U}}$ with window size $N$, and zero-padding is applied for the end subvector if necessary. Then, $\mathsf{P_0}$ encodes $\overline{\mathbf{U}}$ as $\mathsf{SV}_{\overline{\mathbf{U}}} \leftarrow \mathsf{SV.En}(\overline{\mathbf{U}})$.

  \item On receiving $\mathbf{loc}_{\mathbf{U}}$, $\mathsf{P_1}$ deletes the elements $\mathbf{D}[i,i]$ if $i$-th column in $\mathbf{U}$ is sparse. The compressed diagonal vector of $\mathbf{D}$ is written as $\overline{\mathbf{d}}$. Then $\mathsf{P_1}$ encodes and encrypts it as : $\mathsf{RCT}_{\overline{\mathbf{d}}} \leftarrow \mathsf{R.Enc}(\mathsf{pk_R}, \mathsf{SV.En}(\overline{\mathbf{d}}))$. The ciphertext $\mathsf{RCT}_{\overline{\mathbf{d}}}$  is then forwarded to $\mathsf{P_0}$.

  \item Given $\mathsf{RCT}_{\overline{\mathbf{d}}}$, $\mathsf{P_0}$ operates $\mathsf{RCT}_{\overline{\mathbf{U}} \star \overline{\mathbf{d}}} \leftarrow \mathsf{SV}_{\overline{\mathbf{U}}}\boxtimes \mathsf{RCT}_{\overline{\mathbf{d}}}$. Then $\mathsf{P_0}$ uniformly samples a random matrix $\mathbf{R}$ with exactly the same scale and domain as $\overline{\mathbf{U}}$. $\mathsf{P_0}$ encodes $\mathbf{R}$ as $\mathsf{SV}_{\mathbf{R}} \leftarrow \mathsf{SV.En}(\mathbf{R})$. $\mathsf{P_0}$ masks  $\mathsf{RCT}_{\overline{\mathbf{U}} \star \overline{\mathbf{d}}}$ by computing $\mathsf{RCT}_{\overline{\mathbf{U}} \star \overline{\mathbf{d}}}^{\prime} \leftarrow \mathsf{RCT}_{\overline{\mathbf{U}} \star \overline{\mathbf{d}}}\boxminus \mathsf{SV}_{\mathbf{R}}$. Afterwards, $\mathsf{P_0}$ keeps $\mathbf{R}$ as its own share $\langle \mathbf{Z}_1\rangle_{0}$, and sends the masked ciphertexts $\mathsf{RCT}_{\overline{\mathbf{U}} \star \overline{\mathbf{d}}}^{\prime}$ to $\mathsf{P_1}$.

  \item Upon getting $\mathsf{RCT}_{\overline{\mathbf{U}} \star \overline{\mathbf{d}}}^{\prime}$, $\mathsf{P_1}$ decrypts and decodes it as its share $\langle \mathbf{Z}_1\rangle_{1} \leftarrow \mathsf{SV.De}(\mathsf{R.Dec}(\mathsf{sk_R}, \mathsf{RCT}_{\overline{\mathbf{U}} \star \overline{\mathbf{d}}}^{\prime}))$.
\end{enumerate}
$\blacksquare$ \textbf{Securely compute $\mathbf{U}\mathbf{S}^{T}$:}
\begin{enumerate}[1:]

%\item $\mathsf{P_1}$ first sends the non-sparse locations $\mathbf{loc}_{\mathbf{S}}$ of $\mathbf{S}$ to $\mathsf{P_0}$. Then, $\mathsf{P_1}$ compress the matrix similarly by removing the sparse values. Assume that the transferred $\mathbf{S}$ is $\mathbf{S}^{*}$, then the compressed matrix can be written as $\overline{\mathbf{S}^{*}}$. Afterwards, $\mathsf{P_1}$ partitions $\overline{\mathbf{S}^{*}}$ into block matrices $\overline{\mathbf{S}^{*}_{\rho}} \in \mathbb{Z}_{p}^{m_w\times m_w}$ for $\rho \in [m^{\prime}]$. $\mathsf{P_1}$ maps all the matrices to polynomials $\widehat{s}_{\rho} = \pi_2(\overline{\mathbf{S}^{*}_{\rho}})$. Then, $\mathsf{P_1}$ encrypts the polynomials $\mathsf{SV}_{\rho} \leftarrow \mathsf{R.Enc}(\mathsf{pk_R}, \widehat{s}_{\rho})$ and sends them to $\mathsf{P_0}$.

\item $\mathsf{P_1}$ sends the non-sparse locations $\mathbf{loc}_{\mathbf{S}}$ of $\mathbf{S}$ to $\mathsf{P_0}$. Then, $\mathsf{P_1}$ compresses the matrix similarly by removing the sparse values. Let the transferred $\mathbf{S}$ be $\mathbf{S}^{*}$, the compressed matrix be $\overline{\mathbf{S}^{*}}$, and the $j$-th column vector in $\overline{\mathbf{S}^{*}}$ is denoted as $\mathbf{s}_{j}^{*}$.

\item $\mathsf{P_1}$ partitions $\mathbf{s}_{j}^{*}$ into subvectors $\mathbf{s}_{j,\rho}^{*}$ for $j\in [m]$ (with zero-padding if necessary). The window size $m_w$ and a number of subvectors are set dynamically according to $\mathbf{loc}_{\mathbf{S}}$. $\mathsf{P_1}$ maps all the subvectors into polynomials $\widehat{s}_{\rho} = \pi_2(\mathbf{s}_{j,\rho}^{*})$. At last, $\mathsf{P_1}$ encrypts all the polynomials $\mathsf{RCT}_{\rho} \leftarrow \mathsf{R.Enc}(\mathsf{pk_R}, \widehat{s}_{\rho})$ and sends them to $\mathsf{P_0}$.

%\item $\mathsf{P_0}$ compress the matrix according to $\mathbf{loc}_{\mathbf{S}}$. Assume that $\mathbf{U}$ is compressed to $\overline{\mathbf{U}}$, then $\mathsf{P_0}$ partitions it into block matrices $\overline{\mathbf{U}}_{\delta, \rho} \in \mathbb{Z}_{p}^{l_w\times m_w}$ for $\delta \in [l^{\prime}], \rho \in [m^{\prime}]$. Then all the matrices are mapped to polynomials $\widehat{u}_{\delta, \rho} = \pi_{1}(\overline{\mathbf{U}}_{\delta,\rho})$. $\mathsf{P_0}$ uniformly sample a random matrix $\mathbf{Q}$ with the same scale and domain as $\overline{\mathbf{U}}$.

\item $\mathsf{P_0}$ receives the encrypted polynomials $\mathsf{RCT}_{\rho}$ for all $m$ columns in $\overline{\mathbf{S}^{*}}$, and $\mathbf{loc}_{\mathbf{S}}$ from $\mathsf{P_1}$. For $j$-th column in $\overline{\mathbf{S}^{*}}$, $\mathsf{P_0}$ first compresses $\mathbf{U}$ to $\overline{\mathbf{U}}_j$. Then $\mathsf{P_0}$ partitions it into block matrices $\overline{\mathbf{U}}_{\delta, \rho}$, where the window size $l_w\times m_w$ and number of block matrices are set dynamically according to $\mathbf{loc}_{\mathbf{S}}$. $\mathsf{P_0}$ maps all the matrices to polynomials $\widehat{u}_{\delta, \rho} = \pi_{1}(\overline{\mathbf{U}}_{\delta,\rho})$.

\item $\mathsf{P_0}$ operates $\mathsf{RCT}_{\delta} \leftarrow \boxplus_{\rho \in [m^{\prime}]} (\widehat{u}_{\delta, \rho}\boxtimes \mathsf{RCT}_{\rho})$ for all $\delta \in [l^{\prime}]$. To remask the multiplication results, $\mathsf{P_0}$  first uniformly sample a random vector $\mathbf{q}$ according to $\mathbf{loc}_{\mathbf{S}}$, and map it as a polynomial $\widehat{q} = \pi_2(\mathbf{q})$, then operates $\mathsf{RCT}_{\delta}^{\prime}\leftarrow \mathsf{RCT}_{\delta} \boxminus \widehat{q}$ for $\delta \in l^{\prime}$. Here $l^{\prime}$ and $m^{\prime}$ are the number of windows that are set dynamically according to $\mathbf{loc}_{\mathbf{S}}$ and window size $l_w, m_w$. Similarly, $\mathsf{P_0}$ repeats the above operation for every column in $\overline{\mathbf{S}^{*}}$. The set of random vectors are arranged with the same format as $\overline{\mathbf{U}}$, which is written as $\mathbf{Q}$. At last, $\mathsf{P_0}$ keeps $\mathbf{Q}$ as its own share $\langle \mathbf{Z}_2\rangle_{0}$, and sends all the masked multiplication ciphertexts $\mathsf{RCT}_{\delta}^{\prime}$ to $\mathsf{P_1}$.

\item On receiving all the ciphertexts $\mathsf{RCT}_{\delta}^{\prime}$, $\mathsf{P_1}$ first extract the LWE ciphertexts by invoking $\mathsf{LCT}_{i}^{\prime}\leftarrow \mathsf{HE.Extract}(\mathsf{RCT}_{j}^{\prime}, \textsf{ind})$. The index $j$ and $\mathsf{ind}$ can be computed with the window size, $\mathbf{loc}_{\mathbf{S}}$ according to \textbf{Theorem} \ref{theorem:matrix multiplication using cheetah}. For each LWE ciphertext, $\mathsf{P_1}$ decrypts it by invoking $\mathsf{L.Dec}(\mathsf{sk_L}, \mathsf{LCT}_{i}^{\prime})$. Then, $\mathsf{P_1}$ arranges each plaintext into the appropriate location of a matrix according to $\mathbf{loc}_{\mathbf{S}}$, and keeps the matrix as its share $\langle \mathbf{Z}_2 \rangle_1$.
\end{enumerate}
$\blacksquare$ \textbf{Compute and return the shares for $\mathbf{Z}$:}
\begin{enumerate}[1:]

  \item $\mathsf{P_0}$ operates $\bar{\langle \mathbf{Z}\rangle}_0 \leftarrow \frac{\alpha}{2}(\langle \mathbf{Z}_1 \rangle_0 + \langle \mathbf{Z}_1 \rangle_1) \ \text{mod}\ p$. Then, $\mathsf{P_0}$ expands $\bar{\langle \mathbf{Z}\rangle}_0$ to meets the format $\mathbb{Z}_{p}^{l\times m}$, that the values in sparse locations are set to $0$ according to $\mathbf{loc}_{\mathbf{S}}$. At last, $\mathsf{P_0}$ takes the expanded share $\langle \mathbf{Z}\rangle_0$ as the output.

   \item $\mathsf{P_1}$ operates $\bar{\langle \mathbf{Z}\rangle}_1 \leftarrow -\alpha(\langle \mathbf{Z}_2 \rangle_0 + \langle \mathbf{Z}_2 \rangle_1) \ \text{mod}\ p$. Then, $\mathsf{P_1}$ expands $\bar{\langle \mathbf{Z}\rangle}_1$ to meets the format $\mathbb{Z}_{p}^{l\times m}$, that the values in sparse locations are set to $0$ according to $\mathbf{loc}_{\mathbf{S}}$. At last, $\mathsf{P_1}$ takes the expanded share $\langle \mathbf{Z}\rangle_1$ as the output.

\end{enumerate} \\
\hline
\end{tabular}
\vspace{2pt}
\caption{Implementation of  $\mathsf{\Pi}_\mathrm{ins}$.}
\label{figure:implementation of insensitive scheme}
\end{table*}

%简述给出的insensitive协议的基本流程。并给出一个具体的协议图。
As shown in Fig. \ref{figure:implementation of insensitive scheme}, we give the detailed implementation for our location insensitive scheme $\mathsf{\Pi}_\mathrm{ins}$. To initiate the protocol, party $\mathsf{P_0}$ and $\mathsf{P_1}$ collaboratively generate the public parameters for RLWE/LWE HE scheme and the training related parameter $\alpha$. Note that, since the input matrices are too large to be taken as the plaintext, we need to partition them to obtain block matrices or subvectors that are compatible with packing and encryption algorithms. For computing $\mathbf{U}\mathbf{D}$, the window size is fixed to $N$. $\mathsf{P_0}$ and $\mathsf{P_1}$ just trivially segment the input matrix and vector into subvectors with $N$ elements. Thus, in Fig. \ref{figure:implementation of insensitive scheme}, we omit the description of partition operation. For computing $\mathbf{U}\mathbf{S}^{T}$, the partition window sizes $l_w$ and $m_w$ need to be dynamically appointed according to the sparsity level of each column in $\mathsf{S}$ (i.e, $\mathbf{loc}_{\mathbf{S}}$). In another word, the shape of the compressed matrix/vector is uncertain, which results in the dynamic nature of window size. The selection of $l_w, m_w$ can be formalized as an optimization problem. We defer the analysis on this issue to the performance evaluation section. Note that in order to avoid message overflow when conducting polynomial multiplication in a ring $R_{N, q}$, the window size parameters should meet $l_w\times m_w\leq N$.

$\mathsf{\Pi}_\mathrm{ins}$ breaks down the entire computing task $\mathbf{Z} = \alpha\mathbf{U}\mathbf{D}/2-\alpha\mathbf{U}\mathbf{S}^{T}$ into three steps, that are securely computing $\mathbf{U}\mathbf{D}$, securely computing $\mathbf{U}\mathbf{S}^{T}$, and reconstructing the two shares $\langle \mathbf{Z}\rangle_0, \langle \mathbf{Z}\rangle_1$, respectively. As the calculation of $\mathbf{U}\mathbf{D}$ is transferred to Hadamard product, we can not only take the advantage of efficient SV packing method but also eliminate heavy rotation operations. The entire processing basically follows the tune of work flow described in Section \ref{subsec:scheme overview}. $\mathsf{P_0}$ first shares the sparsity with $\mathsf{P_1}$. Then $\mathsf{P_1}$ compresses, packs and encrypts the diagonal vector $\mathbf{d}$ for $\mathbf{D}$ accordingly. Once getting ciphertext from $\mathsf{P_1}$, $\mathsf{P_0}$ conducts homomorphic multiplication evaluation, and remasks the results before sending it to $\mathsf{P_1}$. $\mathsf{P_0}$ keeps the random masking matrix $\mathbf{R}$ as its secret share. $\mathsf{P_1}$ can simply decrypt and unpack the masked ciphertext as the share.

When the problem becomes matrix multiplication, SV packing method \cite{acar2018survey} is often plagued by the seesaw effect between security and efficiency. Therefore, the proposed $\mathsf{\Pi}_\mathrm{ins}$ seeks to explore rotation free packing method \cite{huang2022cheetah} (see Equation \ref{equation:packing method}). Similarly, since the input matrix $\mathbf{S}$ is extremely spares, $\mathsf{P_1}$ first share the sparse locations $\mathbf{loc}_{\mathbf{S}}$ with $\mathsf{P_0}$. Then both parties compress their input matrices accordingly. Since each row in $\mathbf{S}$ has different sparse locations, $\mathsf{P_0}$ needs to generate corresponding input matrices for each row. For instance, if the $i$-th element in vector $\mathbf{S}^{*}$ is sparse, then $\mathsf{P_0}$ just delete the $i$-th column. This operation almost brings no additional computational load. In specific, $\mathbf{U}\mathbf{S}^{T}$ is solved by computing $\mathbf{U}\mathbf{s}_i^{*}$ for $i\in [m]$. In general, $\mathsf{P_0}$ and $\mathsf{P_1}$ collaboratively generate the two secret shares $\langle \mathbf{Z}_2 \rangle_{0}, \langle \mathbf{Z}_2 \rangle_{1}$ by applying the similar secure two-party computation method. As shown in Fig. \ref{figure:implementation of insensitive scheme}, when computing $\mathbf{U}\mathbf{S}^{T}$, $\mathsf{P_0}$ and $\mathsf{P_1}$ also use RLWE HE to encrypt the packed inputs; conduct homomorphic evaluations to obtain the ciphertexts for matrix-vector multiplication, and sample a random matrix $\mathbf{Q}$ to remask the ciphertext. $\mathbf{P_0}$ simply takes random matrix $\mathbf{Q}$ as its share. $\mathsf{P_1}$ needs to extract the coefficients from the RLWE ciphertexts and decrypt them as its own secret share. Note that, the extracted ciphertexts are in LWE format. Thus, $\mathsf{P_1}$ needs to decrypt them by invoking $\mathsf{L.Dec}(\mathsf{sk_L}, \cdot)$. At last, $\mathsf{P_0}$ and $\mathsf{P_1}$ return two secrete shares $\langle \mathbf{Z}\rangle_0, \langle \mathbf{Z}\rangle_0$ as the outputs for $\mathsf{\Pi}_\mathrm{ins}$, which will be fed into the next step in Fig. \ref{figure:work flow}.

%此章节主题内容结束。酌情看是否需要remark一些内容.就是为啥UD中只考虑U列稀疏的原因，这有很多考量和分析，但是也是个局限性的问题。暂定不分析原因。

\subsection{Sparse Location Sensitive Scheme $\mathsf{\Pi}_\mathrm{sen}$}
\label{subsec:Sparse Location Sensitive Scheme}

%US方案是可以用SV打包的，只需要将查询方的明文除了要查询的位置外的其他数据都置0（one hot coding））.U矩阵必须加密，所以，从消息长度的角度看是不能用LWE/RLWE HE，只能用PAILLIER这种半同态。

%第一部分是UD，这个部分在这个场景里是没法利用稀疏性的， 所以只能是直接RLWE方案下相乘。具体原因：首先是U不好直接全部加密，这样开销会大很多。然后如果加密D，则需要用PIR逐一查询，然后D的对角不断变化，需要使用不同的打包方案。开销也非常大。加上PIR，如果U不是极端稀疏的话是没有意义的。所以方案就直接把位置泄露方案中的loc去掉，使用相同的方式来一遍即可。

%第二部分是引入PIR的。其中U用PHE加密，然后利用打包加密来减少开销。S则使用one hot编码后用SV编码。然后两方还是一对一计算。但是U的计算开销小很多。由S进行PIR查询，并获得U的密文。整体的计算流程和位置泄露方案相似。

%此处给出位置敏感方案的具体算法。
% \renewcommand\tablename{Figure}
% \renewcommand \thetable{\arabic{table}}
% \setcounter{table}{4}
\begin{table*}
\centering
\begin{tabular}{|p{17cm}|}
\hline
\begin{center}
  Implementation of $\mathsf{\Pi}_\mathrm{sen}$
\end{center}
\textbf{Public Parameters:} $\mathsf{pp} = \{\alpha, \mathsf{HE.pp}, \mathsf{pk_R}, \mathsf{pk_P}, l, m, s\}$.   \\
$\bullet$ $\{l,m\}$ are the input matrix dimensions, and $s$ is the partition window size (i.e., the packing size for PHE crypto-system). \\
\textbf{Input:} $\mathsf{P_1}$ holds the social matrix $\mathbf{S} \in \mathbb{Z}_p^{m\times m}$, and the diagonal matrix $\mathbf{D}\in \mathbb{Z}_p^{m\times m}$, $\mathsf{P_0}$ holds the matrix $\mathbf{U}\in \mathbb{Z}_{p}^{l\times m}$.  \\
\textbf{Output:} $\mathsf{P_0}$ and $\mathsf{P_1}$ obtain two shares $\langle \mathbf{Z}\rangle_0, \langle \mathbf{Z}\rangle_1\in \mathbb{Z}_{p}^{l\times m}$, respectively, where $\mathbf{Z} = \alpha\mathbf{U}\mathbf{D}/2-\alpha\mathbf{U}\mathbf{S}^{T}$.   \\
\vspace{-2pt} \\
\hline
\vspace{-2pt} \\
$\blacksquare$ \textbf{Securely compute $\mathbf{U}\mathbf{D}$:}
\begin{enumerate}[1:]

  \item $\mathsf{P_1}$ first partitions the diagonal vector $\mathbf{d}$ of the input matrix $\mathbf{D}$ into subvectors with $N$ (fetched from $\mathsf{HE.pp}$) elements. Zero-padding is applied for the end subvector if necessary. Then for each subvector, $\mathsf{P_1}$ packs it using SV method and encrypts it by invoking RLWE HE scheme. Same as $\mathsf{\Pi}_\mathrm{ins}$, considering the partition size is fixed as $N$, we omit this processing. The ciphertext of vector $\mathbf{d}$ is generated as $\mathsf{RCT}_{\mathbf{d}} \leftarrow \mathsf{R.Enc}(\mathsf{pk_R}, \mathsf{SV.En} (\mathbf{d}))$. Afterward, $\mathsf{P_1}$ sends $\mathsf{RCT}_{\mathbf{d}}$ to party $\mathsf{P_0}$.

  \item Upon receiving $\mathsf{RCT}_{\mathbf{d}}$, $\mathsf{P_0}$ partitions all the row vectors in matrix $\mathbf{U}$ in the same way as $\mathsf{P_1}$. The partition size (i.e., packing size) is also set as $N$. Then $\mathsf{P_0}$ packing the input matrix using SV method as $\mathsf{SV}_{\mathbf{U}} \leftarrow \mathsf{SV.En}(\mathbf{U})$. Afterward, $\mathsf{P_0}$ operates $\mathsf{RCT}_{\mathbf{U} \star \mathbf{d}} \leftarrow \mathsf{SV}_{\mathbf{U}} \boxtimes \mathsf{RCT}_{\mathbf{d}}$. To remask the ciphertext, $\mathsf{P_0}$ uniformly samples a random matrix $\mathbf{R}\in \mathbb{Z}_{p}^{l\times m}$ and partitions it in the same way as $\mathbf{U}$. To keep the format consistent, the partitioned $\mathbf{R}$ is also packed using SV as $ \mathsf{SV}_{\mathbf{R}} \leftarrow \mathsf{SV.En}(\mathbf{R})$. Then $\mathsf{P_0}$ operates $\mathsf{RCT}_{\mathbf{U} \star \mathbf{d}}^{\prime} \leftarrow \mathsf{RCT}_{\mathbf{U} \star \mathbf{d}}\boxminus \mathsf{SV}_{\mathbf{R}}$. $\mathsf{P_0}$ keeps $\mathbf{R}$ as its own share $\langle \mathbf{Z}_1\rangle_{0}$, and sends the remasked ciphertexts $\mathsf{RCT}_{\mathbf{U} \star \mathbf{d}}^{\prime}$ to $\mathsf{P_1}$.

  \item Upon receiving $\mathsf{RCT}_{\mathbf{U} \star \mathbf{d}}^{\prime}$, $\mathsf{P_1}$ decrypts and decodes it as its share $\langle \mathbf{Z}_1\rangle_{1} \leftarrow \mathsf{SV.De}(\mathsf{R.Dec}(\mathsf{sk_R}, \mathsf{RCT}_{\mathbf{U} \star \mathbf{d}}^{\prime}))$.

\end{enumerate}
$\blacksquare$ \textbf{Securely compute $\mathbf{U}\mathbf{S}^{T}$:}
\begin{enumerate}[1:]
%这里要用PIR加打包的方式做矩阵乘积。基本流程（以向量乘以向量为例）：
%1. P0将U以尺寸s的大小分割，打包，然后在用PHE加密，并和P1协商好一个适合的index.
%2. P1 将所有需要查询的稀疏位置发送PIR查询到P0. 这个查询的位置如果在同一个包中则用相同的index。这个操作是可以降低查询次数的，后面可具体分析。
%3. P0收到PIR查询，然后调用PIR将密文消息发送返回给P1.

\item $\mathsf{P_0}$ partitions the matrix $\mathbf{U}$ into subvectors $\mathbf{u}_{\delta, \rho}$  (using zero-padding for end subvectors if necessary) with the window size $s$, where $\rho \in [l], \delta \in \left [\lceil m/s  \right \rceil]$. Then $\mathsf{P_0}$ packs and encrypts all the subvectors as $\mathsf{PCT}_{\mathbf{u}_{\delta, \rho}}\leftarrow \mathsf{P.Enc}(\mathsf{pk_P}, \mathsf{SV.En}(\mathbf{u}_{\delta, \rho}))$. The window size $s$ is negotiated by $\mathsf{P_0}$ and $\mathsf{P_1}$ according to the data distribution in $\mathbf{U}$ and $\mathbf{S}$, the PHE parameter setting, and the applied SV packing method. In addition, the query index needs to be set as the PIR parameter shared between $\mathsf{P_0}$ and $\mathsf{P_1}$.

\item $\mathsf{P_1}$ partitions the matrix $\mathbf{S}$ into subvectors $\mathbf{s}_{\mu, \nu}$ using exactly the same way as $\mathbf{U}$, where $\mu \in [m], \nu \in \left [\lceil m/s  \right \rceil]$. Then $\mathsf{P_1}$ first checks the query history and fetches the needed results from the records. Otherwise, $\mathsf{P_1}$ issues a PIR query to $\mathsf{P_0}$ for the non-sparse values in $\mathbf{S}$. Given the non-sparse values locate within the same subvector $\mathbf{s}_{\mu, \nu}$, $\mathsf{P_1}$ invokes $q_{\mu, \nu} \leftarrow \mathsf{MulPIR.Query}(\mu, \nu)$. Then $q_{\mu, \nu}$ is sent to $\mathsf{P_0}$.

\item Upon receiving $q_{\mu, \nu}$, $\mathsf{P_0}$ operates $r_{\mu, \nu} \leftarrow \mathsf{MulPIR.Response}(q_{\mu, \nu}, \mathbf{U})$, where the matrix $\mathbf{U}$ is the database. Afterward, $\mathsf{P_0}$ returns $r_{\mu, \nu}$ to $\mathsf{P_1}$.

%4.P1收到消息后，先用PIR抽取算法恢复出U中的消息密文。然后用打包后的明文和其逐一先相乘，然后再累加（注意：此处有个很重要的技巧，累加后的向量是不能做旋转的，将累加后密文当做向量，是内积的分割而非内积。）。最后再取随机数向量并打包，用于加噪。随机噪音的和为P1的share矩阵中的一个元素。加噪后密文向量返回给P0.
%5. P0将密文向量解密，然后解包。得到所有的元素，累加后的结果就是向量的内积加上噪音。

\item On obtaining the query result $r_{\mu, \nu}$, $\mathsf{P_1}$ recovers the queried value by invoking $d_{\mu, \nu} \leftarrow \mathsf{MulPIR.Extract}(r_{\mu, \nu})$. Here, $d_{\mu, \nu}$ is a packed and encrypted subvector fetched from matrix $\mathbf{U}$. Then $\mathsf{P_1}$ operates $\mathsf{PCT}_{\mathbf{U}\cdot \mathbf{S}^{T}}\leftarrow \oplus_{\nu \in \left [\lceil m/s  \right \rceil]}d_{\mu, \nu}\otimes \mathsf{SV}_{\mathbf{s}_{\mu, \nu}}$, for all all queried index $(\mu,\nu)$ where $\mu \in [m]$. If several non-spare elements appear in the same subvector, only one PIR query is needed and the processing remains the same.

%得到密文结果后，下面对密文进行remask。由于方案和前面都不同，单独一个步骤描述。

\item $\mathsf{P_1}$ arranges the encrypted results $\mathsf{PCT}_{\mathbf{U}\cdot \mathbf{S}^{T}}$ into an $l\times m$ empty temporal matrix $\mathbf{T}$, and the sparse locations in $\mathbf{T}$ are all set to $0$. Then, $\mathsf{P_1}$ uniformly samples a random tensor $\mathbf{Q_t}$ from $\mathbb{Z}_{p}^{l\times m \times s}$. $\mathsf{P_1}$ computes $\mathsf{PCT}_0\leftarrow \mathsf{P.Enc}(\mathsf{pk_P}, \mathsf{SV.En}(\mathbf{\phi}))$, where $\mathbf{\phi} = \{0\}^{s}$. All the sparse locations in $\mathbf{T}$ are set as $\mathsf{PCT}_0$. $\mathsf{P_1}$ operates $\mathsf{PCT}_{\mathbf{U}\cdot \mathbf{S}^{T}}^{\prime} \leftarrow \mathbf{T}[i,j]\oplus \mathsf{SV.En}(\mathbf{Q_t}[i,j,*])$, for all $i\in [l], j \in [m]$. $\mathsf{P_1}$ computes $\mathbf{Q}\leftarrow \sum_{k\in [s]}-\mathbf{Q_t}[i,j,k]$ for all $i\in [m], j\in [n]$. At last, $\mathsf{P_1}$ keeps $\mathbf{Q}$ as the secret share $\langle \mathbf{Z}_2 \rangle_1$, and sends $\mathsf{PCT}_{\mathbf{U}\cdot \mathbf{S}^{T}}^{\prime}$ to $\mathsf{P_0}$.

\item On receiving $\mathsf{PCT}_{\mathbf{U}\cdot \mathbf{S}^{T}}^{\prime}$, $\mathsf{P_0}$ first recovers the encrypted tensor as $\mathbf{Z_t} \leftarrow \mathsf{SV.De}(\mathsf{P.Dec}(\mathsf{sk_P}, \mathsf{PCT}_{\mathbf{U}\cdot \mathbf{S}^{T}}^{\prime}))$. Then $\mathsf{P_0}$ obtain its share as $\langle \mathbf{Z}_2 \rangle_0 \leftarrow \sum_{k\in [s]} \mathbf{Z_t}[i,j,k]$, where $i\in[l], j\in [m]$.

\end{enumerate}
$\blacksquare$ \textbf{Compute and return the shares for $\mathbf{Z}$:}
\begin{enumerate}[1:]
%本方案没有压缩，分享的格式是一样的。
  \item $\mathsf{P_0}$ operates $\langle \mathbf{Z}\rangle_0 \leftarrow \frac{\alpha}{2}(\langle \mathbf{Z}_1 \rangle_0 + \langle \mathbf{Z}_1 \rangle_1) \ \text{mod}\ p$. Then, $\mathsf{P_0}$ takes the share $\langle \mathbf{Z}\rangle_0$ as the output.

   \item $\mathsf{P_1}$ operates $\langle \mathbf{Z}\rangle_1 \leftarrow -\alpha(\langle \mathbf{Z}_2 \rangle_0 + \langle \mathbf{Z}_2 \rangle_1) \ \text{mod}\ p$. Then, $\mathsf{P_1}$ takes the share $\langle \mathbf{Z}\rangle_1$ as the output.
\end{enumerate} \\
\hline
\end{tabular}
\vspace{2pt}
\caption{Implementation of  $\mathsf{\Pi}_\mathrm{sen}$.}
\label{figure:implementation of sensitive scheme}
\end{table*}

In this part, we elaborate on the technical details of the sparse location sensitive scheme $\mathsf{\Pi}_\mathrm{sen}$. In specific, compare with $\mathsf{\Pi}_\mathrm{ins}$, the key additional privacy enhancing measurement is concealing the sparse locations in the input matrices $\mathbf{U}, \mathbf{S}$. To achieve this goal while utilizing the input sparsity for efficiency promotion, we introduce to use PIR \cite{angel2018pir} to fetch the values in $\mathbf{U}$ without disclosing the sparse locations (i.e., query indexes) in $\mathbf{S}$ and the plaintexts in $\mathbf{U}$. Similarly, we also solve the problem by proposing two secure two-party computation protocols. In specific, one is for $\mathbf{U}\mathbf{D}$ and the other is for $\mathbf{U}\mathbf{S}^{T}$. Once the intermediate shares are obtained, the two participants jointly output the final shares for $\mathbf{Z}$. In the following paragraphs, we describe the design rationales and implementation details.

The first core task in $\mathsf{\Pi}_\mathrm{sen}$ is computing $\mathbf{U}\mathbf{D}$. Recall that matrix $\mathbf{D}$ is a diagonal matrix and encompasses no sparse locations. In addition, the computing processing of $\mathbf{U}\mathbf{D}$ can be decomposed by calculating certain times of Hadamard products as shown in Equation \ref{entrywise multiplication of UD}. Thus, if the sparse locations in $\mathbf{U}$ need to be concealed, we have to let party $\mathsf{P_0}$ who holds $\mathbf{U}$ to send PIR queries to party $\mathsf{P_1}$ to fetch element values in $\mathbf{D}$. However, intuitively, the costs brought by invoking PIR protocols should be higher than directly encrypting the entire diagonal vector $\mathbf{d}$ (i.e., $\mathbf{D}$) and sharing it with $\mathsf{P_0}$ for ciphertext-plaintext HE evaluation. Some other secure data transfer protocols such as OT \cite{yadav2022survey} can achieve the same function yet still incuse more overheads than straightforward encryption. Even in the sparse location insensitive case, only column sparsity $\mathbf{loc}_{\mathbf{S}}$ can be utilized to compress the costs. Because if different rows in $\mathbf{U}$ report different sparse locations, to be computable, the vector $\mathbf{d}$ has to be packed and encrypted accordingly. Therefore, the increased costs in party $\mathsf{P_1}$ will be much higher than decreased costs in party $\mathsf{P_0}$. Moreover, the ciphertext volume sent from $\mathsf{P_1}$ to $\mathsf{P_0}$ will expands by $l\times$ (the number of rows in $\mathbf{U}$). To this end, in scheme $\mathsf{\Pi}_\mathrm{sen}$, we choose to compute $\mathbf{U}\mathbf{D}$ without utilizing data sparsity. At a high level, we follow the basic work flow of $\mathsf{\Pi}_\mathrm{ins}$ to implement the secure computing protocol for $\mathbf{U}\mathbf{D}$. The efficiency boosting trick, SV packing method for Hadamard product, is applied.

%US是咋用PIR算的，然后简述一下优化方案：1. 打包和PIR结合。 2. 用新的高效率的PIR方案。3.每个部分用的加密方案以及为啥用该方案。

The second core task in $\mathsf{\Pi}_\mathrm{sen}$ is securely computing $\mathbf{U}\mathbf{S}^{T}$. Recall that the matrix $\mathbf{S}$ is extreme sparse \cite{cui2021exploiting} ($\leq 0.02\%$). Straightforward encryption of $\mathbf{S}$ leads to prohibitive costs. To alleviate this issue and exploit data sparsity, PIR is employed by $\mathsf{P_1}$ to fetch values in matrix $\mathbf{U}$ corresponding to the sparse locations in $\mathbf{S}$. For instance, to compute inner product $\mathbf{U}[0,*]\cdot \mathbf{S}[*,0]$, $\mathsf{P_1}$ first issues PIR queries with non-sparse locations in $\mathbf{S}^{T}[*,0]$ as the index to $\mathsf{P_0}$. Upon receiving the returned values, $\mathsf{P_0}$ and $\mathsf{P_1}$ can directly compute the inner product without considering the sparse values. Recall that $\mathbf{S}\in \mathbb{Z}_{p}^{m\times m}$, where $m$ indicates the number of users in the social platform, which is commonly large. Thus, the computational cost will be significantly reduced if $\mathbf{S}$ is extremely sparse. In addition, we further compress the computation/communication costs by proposing the following optimizations.

\begin{itemize}
  \item \textbf{Compress the encryption cost on $\mathsf{P_0}$.} \textit{Recall that the PIR protocol cannot preserve the privacy of queried data.} To protect the privacy of $\mathbf{U}$ and support secure matrix multiplication, recent work \cite{cui2021exploiting} applied PHE to encrypt the entire matrix $\mathbf{U}$. This operation imposes heavy encryption overheads on $\mathsf{P_0}$.
  We compress the encryption cost by designing a protocol that is compatible with SV packing method. It is non-trivial to make this idea workable. First, on the $\mathsf{P_0}$ side, we reorganize the query index to fit the packing operation. Second, if the packing size is $s$, $\mathsf{P_0}$ partitions the rows in $\mathbf{U}$ and packs them using SV method. Third, on the $\mathsf{P_1}$ side, the sparse matrix $\mathbf{S}$ is partitioned and packed in the same way as $\mathbf{U}$. Forth, the random factors used for remasking the encrypted result need to be carefully designed to guarantee correctness and security simultaneously. To achieve this goal, the encrypted results are extended from a $l\times m$ matrix to a $l\times m \times s$ tensor. In doing so, the encryption costs on $\mathsf{P_0}$ are roughly compressed by $s$.

  \item \textbf{Compress the communication cost.} \\
  1). The query history is recorded as a table $\mathsf{T}$ and used to avoid repeat PIR processing with the same index. $\mathsf{P_1}$ refers to $\mathsf{T}$ before issuing PIR query. \\
  2). We propose to apply a fast and compact PIR protocol $\mathsf{MulPIR}$ \cite{ali2021communication} to further compress the upload and download costs by adopting the following two tricks. \\
  \textbf{[Compress the upload].} In the context of PIR \cite{angel2018pir}, the query issuer needs to encrypt (i.e, FV encryption \cite{fan2012somewhat}) the index with the public key. In concrete, the FV ciphertext is a tuple $\{\mathsf{CT_0},\mathsf{CT_1}\}$ in $R_{N, q}^{2}$. A key insight is that we can treat element $\mathsf{CT_0}$ as a random factor sampled from $R_{N,q}$. If the query issuer directly shares a random seed $\lambda \in \{0,1\}^{\kappa}$ in advance with the server, the server can locally reconstruct $\mathsf{CT_0}$. In doing so, the size of the encrypted query index is compressed by a factor $2\times$. \\
  \textbf{[Compress the download].} In \cite{angel2018pir}, the returned query result is FV ciphertexts that no further processing is needed that are decrypted by the query issuer. Therefore, we can use the modulus switching \cite{fan2012somewhat} method to reduce the ciphertext size. Given a ciphertext $\mathsf{CT} \in R_{N,q}^{2}$ from the query response, the server can apply modulus switching to transfer $\mathsf{CT}$ to a new ciphertext $\mathsf{CT}^{\prime} \in  R_{N,q^{\prime}}^{2}$. In practice, $q^{\prime} \ge p^{2}$ is chosen large enough for correct decryption, where $p$ is the plaintext space. Thus, the download size is reduced roughly by $\log_{2}{q}/(2\log{}{p})$. For instance, if the prime $q^{\prime}$ is set around $2^{25}$, the download cost will be reduced by a factor $2.4\times$.
\end{itemize}

In Figure \ref{figure:implementation of sensitive scheme}, we have described the implementation details for $\mathsf{\Pi}_\mathrm{sen}$. As aforementioned, the computation of $\mathbf{U}\mathbf{D}$ is similar to $\mathsf{\Pi}_\mathrm{ins}$, we also use SV packing method and RLWE HE scheme to pack and encrypt the input matrices $\mathbf{U}$ and $\mathbf{D}$. When computing $\mathbf{U}\mathbf{S}^{T}$, in order to adopt the packing method on $\mathsf{P_0}$ for the encryption of $\mathbf{U}$, we propose to packing $\mathbf{S}$ in the same way. Thus, each non-sparse subvector in $\mathbf{S}$ is a $s$-length vector (same as the packing size on $\mathbf{U}$). Recall that SV packing for PHE encrypted ciphertext cannot support rotation operation. To compute the inner product over ciphertext, we design a new and efficient SV-compatible secure two-party vector inner product method. To ease understanding, we give a toy example as follows.

\textbf{A Toy Example of computing inner product.} Assume that party $\mathsf{P_0}$ holds input vector $\mathbf{x} = (1,2,3,4,5,6,7,8,9)$, and party $\mathsf{P_1}$ holds sparse input vector $\mathbf{y} = (1,2,0,0,0,0,0,8,0)$. The packing size is set to $3$. Then $\mathsf{P_0}$ packs $\mathbf{x}$ into three subvectors: $\mathsf{SV}_{\mathbf{x_{0}}} \leftarrow \mathsf{SV.En}(1,2,3), \mathsf{SV}_{\mathbf{x_{1}}} \leftarrow \mathsf{SV.En}(4,5,6), \mathsf{SV}_{\mathbf{x_{2}}} \leftarrow \mathsf{SV.En}(7,8,9)$. The encrypted subvectors are written as $\mathsf{PCT}_{\mathbf{x_{0}}}, \mathsf{PCT}_{\mathbf{x_{1}}}, \mathsf{PCT}_{\mathbf{x_{2}}}$. On $\mathsf{P_1}$ side, $\mathbf{y}$ is partitioned into three subvectors $(1,2,0),(0,0,0),(0,8,0)$ denoted as $\mathbf{y_{0}},\mathbf{y_{1}},\mathbf{y_{2}}$, respectively. The non-sparse subvectors are then packed using SV, which are denoted as $\mathsf{SV}_{\mathbf{y_{0}}}, \mathsf{SV}_{\mathbf{y_{2}}}$. Then, $\mathsf{P_1}$ issues PIR queries to $\mathsf{P_0}$ to fetch the corresponding subvectors $\mathsf{PCT}_{\mathbf{x_{0}}}, \mathsf{PCT}_{\mathbf{x_{2}}}$. Upon receiving the results, $\mathsf{P_1}$ operates $\mathsf{PCT}_{\mathbf{x}\cdot \mathbf{y}} \leftarrow  (\mathsf{PCT}_{\mathbf{x_{0}}}\otimes \mathsf{SV}_{\mathbf{y_{0}}})\oplus (\mathsf{PCT}_{\mathbf{x_{2}}}\otimes \mathsf{SV}_{\mathbf{y_{2}}})$. We interpret this operation in the view of plaintext domain as $(1, 68, 0) \leftarrow (1\times 1, 2\times 2, 0)+(0,8\times 8,0)$. In another word, $\mathsf{PCT}_{\mathbf{x}\cdot \mathbf{y}}$ is a ciphertext of vector $(1, 68, 0)$. To remask $\mathsf{PCT}_{\mathbf{x}\cdot \mathbf{y}}$, $\mathsf{P_1}$ uniformly samples a random vector $\mathbf{r} = (r_0, r_1, r_2)$, where $r = r_0 + r_1 + r_2$, and operates $\mathsf{PCT}_{\mathbf{x}\cdot \mathbf{y}}^{\prime} \leftarrow \mathsf{PCT}_{\mathbf{x}\cdot \mathbf{y}} \oplus \mathsf{SV.En}(\mathbf{r})$. The masked ciphertext $\mathsf{PCT}_{\mathbf{x}\cdot \mathbf{y}}^{\prime}$ is then returned to $\mathsf{P_0}$, which is a ciphertext of vector $(1+r_0, 68+r_1, r_2)$. $\mathsf{P_0}$ can recover this vector and sum all the elements to obtain the masked inner product $\mathbf{x}\cdot \mathbf{y}+r = 69+r$. Note that the modulo operations on the plaintext domain are omitted for simplicity.

By using our proposed SV-compatible secure inner product method, $\mathbf{U}\mathbf{S}^{T}$ can be correctly and efficiently computed without any decryption operation in the middle. Moreover, the lightweight character of PHE (compared to RLWE HE) and the encryption acceleration technique SV are well leveraged without adopting any rotation operation. The random factor in this method is expanded to a tensor rather than a matrix to guarantee input privacy. With such efficiency-boosting processing, the additional overhead brought by random tensor generation and SV packing is negligible.

\section{Security Analysis}
\label{sec:security analysis}
In this section, we prove the security of the proposed two schemes $\mathsf{\Pi}_\mathrm{ins}$, $\mathsf{\Pi}_\mathrm{sen}$ against the semi-honest probabilistic polynomial time (PPT) adversaries $\mathcal{A}$. In specific, we use the simulation paradigm \cite{lindell2017simulate} to construct simulators that make the simulated views indistinguishable from the real views. We first define the ideal functionalities for $\mathsf{\Pi}_\mathrm{ins}$ and $\mathsf{\Pi}_\mathrm{sen}$ to specify the inputs and outputs. Then we elaborate on the simulator construction details by bulleting the hybrid arguments.

%%%%%%%%%%%%%%%%%%%%%%%%%%%%%%%%%%%%方案一的安全证明
\subsection{Security of $\mathsf{\Pi}_\mathrm{ins}$}
\label{subsec: Security of ins}

%先给出ideal function
% \renewcommand\tablename{Figure}
% \renewcommand \thetable{\arabic{table}}
% \setcounter{table}{5}
\begin{table}
\centering
\begin{tabular}{|p{8cm}|}
\hline
\begin{center}
  $\mathcal{F}_{\mathrm{ins}}$: Functionality of $\mathsf{\Pi}_\mathrm{ins}$
\end{center}
%输入和输出
%\textbf{Public Parameters:} $\mathsf{pp} = \{\alpha, \mathsf{HE.pp}, \mathsf{pk_R}, l, m, l_w, m_w\}$.   \\
%$\bullet$ $\{l,m\}$ are the input matrix dimensions, and $\{l_w,m_w\}$ are the partition window size, where %$0<l_w\leq l$, $0<m_w\leq m$, and $l_w {m_w}\leq N$ holds. \\
\textbf{Input:} $\mathsf{P_1}$ holds the social matrix $\mathbf{S} \in \mathbb{Z}_p^{m\times m}$, and the diagonal matrix $\mathbf{D}\in \mathbb{Z}_p^{m\times m}$, $\mathsf{P_0}$ holds the matrix $\mathbf{U}\in \mathbb{Z}_{p}^{l\times m}$. $\mathsf{P_0}, \mathsf{P_1}$ shares the sparse locations to each other in matrices $\mathbf{U}, \mathbf{S}$, and the public parameters $\mathsf{pp}$ as defined in Figure \ref{figure:implementation of insensitive scheme}. \\
\textbf{Output:} $\mathsf{P_0}$ and $\mathsf{P_1}$ obtain two shares $\langle \mathbf{Z}\rangle_0, \langle \mathbf{Z}\rangle_1\in \mathbb{Z}_{p}^{l\times m}$, respectively, where $\mathbf{Z} = \alpha\mathbf{U}\mathbf{D}/2-\alpha\mathbf{U}\mathbf{S}^{T}$.   \\
\hline
\end{tabular}
\vspace{2pt}
\caption{Functionality of $\mathsf{\Pi}_\mathrm{ins}$.}
\label{figure:ideal functionality of ins}
\end{table}

%%%%%%%%%%%%%%%%%%%%%%%%%%%%%%%%%%%%%%%%%%%%%%%%%%%%%%%%%%%%%%%%%%%%%%%%%%%%%%%%%%%%%%%%%%%%%%
$\mathsf{\Pi}_\mathrm{ins}$ is secure against the semi-honest PPT $\mathcal{A}$, which is formalized as following theorem.
\begin{Theorem}[Security of $\mathsf{\Pi}_\mathrm{ins}$]
\label{theorem:security prove of ins}
If the crypto-system RLWE HE used in $\mathsf{\Pi}_\mathrm{ins}$ are semantically secure against the semi-honest adversaries, then the proposed protocol $\mathsf{\Pi}_\mathrm{ins}$ is secure against the semi-honest PPT $\mathcal{A}$.
\end{Theorem}

\textit{Proof:} Given $\mathcal{F}_{\mathrm{ins}}$ defined in Figure \ref{figure:ideal functionality of ins} for $\mathsf{\Pi}_\mathrm{ins}$, we prove the security of $\mathsf{\Pi}_\mathrm{ins}$ against PPT semi-honest $\mathcal{A}$ as follows.

\noindent \textbf{Semi-honest $\mathsf{P_0}$ security.} In this setting, it is assumed that a semi-honest adversary $\mathcal{A}$ compromises the participant $\mathsf{P_0}$. In the following content, the existence of a simulator $\mathsf{Sim_0}$ is demonstrated by constructing a list of hybrid arguments. $\mathsf{Sim_0}$ is allowed to access the inputs and outputs of $\mathsf{P_0}$. The goal of simulator $\mathsf{Sim_0}$ is to make the simulated $\mathsf{P_0}$'s view indistinguishable from the real view.

\begin{itemize}
    \item $\mathrm{Hyb_1}:$ This hybrid follows the real execution of $\mathsf{\Pi}_\mathrm{ins}$.

    \item $\mathrm{Hyb_2}:$ In this hybrid, $\mathsf{Sim_0}$ receives the sparse location $\mathbf{loc}_{\mathbf{U}}$ from the output of $\mathsf{P_0}$. instead of taking the original matrix $\mathbf{S}$ as the input, $\mathsf{Sim_0}$ randomly samples a sparse matrix from $\mathbb{Z}_{p}^{m\times m}$. A diagonal matrix is generated accordingly by using the random matrix. Afterward, $\mathsf{Sim_0}$ compresses the diagonal vector according to $\mathbf{loc}_{\mathbf{U}}$. Then $\mathsf{Sim_0}$ encodes and encrypts it with public key $\mathsf{pk_R}$. Afterward, the ciphertext is sent to $\mathsf{P_0}$. Since $\mathsf{P_0}$ cannot access the private key $\mathsf{sk_R}$ of RLWE HE, the indistinguishability between this hybrid and the real view is guaranteed by the semantic security of the RLWE HE.

    \item $\mathrm{Hyb_3}:$ In this hybrid, $\mathsf{Sim_0}$ uses the generated random matrix in $\mathrm{Hyb_2}$ as the input. $\mathsf{Sim_0}$ sends its sparse locations to $\mathsf{P_0}$. It is possible because the sparse location is considered as the public information between the two participants. Then $\mathsf{Sim_0}$ uses the sampled matrix as the input to compress, pack and encrypt it using $\mathsf{pk_R}$. The obtained ciphertexts are sent to $\mathsf{P_0}$. Same as $\mathrm{Hyb_2}$, the view between this hybrid and the real world is indistinguishable due to the use of semantic secure RLWE HE.

    \item $\mathrm{Hyb_4}:$ In this hybrid, instead of repeating the real execution of $\mathsf{\Pi}_\mathrm{ins}$, $\mathsf{Sim_0}$ firstly randomly samples a random matrix from $\mathbb{Z}_{p}^{l\times m}$ and the spare column is set to $0$ according to $\mathbf{loc}_{\mathbf{U}}$. This operation aims to simulate a secret share of $\mathbf{U}\mathbf{D}$. Then $\mathsf{Sim_0}$ randomly samples a matrix from $\mathbb{Z}_{p}^{l\times m}$ and the sparse locations are set to $0$ according to $\mathbf{loc}_{\mathbf{U}}$ and $\mathbf{loc}_{\mathbf{S}}$. This operation aims to simulate a secret share of $\mathbf{U}\mathbf{S}^{T}$. At last, $\mathsf{Sim_0}$ adds these two shares as the output share. Since the encryption of the generated random secrete share is indistinguish to the ciphertext generated by $\mathsf{P_0}$, which is guaranteed by the randomness of RLWE HE ciphertext. Thus, this hybrid is indistinguishable from the real view, which finishes the construction of $\mathsf{Sim_0}$.
\end{itemize}

\noindent \textbf{Semi-honest $\mathsf{P_1}$ security.} In this setting, it is assumed that a semi-honest adversary $\mathcal{A}$ compromises the participant $\mathsf{P_1}$. Similarly, the existence of simulator $\mathsf{Sim_1}$ is demonstrated by hybrid arguments. $\mathsf{Sim_1}$ is allowed to access the inputs and outputs of $\mathsf{P_1}$. The goal of simulator $\mathsf{Sim_1}$ is to make the simulated $\mathsf{P_1}$'s view indistinguishable from the real view.

\begin{itemize}
    \item $\mathrm{Hyb_1}:$ This hybrid follows the real execution of $\mathsf{\Pi}_\mathrm{ins}$.

    \item $\mathrm{Hyb_2}:$ In this hybrid, instead of taking original matrix $\mathbf{U}$ as the input, $\mathsf{Sim_1}$ randomly samples a matrix $\mathbf{U}^{*}$ from $\mathbb{Z}_{p}^{l\times m}$ and shares the sparse locations $\mathbf{loc}_{\mathbf{U}}$ with $\mathsf{P_1}$. Then $\mathsf{Sim_1}$ packs the compressed matrix and follows the real execution. $\mathsf{Sim_1}$ randomly samples another matrix $\mathbf{R}\in \mathbb{Z}_{p}^{l\times m}$ with same sparse locations as $\mathbf{U}^{*}$. Then, $\mathsf{Sim_1}$ uses  $\mathbf{U}^{*}$ and $\mathbf{R}$ to generates and remasks the result. Since two randomly masked matrices with the same shape are indistinguishable, that indicates the indistinguishability between this hybrid and real view.

    \item $\mathrm{Hyb_3}:$ In this hybrid, $\mathsf{Sim_1}$ first compresses the random matrix $\mathbf{U}^{*}$ by checking $\mathbf{loc}_{\mathbf{S}}$. Then it is packed by invoking mapping function $\pi_{1}$. Afterward, $\mathsf{Sim_1}$ randomly samples another matrix $\mathbf{Q}\in \mathbb{Z}_{p}^{l\times m}$ with same format as $\mathbf{U}^{*}$. Given these two matrices, $\mathsf{Sim_1}$ generates and remasks the results in the same way as real execution. At last, $\mathsf{Sim_1}$ returns the masked results to $\mathsf{P_0}$. Similarly, the masking operation guarantees the indistinguishability between this hybrid and the real view.

    \item $\mathrm{Hyb_4}:$ In this hybrid, $\mathsf{Sim_1}$ outputs the addition of two random matrices $\mathbf{R}, \mathbf{Q}$ as the secret share. Since $\mathsf{P_0}$ cannot access $\mathbf{R}$ or $\mathbf{Q}$, the output matrix is indistinguishable from any other random matrix with the same shape. Thus, this finishes the construction of simulator $\mathsf{Sim_1}$.
\end{itemize}

%%%%%%%%%%%%%%%%%%%%%%%%%%%%%%%%%%方案二的安全证明
\subsection{Security of $\mathsf{\Pi}_\mathrm{sen}$}
\label{subsec: Security of sen}

%先给出ideal $\mathrm{Hyb_4}:$
% \renewcommand\tablename{Figure}
% \renewcommand \thetable{\arabic{table}}
% \setcounter{table}{6}
\begin{table}
\centering
\begin{tabular}{|p{8cm}|}
\hline
\begin{center}
  $\mathcal{F}_{\mathrm{sen}}$: Functionality of $\mathsf{\Pi}_\mathrm{sen}$
\end{center}
%输入和输出
%\textbf{Public Parameters:} $\mathsf{pp} = \{\alpha, \mathsf{HE.pp}, \mathsf{pk_R}, \mathsf{pk_P}, l, m, s\}$.   \\
%$\bullet$ $\{l,m\}$ are the input matrix dimensions, and $s$ is the partition window size (i.e., the packing size for PHE crypto-system). \\
\textbf{Input:} $\mathsf{P_1}$ holds the social matrix $\mathbf{S} \in \mathbb{Z}_p^{m\times m}$, and the diagonal matrix $\mathbf{D}\in \mathbb{Z}_p^{m\times m}$, $\mathsf{P_0}$ holds the matrix $\mathbf{U}\in \mathbb{Z}_{p}^{l\times m}$, and the public parameters $\mathsf{pp}$ as defined in Figure \ref{figure:implementation of sensitive scheme}.  \\
\textbf{Output:} $\mathsf{P_0}$ and $\mathsf{P_1}$ obtain two shares $\langle \mathbf{Z}\rangle_0, \langle \mathbf{Z}\rangle_1\in \mathbb{Z}_{p}^{l\times m}$, respectively, where $\mathbf{Z} = \alpha\mathbf{U}\mathbf{D}/2-\alpha\mathbf{U}\mathbf{S}^{T}$.   \\
\hline
\end{tabular}
\vspace{2pt}
\caption{Functionality of $\mathsf{\Pi}_\mathrm{sen}$.}
\label{figure:ideal functionality of sen}
\end{table}

$\mathsf{\Pi}_\mathrm{sen}$ is secure against the semi-honest PPT $\mathcal{A}$, which is formalized as following theorem.
\begin{Theorem}[Security of $\mathsf{\Pi}_\mathrm{sen}$]
\label{theorem:security prove of sen}
If RLWE HE, PHE, and PIR protocol used in $\mathsf{\Pi}_\mathrm{sen}$ are semantically secure against the semi-honest adversaries, then the proposed protocol $\mathsf{\Pi}_\mathrm{sen}$ is secure against semi-honest PPT $\mathcal{A}$.
\end{Theorem}

\textit{Proof:} Given $\mathcal{F}_{\mathrm{sen}}$ defined in Figure \ref{figure:ideal functionality of sen} for $\mathsf{\Pi}_\mathrm{sen}$, we prove the security of $\mathsf{\Pi}_\mathrm{sen}$ against PPT semi-honest $\mathcal{A}$ as follows.

\noindent \textbf{Semi-honest $\mathsf{P_0}$ security.} Assume that a semi-honest adversary $\mathcal{A}$ compromises $\mathsf{P_0}$. The existence of a simulator $\mathsf{Sim_0}$ is demonstrated by a list of hybrid arguments. $\mathsf{Sim_0}$ can access the inputs and outputs of $\mathsf{P_0}$. The goal of $\mathsf{Sim_0}$ is to make the simulated $\mathsf{P_0}$'s view indistinguishable from the real view.

\begin{itemize}
    \item $\mathrm{Hyb_1}:$ This hybrid follows the real execution of $\mathsf{\Pi}_\mathrm{ins}$.

   \item $\mathrm{Hyb_2}:$ In this hybrid, instead of taking the original matrix $\mathbf{S}$ as the input,
    $\mathsf{Sim_0}$ randomly generates a sparse matrix from $\mathbb{Z}_{p}^{m\times m}$. Then the corresponding diagonal matrix is generated upon the random matrix. Afterward, its diagonal vector is packed and encrypted with the public key $\mathsf{pk_R}$ that is exactly the same as the real execution. The generated ciphertext is then sent to $\mathsf{P_0}$. Since the private key $\mathsf{sk_R}$ is kept confidential from $\mathsf{P_0}$, it cannot distinguish this hybrid from the real view due to the semantic security of used RLWE HE.

    \item $\mathrm{Hyb_3}:$ In this hybrid, $\mathsf{Sim_0}$ issues PIR queries to $\mathsf{P_0}$ for non-sparse values in the random input matrix. The semantic security of the underlying PIR protocol indicates that this hybrid is indistinguishable from the real view.

    \item $\mathrm{Hyb_4}:$ In this hybrid, $\mathsf{Sim_0}$ just follow the real execution to extract the PIR query results. Then $\mathsf{Sim_0}$ randomly sample a tensor from $\mathbb{Z}_{p}^{l\times m\times s}$, and use it to remask the corresponding encrypted results. The masked ciphertexts will be returned to $\mathsf{P_0}$. Upon receiving the ciphertext, $\mathsf{P_0}$ can decrypt it and obtain the masked plaintexts. Since the random tensor is kept private to $\mathsf{P_0}$, this hybrid is indistinguishable from the real view.

   \item $\mathrm{Hyb_5}:$ In this hybrid, $\mathsf{Sim_0}$ decrypts the masked ciphertexts received from $\mathsf{P_0}$ using $\mathsf{sk_R}$ as the first part of the secret share (for $\mathbf{U}\mathbf{D}$). Then the random tensor is aggregated to the matrix in $\mathbb{Z}_{p}^{l\times m}$ as the second part of the share (for $\mathbf{U}\mathbf{S}^{T}$). These two secret shares are then combined as the output. The randomness of the share is preserved if the random tensor is confidential to $\mathsf{P_0}$. Thus, this hybrid is indistinguishable from the real view. This finishes the construction of $\mathsf{Sim_0}$.
\end{itemize}

\noindent \textbf{Semi-honest $\mathsf{P_1}$ security.} Assume that a semi-honest adversary $\mathcal{A}$ compromises $\mathsf{P_1}$. The existence of a simulator $\mathsf{Sim_1}$ can be proved by a list of hybrid arguments. $\mathsf{Sim_1}$ can access the inputs and outputs of $\mathsf{P_1}$. The goal of $\mathsf{Sim_1}$ is to simulate $\mathsf{P_1}$'s view and make it indistinguishable from the real view.
%simulator模拟P0的行为。对P1进行欺骗。
\begin{itemize}
    \item $\mathrm{Hyb_1}:$ This hybrid follows the real execution of $\mathsf{\Pi}_\mathrm{ins}$.

    \item $\mathrm{Hyb_2}:$ In this hybrid, instead of using the original matrix $\mathbf{U}$, $\mathsf{Sim_1}$ randomly samples a matrix from $\mathbb{Z}_{p}^{l\times m}$ as the input. Afterward, $\mathsf{Sim_1}$ randomly samples another matrix $\mathbf{R} \in \mathbb{Z}_{p}^{l\times m}$. The first random matrix is fed into the packing and homomorphic operations to obtain the encrypted result. Then $\mathbf{R} $ is used to remask the result in the ciphertext domain. The masked results are returned to $\mathsf{P_1}$. Due to the semantic security provided by RLWE HE, $\mathsf{P_1}$ cannot distinguish the output of this hybrid from the real view.

    \item $\mathrm{Hyb_3}:$ In this hybrid, $\mathsf{Sim_1}$ continues to use the same random matrix as the input and follows the real execution to generate the encrypted database for PIR query. Once receiving a PIR query, $\mathsf{Sim_1}$ response it to $\mathsf{P_1}$ accordingly. The returned queried values are all encrypted by PHE, which guarantees that it is indistinguishable from the ciphertext of a real message. That is to say, this hybrid is indistinguishable from the real view.

    \item $\mathrm{Hyb_4}:$ In this hybrid, $\mathsf{Sim_1}$ takes the random matrix $\mathbf{R}$ as the first secret share (for $\mathbf{U}\mathbf{D}$). Then it decrypts the received results as the second part of the secret share (for $\mathbf{U}\mathbf{S}^{T}$). Thus the addition of these two parts is shared with $\mathsf{P_1}$ as the output. The randomness of the share is preserved since $\mathbf{R}$ is confidential to $\mathsf{P_1}$. Thus, this hybrid is indistinguishable from the real view. This finishes the construction of $\mathsf{Sim_1}$.
\end{itemize}

%%%%%%%%%%%%%%%%%%%%%%%%%%%%%%%%%%%%%%%%%%%%%%%%%%%%%%%%%%%%%%%%%%%%%%%%%%%%%%%%%%%%%%%%%%%%%%%%%%%%%%%%%

\section{Performance Evaluation}
\label{sec:performance evaluation}
In this section, we elaborate on the performance of our proposed two constructs $\mathsf{\Pi}_\mathrm{ins}$, $\mathsf{\Pi}_\mathrm{sen}$, and compare the experimental results with the state-of-the-art scheme $\mathsf{S^{3}Rec}$ \cite{cui2021exploiting}. In specific, both the sparse location insensitive and sensitive schemes are comprehensively evaluated in terms of computation, communication, storage, and accuracy. The experiments are conducted over two popular benchmark datasets, that are Epinions \cite{massa2007trust} and LibraryThing (LiThing) \cite{zhao2015improving}. In addition, since social recommendation data is highly private and hard to be acquired from commercial organizations subject to legal requirements, we synthetic two large-scale datasets to simulate the real-world performance. The impact of social data sparsity is evaluated by varying the data density.

\subsection{Implementation Settings}
\label{subsec:implementation settings}
The experiments are conducted on the computing machine with Inter(R) Xeon(R) E5-2697 v3 2.6GHz CPUs with 28 threads on 14 cores and 64GB memory. The programming language is C++. The tests are carried out in a local network with on average roughly 3ms latency. We use mainstream open-source libraries to implement cryptographical tools. For RLWE/LWE HE scheme, the SEAL \cite{seal} library is used. The cyclotomic ring dimension is chosen as $2^{13}$ (i.e., $N=2^{13}$) and the ciphertext space is chosen as $2^{47}$. It guarantees 80-bit security. For PHE scheme (Paillier) \cite{paillier1999public}, we adopt libpaillier \cite{libpaillier} and choose 128-bit security. The public parameters in underlying building blocks including the social recommendation system and the used PIR scheme are all set exactly the same as the original papers \cite{ma2011recommender,ali2021communication}. When implementing the comparison scheme $\mathsf{S^{3}Rec}$ \cite{cui2021exploiting}, the general MPC library ABY \cite{demmler2015aby} and SealPIR \cite{angel2018pir} are applied by the same parameter settings. In all the experiments, the length of secret sharing is chosen to be 64 bits. To be clear, the sparse location insensitive/sensitive schemes of $\mathsf{S^{3}Rec}$ are denoted as $\mathsf{S^{3}Rec}_{\mathrm{ins}}$ and $\mathsf{S^{3}Rec}_{\mathrm{sen}}$, respectively. $\mathsf{Nospa}$ stands for the simulated scheme without considering the data sparsity. The remaining details will be given in the corresponding subsections.

\renewcommand\tablename{TABLE}
\renewcommand \thetable{\Roman{table}}
\setcounter{table}{0}
\setcounter{figure}{8}
\begin{table}[htb]
\centering
\caption{Testing dataset statitics}
\label{table:dataset}\
\setlength{\tabcolsep}{2mm}{
\begin{tabular}{c|c|c|c|c}
\hline
Dataset&user&item &social relation  &social density     \\
\hline
Epinions & 11,500 & 7,596 & 275,117 & 0.21\%       \\
\hline
LiThing & 15,039  & 14,957 & 44,710 &  0.02\% \\
\hline
\end{tabular}}
\end{table}

\textbf{Dataset.} To be consistent with the comparison scheme, the same testing datasets Epinions \cite{massa2007trust} and LibraryThing (LiThing) \cite{zhao2015improving} are adopted. Similar to $\mathsf{S^{3}Rec}$, if the interactions are less than 15, the corresponding users and items will be removed. However, as shown in Table \ref{table:dataset}, the scale of the testing data is insufficient to simulate the real-world situation. Up to now, the well-known E-commerce Amazon \cite{user_number_amazon} and social media giant Facebook \cite{user_number_facebook} are serving more than $1.5\times 10^{9}$ users. To make the performance evaluation results more convincing, we synthetic two large-scale datasets by expanding the user number with factor $10^{2}$ for the real datasets Epinions and LiThing, respectively. The simulated datasets for Epinions, and LiThing are written as SynEp, SynLi. In specific, the sparse level, as well as the distribution of the simulated datasets, are fixed exactly the same as the corresponding original datasets. Using synthetic large-scale datasets to simulate the performance is a common method \cite{zengy2022shadewatcher} when the real data is highly private and implies huge commercial interests. In addition, if the input data distribution and sparsity level remain unchanged, the reported results can precisely reflect the real performance.

\begin{figure}[htb]
\centering
\includegraphics[width=0.47\textwidth]{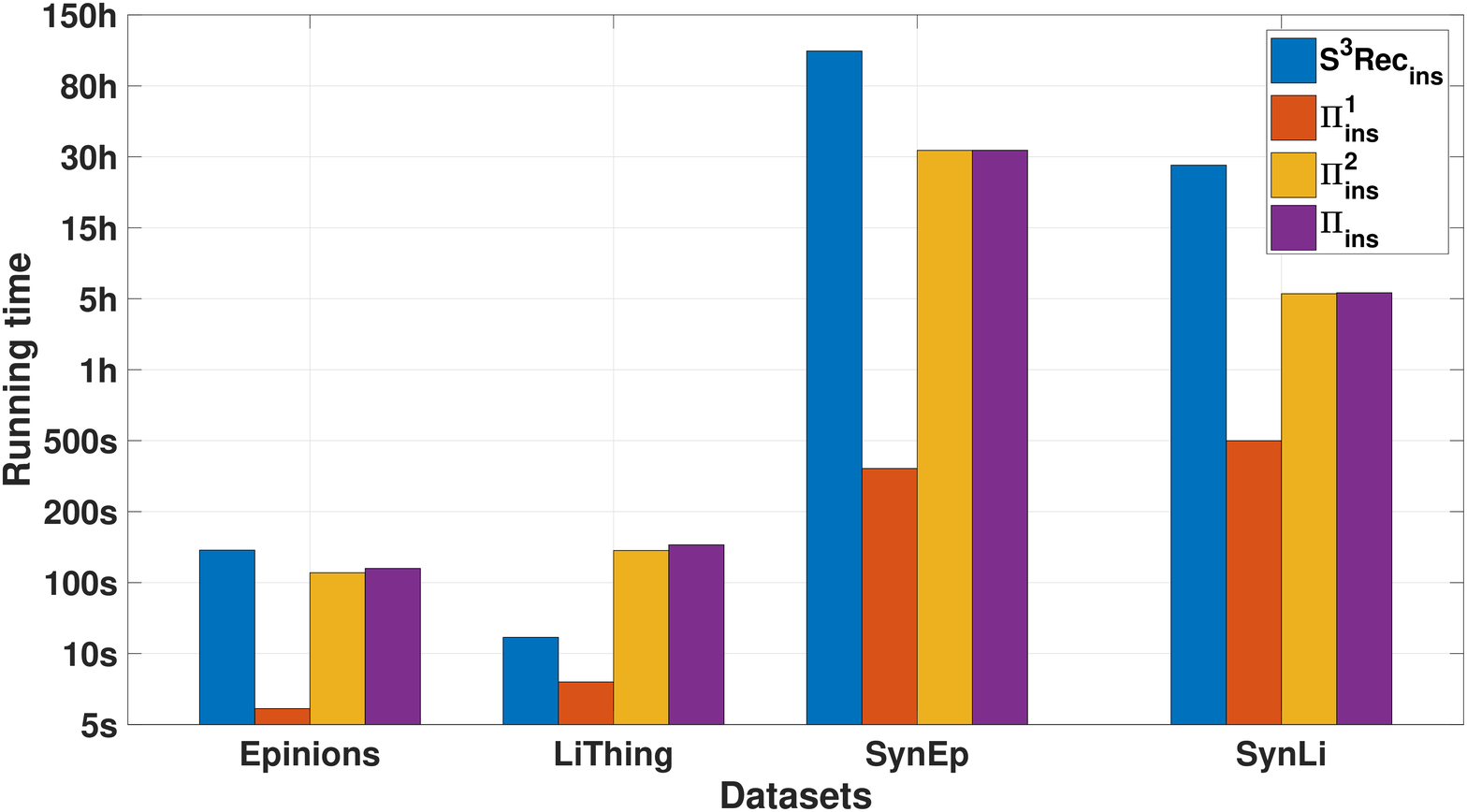}
\caption{Running time of $\mathsf{\Pi}_\mathrm{ins}$ and $\mathsf{S^{3}Rec}_{\mathrm{ins}}$.}
\label{fig.exp-Running time of ins}
\end{figure}

\begin{figure}[htb]
\centering
\includegraphics[width=0.47\textwidth]{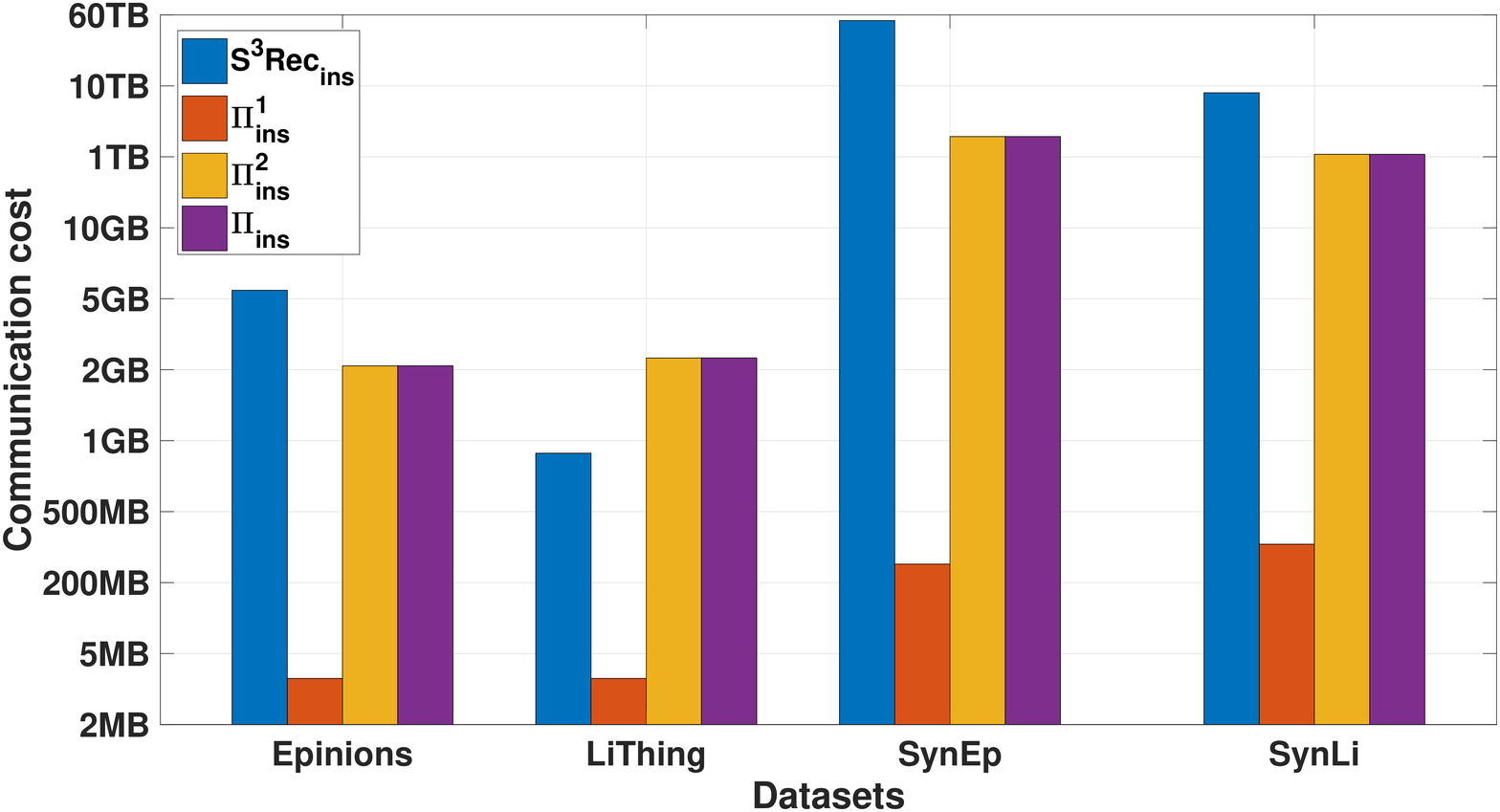}
\caption{Communication cost of $\mathsf{\Pi}_\mathrm{ins}$ and $\mathsf{S^{3}Rec}_{\mathrm{ins}}$.}
\label{fig.exp-commnunication of ins}
\end{figure}

\begin{figure}[htb]
\centering
\includegraphics[width=0.47\textwidth]{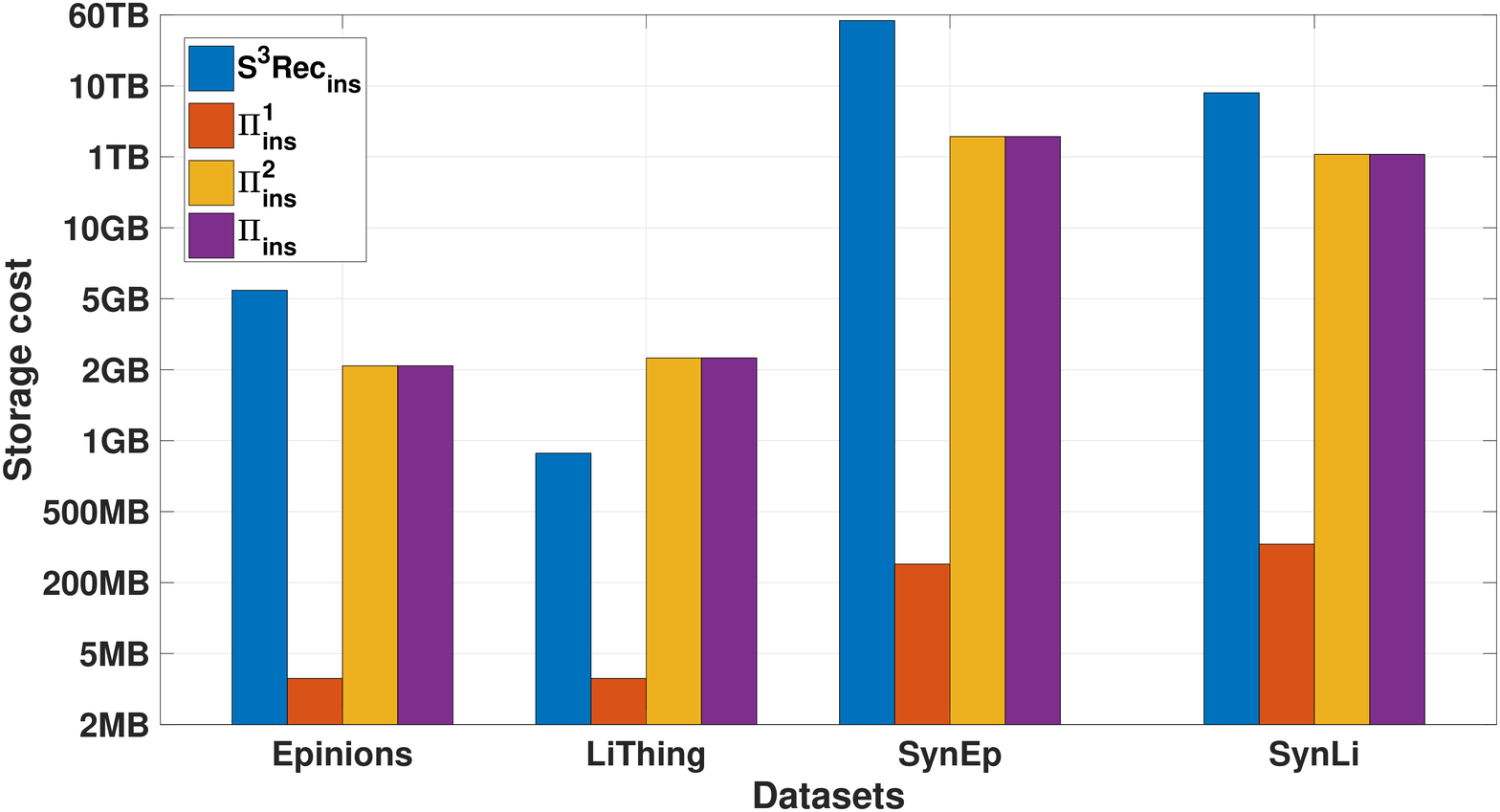}
\caption{Storage cost of $\mathsf{\Pi}_\mathrm{ins}$ and $\mathsf{S^{3}Rec}_{\mathrm{ins}}$.}
\label{fig.exp-storage of ins}
\end{figure}

\subsection{Performance Evaluation on $\mathsf{\Pi}_\mathrm{ins}$ and $\mathsf{S^{3}Rec}_{\mathrm{ins}}$}
\label{subsec:Performance evaluation on insensitive schemes}
%先对两个方案进行一个大概的回顾，给出计算，通信和存储开销的一个理论分析。然后给出具体情况。其中我们方案是UD和US分开做的，所以需要分别给出复杂度分析。后面的部分应该也是这个套路做。

In this part, we report the experimental results for our insensitive sparse location scheme $\mathsf{\Pi}_\mathrm{ins}$ and the comparison scheme $\mathsf{S^{3}Rec}_{\mathrm{ins}}$ \cite{cui2021exploiting}. We first briefly review the technical details of these two schemes and then give an asymptotic analysis of the performance. Finally, the experimental results are reported. Recall that, the main task of $\mathsf{\Pi}_\mathrm{ins}$ and $\mathsf{S^{3}Rec}_{\mathrm{ins}}$ is to securely compute $\alpha\mathbf{U}\mathbf{D}/2-\alpha\mathbf{U}\mathbf{S}^{T}$. In $\mathsf{S^{3}Rec}_{\mathrm{ins}}$, the authors solve this problem by using the existing secure two-party computation protocol ABY \cite{demmler2015aby} without modification. Given the input matrices $\mathbf{U} \in \mathbb{Z}_{p}^{l\times m}, \mathbf{D} \in \mathbb{Z}_{p}^{m\times m}, \mathbf{S} \in \mathbb{Z}_{p}^{m\times m}$ ($l$ is set to $20$), $\mathsf{S^{3}Rec}_{\mathrm{ins}}$ generates $lm^{2}$ Beaver’s triples to support matrix multiplication. To be fair, we also use PHE (Paillier) to implement Beaver’s triple for $\mathsf{S^{3}Rec}_{\mathrm{ins}}$. Note that, in $\mathsf{S^{3}Rec}_{\mathrm{ins}}$, both $\mathbf{U}\mathbf{D}$ and $\mathbf{U}\mathbf{S}^{T}$ are computed with exactly the method. In contrast, $\mathsf{\Pi}_\mathrm{ins}$ computes $\mathbf{U}\mathbf{D}$ and $\mathbf{U}\mathbf{S}^{T}$ with two different acceleration tricks. We use $\mathsf{\Pi}_\mathrm{ins}^{1}$ and $\mathsf{\Pi}_\mathrm{ins}^{2}$ to represent them and evaluate their performance, respectively.

\textbf{Computational costs.} The main cost of $\mathsf{S^{3}Rec}_{\mathrm{ins}}$ is generating the multiplication triples. For one triple, it needs to conduct three-time encryption, one-time decryption, 2 $\oplus$ operations, and two $\otimes$ operation. The SV packing method can also be applied to reduce computational costs for generating triples. However, compared to $\mathsf{S^{3}Rec}_{\mathrm{ins}}$, $\mathsf{\Pi}_\mathrm{ins}^{1}$ only needs one-time encryption for each packed message other than three times. For $\mathsf{\Pi}_\mathrm{ins}^{2}$, the non-sparse elements in matrix $\mathbf{S}$ are mapped directly into the polynomial coefficients. As mentioned in Section \ref{subsec:Sparse Location Insensitive Scheme}, the results (inner product) are implied in the coefficients. The computational cost is then reduced by $O(N/(l_w \times m_w))$, where $N$ is the degree of the polynomial, $l_w, m_w$ are the partition window sizes. Since the datasets Epinions and LiThing are small and extremely sparse, the packing slots (i.e, 8192) cannot be fully used if we choose the RLWE HE for $\mathsf{\Pi}_\mathrm{ins}^{1}$. Instead, we adopt the PHE scheme Paillier \cite{paillier1999public} as the encryption scheme. Note that, Paillier also supports SV packing and ciphertext-plaintext homomorphic operations. Compare to RLWE HE, Paillier provides fewer packing slots ($\approx 128$) and is unable to rotate the packed ciphertexts. Fortunately, $\mathsf{\Pi}_\mathrm{ins}^{1}$ is achieved by computing Hadamard inner products, which can be perfectly supported by Paillier. In contrast, when the input matrices are expanded datasets SynEp and SynLi, we adopt RLWE HE (i.e., FV \cite{fan2012somewhat}) to achieve $\mathsf{\Pi}_\mathrm{ins}^{1}$. As shown in Fig. \ref{fig.exp-Running time of ins}, the specific running time of $\mathsf{S^{3}Rec}_{\mathrm{ins}}$ and $\mathsf{\Pi}_\mathrm{ins}$ are given. For large-scale datasets SynEp and SynLi,  $\mathsf{\Pi}_\mathrm{ins}$ achieves roughly $10 \times$ and $5\times$ running time reduction.

\textbf{Communication costs.} In $\mathsf{S^{3}Rec}_{\mathrm{ins}}$, to generate one multiplication triple, two parties need to exchange three ciphertexts. Given the size of Paillier ciphertext $\omega$ bits, then the communication volume for each triple is $3\omega$ bits. By using the packing method, the communication per triple is reduced to $2\omega+\omega/\left \lfloor \omega /(2\iota +1 + \lambda) \right \rfloor$ \cite{demmler2015aby}, where $\iota$ is the length of a share and $\lambda$ is the security parameter. The total communication cost of $\mathsf{S^{3}Rec}_{\mathrm{ins}}$ is $(\phi lm^{2}+m) (2\omega+\omega/\left \lfloor \omega /(2\iota +1 + \lambda) \right \rfloor)$, where $\phi$ is the data density of input social matrix $\mathbf{S}$. In $\mathsf{\Pi}_\mathrm{ins}^{1}$, if the input data is small real datasets, the total communication volume is $(l \omega +1)\left \lceil m/s \right \rceil $, where $s$ is the packing size. Let $\varphi$ be the size of an RLWE HE ciphertext. $\mathsf{\Pi}_\mathrm{ins}^{1}$ introduces $(l \varphi +1)\left \lceil m/N \right \rceil$ bits communication on large simulated datasets. Assume that the extracted LWE ciphertext has $\gamma$-bit length, then $\mathsf{\Pi}_\mathrm{ins}^{2}$ needs $m\varphi \left \lceil \phi m/m_w \right \rceil+lm\gamma$ bits communication. As depicted in Fig. \ref{fig.exp-commnunication of ins}, for small real datasets Epinions and LiThing, $\mathsf{S^{3}Rec}_{\mathrm{ins}}$ and $\mathsf{\Pi}_\mathrm{ins}$ introduce $5.599$ GB, $2.168$ GB, $0.91$ GB and $2.499$ GB communication costs, respectively. In the large datasets SynEp and SynLi, $\mathsf{\Pi}_\mathrm{ins}$ can decrease the costs roughly by $15 \times$ and $7\times$.

\textbf{Storage costs.} In this paper, we mainly count the total storage costs of two participants introduced by the secure computing protocols. Although the ciphertexts will be decrypted and the used storage space will be released, the computing machine still needs to request sufficient storage space to compress the running time. Otherwise, limited storage space will become the bottleneck. Therefore, it is necessary to review the maximum storage cost. For $\mathsf{S^{3}Rec}_{\mathrm{ins}}$, the size of newly generated ciphertexts is exactly the same as the communication volume. For each multiplication triple, two secret shares, and four temporary parameters with the same length are generated. As aforementioned, the length of each share is set to 64 bits. For our scheme $\mathsf{\Pi}_\mathrm{ins}^{1}$ and $\mathsf{\Pi}_\mathrm{ins}^{2}$, we only needs one share for each participants. We report the maximum storage costs in Fig. \ref{fig.exp-storage of ins}, and the results show that the storage costs of the location insensitive schemes are close to their communication costs.

\begin{figure}[htb]
\centering
\includegraphics[width=0.47\textwidth]{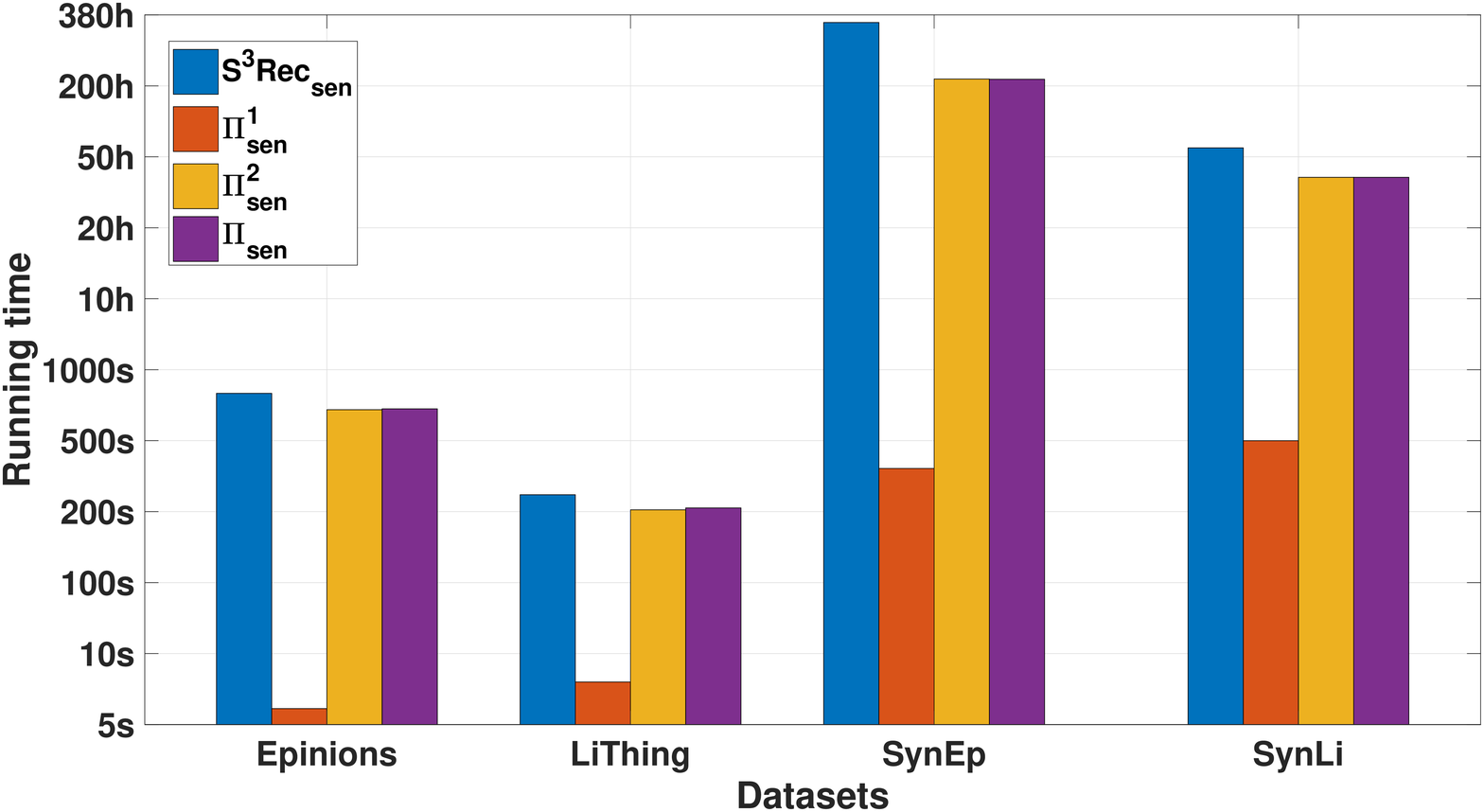}
\caption{Running time of $\mathsf{\Pi}_\mathrm{sen}$ and $\mathsf{S^{3}Rec}_{\mathrm{sen}}$.}
\label{fig.exp-Running time of sen}
\end{figure}

\begin{figure}[htb]
\centering
\includegraphics[width=0.47\textwidth]{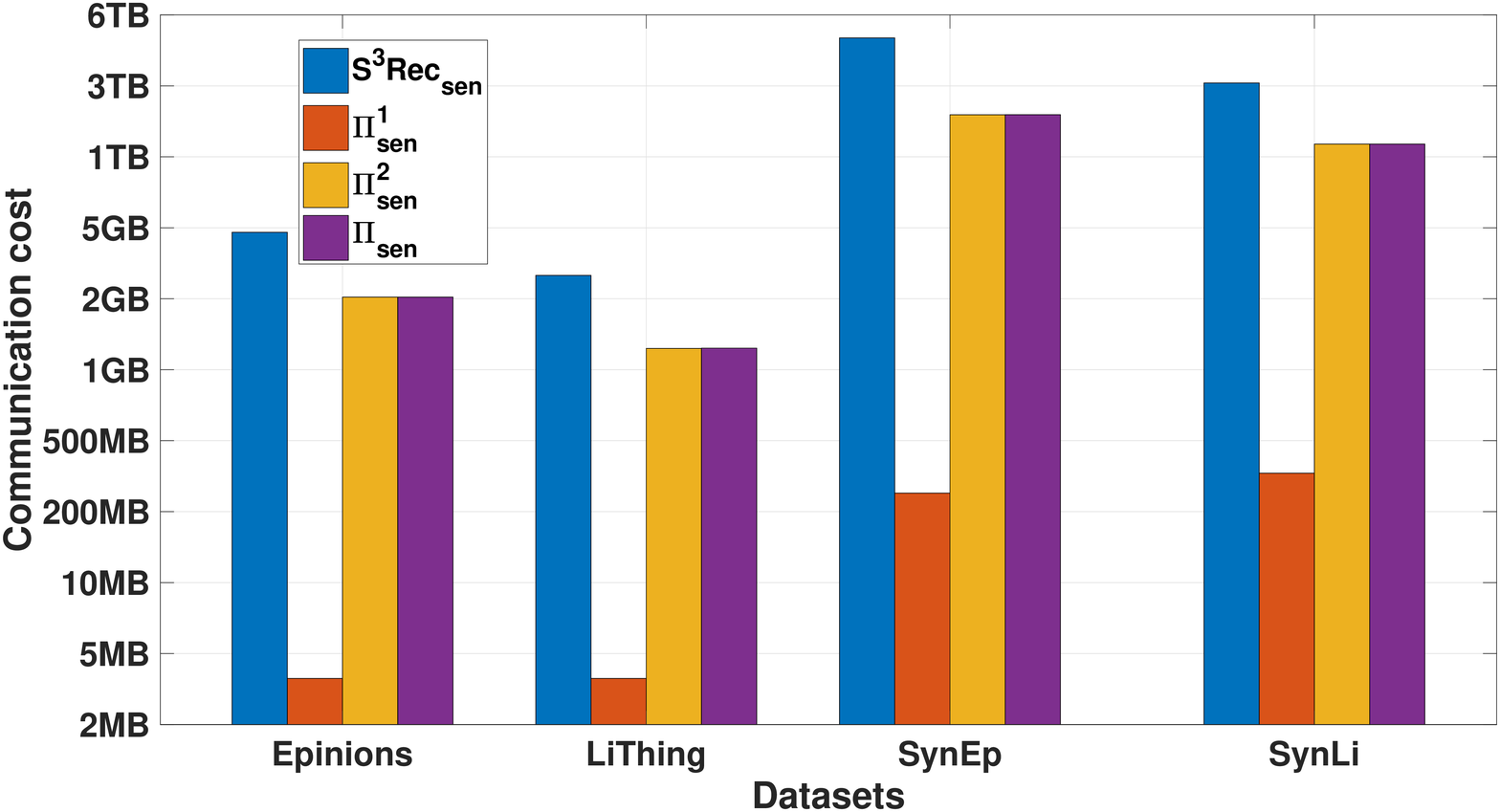}
\caption{Communication cost of $\mathsf{\Pi}_\mathrm{sen}$ and $\mathsf{S^{3}Rec}_{\mathrm{sen}}$.}
\label{fig.exp-commnunication of sen}
\end{figure}

\begin{figure}[htb]
\centering
\includegraphics[width=0.47\textwidth]{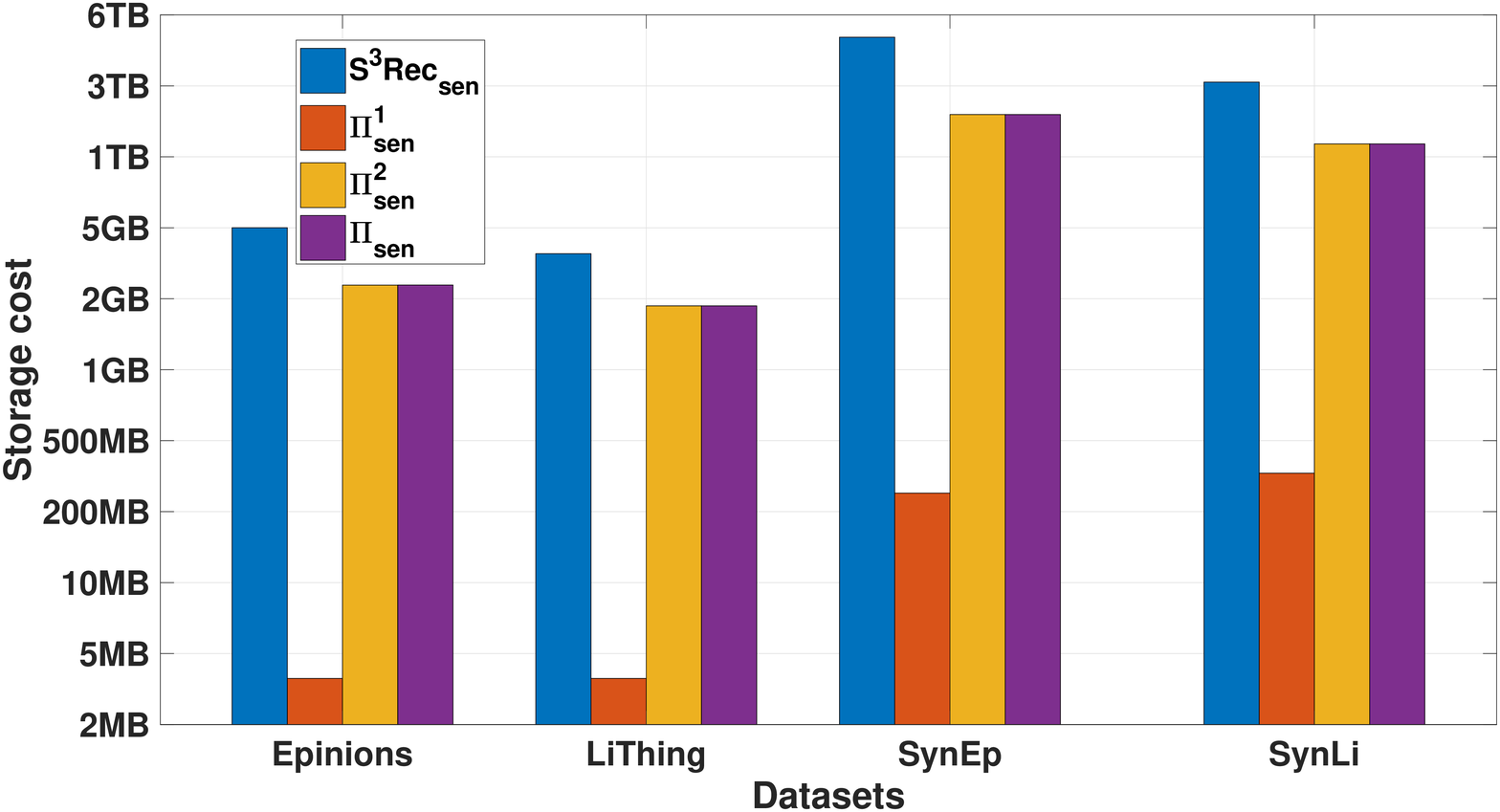}
\caption{Storage cost of $\mathsf{\Pi}_\mathrm{sen}$ and $\mathsf{S^{3}Rec}_{\mathrm{sen}}$.}
\label{fig.exp-storage of sen}
\end{figure}

\subsection{Performance Evaluation on $\mathsf{\Pi}_\mathrm{sen}$ and $\mathsf{S^{3}Rec}_{\mathrm{sen}}$}
\label{subsec:Performance evaluation on sensitive schemes}
In this part, we report the experimental results and give an analysis of the computation, communication, and storage costs for $\mathsf{\Pi}_\mathrm{sen}$ and $\mathsf{S^{3}Rec}_{\mathrm{sen}}$. Similarly, we denote the secure computing of $\mathbf{UD}$ as $\mathsf{\Pi}_\mathrm{sen}^{1}$, and $\mathsf{\Pi}_\mathrm{sen}^{2}$ stands for $\mathbf{U}\mathbf{S}^{T}$. The input datasets remain unchanged. Recall that, in the sparse location sensitive setting, we need to conceal both the original values and their locations. To achieve this goal, the comparison scheme $\mathsf{S^{3}Rec}_{\mathrm{sen}}$ as well our scheme $\mathsf{\Pi}_\mathrm{sen}^{2}$ propose to apply PIR. In doing so, the needed values can be fetched from $\mathbf{U}$ in a privacy-preserving way. To compress the communication, a new and communication-efficient PIR scheme is used in $\mathsf{\Pi}_\mathrm{sen}^{2}$. In addition, we bridge the packing method with PIR to further boost efficiency. Therefore, both computation and communication costs are significantly reduced. Note that, the input matrix $\mathbf{U}$ is a diagonal matrix. However,  $\mathsf{S^{3}Rec}_{\mathrm{sen}}$ did not provide any optimization for computing $\mathbf{UD}$. As a result, $\mathsf{\Pi}_\mathrm{sen}$ outperforms $\mathsf{S^{3}Rec}_{\mathrm{sen}}$ in all aspects.

\textbf{Computational costs.} In $\mathsf{S^{3}Rec}_{\mathrm{sen}}$, all the elements in matrix $\mathbf{U}$ are encrypted one by one as the database for PIR. In contrast, $\mathsf{\Pi}_\mathrm{sen}^{2}$ packs the elements before encryption. Meanwhile, the PIR based vector inner product is still supported without decryption during the processing. Thus the encryption complexity on party $\mathsf{P_0}$ is reduced by $s\times $, where $s$ is the packing size. Moreover, the additional operations in the plaintext domain, including generating $s\times $ more random numbers and aggregating the results, are negligible. As mentioned above, using PIR to compute $\mathbf{UD}$ is time-consuming due to the in-necessary PIR queries and response processing. For each element in the diagonal vector of $\mathbf{D}$, at least one PIR query is needed. Also, $\mathbf{D}$ is extremely sparse. To alleviate heavy PIR operations and fully explore the extreme sparsity of $\mathbf{D}$ (i.e., $1/m$), $\mathsf{\Pi}_\mathrm{sen}^{1}$ uses the same method as $\mathsf{\Pi}_\mathrm{ins}^{1}$. In Fig. \ref{fig.exp-Running time of sen}, the running time of $\mathsf{S^{3}Rec}_{\mathrm{sen}}$ and $\mathsf{\Pi}_\mathrm{ins}^{1}$ on four datasets are clearly shown. The results indicate that our scheme consumes less time. In specific, for SynEp and SynLi, we reduce the time costs roughly by $2.8\times$.

\textbf{Communication costs.} $\mathsf{\Pi}_\mathrm{sen}^{1}$ has significantly compressed the communication cost for the following three reasons. First, the packing method can reduce the number of ciphertexts by the packing size (i.e., $N$) that needs to be exchanged. Second, the comparison scheme $\mathsf{S^{3}Rec}_{\mathrm{sen}}$ has to issue $m$ PIR queries. In particular, it commonly needs $2$ to $3$ RLWE cihpertexts to issue a PIR query. Third, the remasked secret shares need to be returned, which brings $O(l\times m)$ communication complexity. Without the packing process, $\mathsf{S^{3}Rec}_{\mathrm{sen}}$ has to return all unpacked ciphertexts. For each PIR query in $\mathsf{\Pi}_\mathrm{sen}^{2}$, the upload communication is compressed by $2\times$, and the download volume is compressed by $2.4 \times$. In addition, the total query number can be decreased if more than one non-sparse element is located in the same packing slot. Note that the packing operation conducted in $\mathsf{\Pi}_\mathrm{sen}^{2}$ does not introduce an additional communication cost. We report the specific costs in Fig. \ref{fig.exp-commnunication of sen}. Roughly, our scheme $\mathsf{\Pi}_\mathrm{sen}$ achieves $2.3 \times$ communication reduction.

\textbf{Storage costs.} The storage costs of $\mathsf{S^{3}Rec}_{\mathrm{sen}}$ and $\mathsf{\Pi}_\mathrm{sen}^{2}$ mainly comprise the following three parts. First, the ciphertexts generated for PIR queries and responses. Second, the encrypted version of the input matrix $\mathbf{U}$. Third, the remasked encrypted results (encrypted secret shares). The storage cost reduction offered by our scheme $\mathsf{\Pi}_\mathrm{sen}$ stems from the packing operation on the matrix $\mathbf{U}$. We report the detailed costs in Fig. \ref{fig.exp-storage of sen}. For datasets Epinions and LiThing, $\mathsf{\Pi}_\mathrm{sen}^{2}$ needs at most $5.545$ GB and $3.907$ GB storage volumes, yet $\mathsf{\Pi}_\mathrm{sen}$ only requires $2.581$ GB and $1.904$ GB.

\begin{figure}[htb]
\centering
\includegraphics[width=8cm,height=4cm]{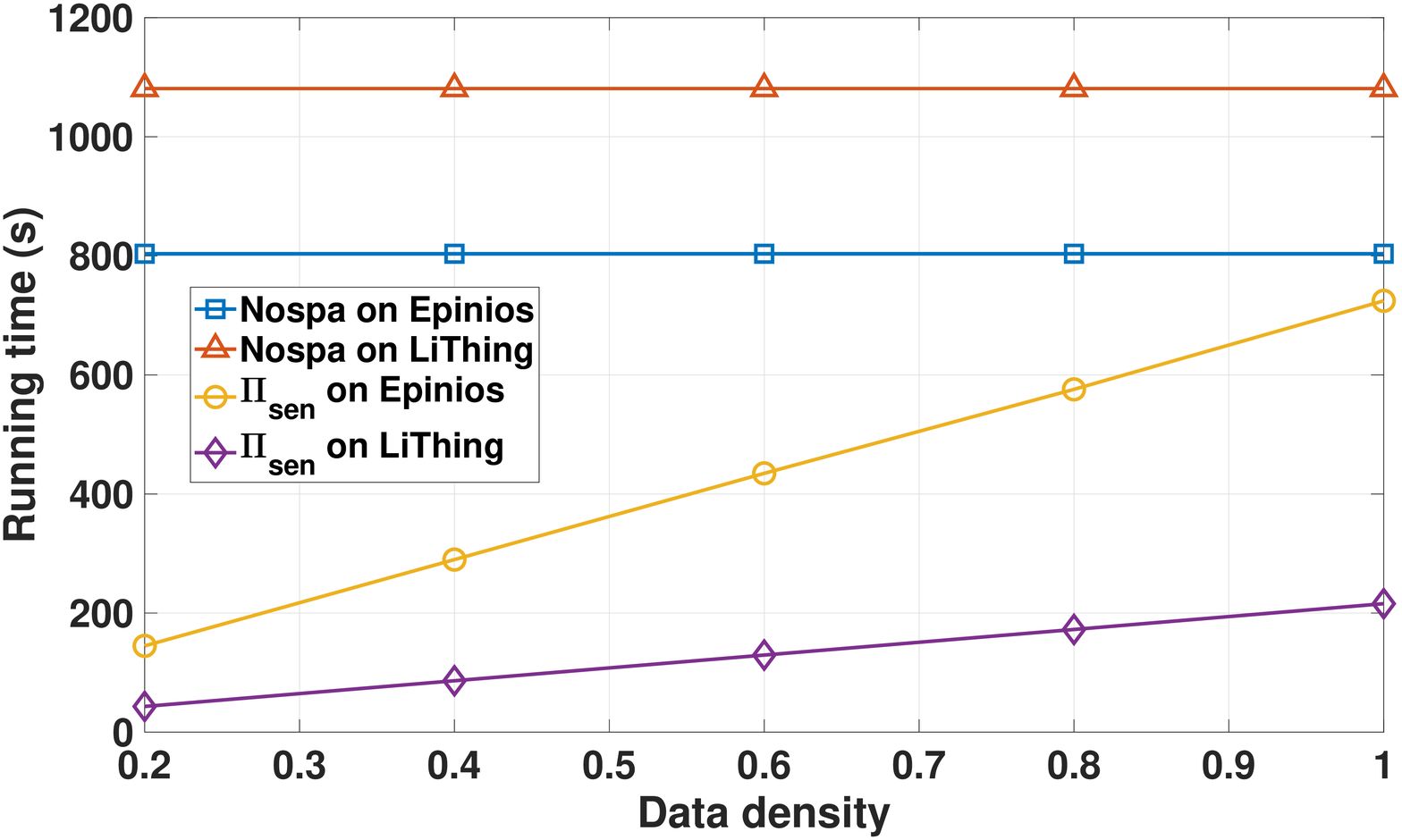}
\caption{Running time of $\mathsf{\Pi}_\mathrm{sen}$ and $\mathsf{Nospa}$.}
\label{fig.exp-Running time on sparsity}
\end{figure}

\begin{figure}[htb]
\centering
\includegraphics[width=8cm,height=4cm]{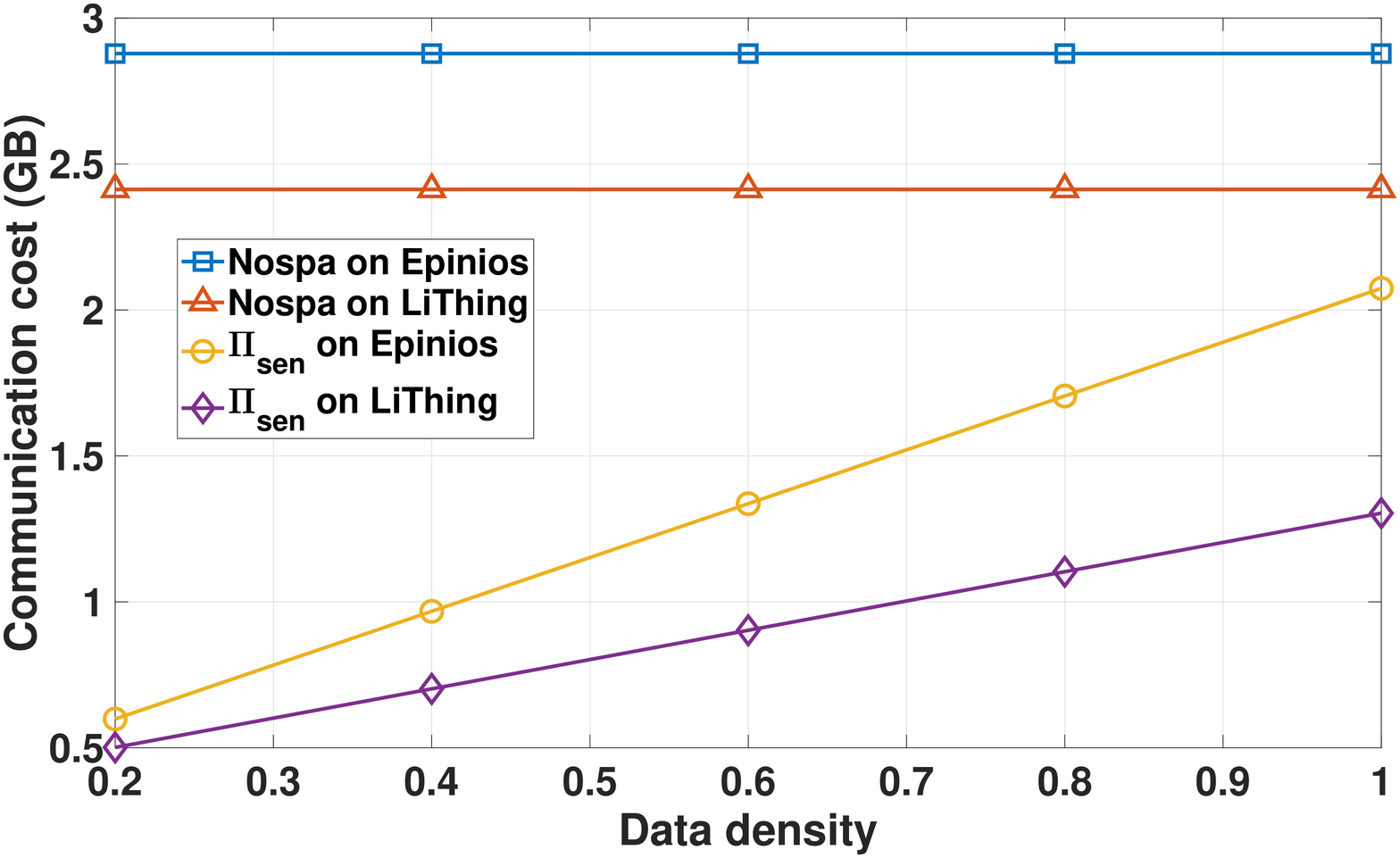}
\caption{Communication cost of $\mathsf{\Pi}_\mathrm{sen}$ and $\mathsf{Nospa}$.}
\label{fig.exp-commnunication on sparsity}
\end{figure}

\begin{figure}[htb]
\centering
\includegraphics[width=8cm,height=4cm]{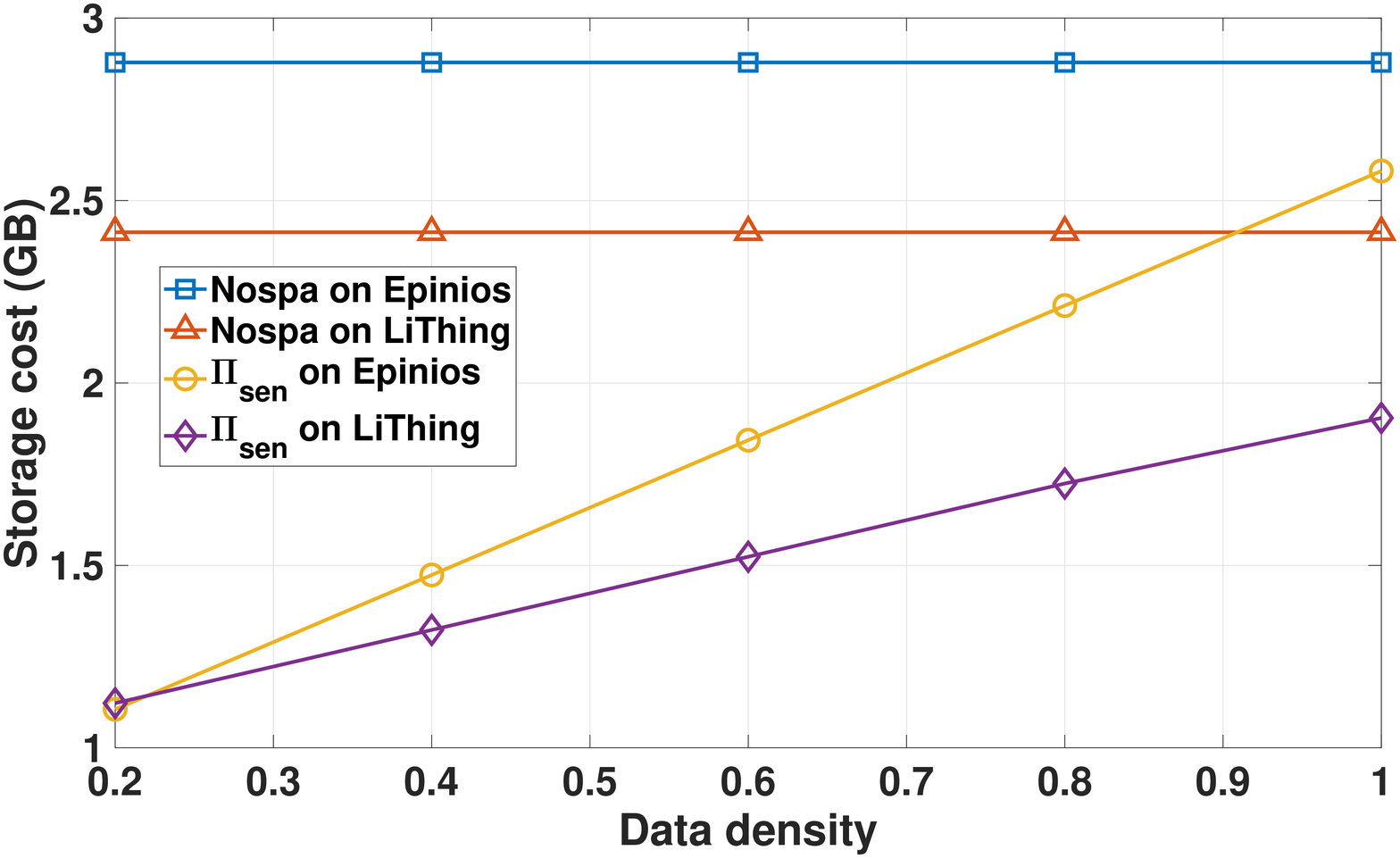}
\caption{Storage cost of $\mathsf{\Pi}_\mathrm{sen}$ and $\mathsf{Nospa}$.}
\label{fig.exp-storage on sparsity}
\end{figure}

\subsection{Effect of Data Sparsity on $\mathsf{\Pi}_\mathrm{sen}$ and $\mathsf{Nospa}$}
%$\mathsf{\Pi}_\mathrm{sen}$ and $\mathsf{Nospa}
\label{subsec:Performance evaluation on sparsity level}
In this part, we study the impacts on the data sparsity of the proposed sparse location sensitive scheme $\mathsf{\Pi}_\mathrm{sen}$ by varying the data density of the datasets Epinions and LiThing. The density of an original simulated dataset is marked as $100\%$. If we uniformly delete $20\%$ non-sparse values, then the density becomes $80\%$. We report the performance by varying the density from $20\%$ to $100\%$ with step length $20\%$. In addition, a simulated scheme $\mathsf{Nospa}$ without considering the data sparsity is used as the baseline to demonstrate the performance gain. Specifically, $\mathsf{Nospa}$ encrypts all the elements of input matrices using the same method as $\mathsf{\Pi}_\mathrm{ins}$. Thus, the costs of $\mathsf{Nospa}$ should be a constant. Since $\mathsf{\Pi}_\mathrm{ins}$ introduces spare location leakage, we choose not to report its performance for fairness. Note that, since data sparsity is not utilized, $\mathsf{Nospa}$ encrypts the input matrix $\mathbf{U}$ rather than the larger input matrix $\mathbf{S}$.

\textbf{Computational costs.} When changing the data density, the sparsity of input matrix $\mathbf{D}$ (diagonal) remains the same. As a result, the computation complexity of $\mathsf{\Pi}_\mathrm{sen}^{1}$ should be a constant. On party $\mathsf{P_0}$, the encryption of matrix $\mathbf{U}$ (as the PIR database) is also irrelevant to the data density. Therefore, the key impact of data sparsity on $\mathsf{\Pi}_\mathrm{sen}^{2}$ is the PIR query scale. In theory, the running time of $\mathsf{\Pi}_\mathrm{sen}$ increase linearly with the data density. The Fig. \ref{fig.exp-Running time on sparsity} has demonstrated that $\mathsf{\Pi}_\mathrm{sen}$ reduced the cost by $10\%$ on Epinions, and at least $5 \times$ on LiThing than $\mathsf{Nospa}$.

\textbf{Communication costs.} Similarly, the communication costs brought by $\mathsf{\Pi}_\mathrm{sen}^{1}$ and $\mathsf{Nospa}$ remain the same when varying the data density. Thus, the number of issued PIR queries becomes the only factor that causes the variation in communication volume. With increasing data density, the communication cost increases linearly. As shown in Fig. \ref{fig.exp-commnunication on sparsity}, the communication costs of $\mathsf{Nospa}$ reach $2.828$ GB and $2.413$ GB on datasets Epinions and LiThing, yet $\mathsf{\Pi}_\mathrm{sen}$ only needs $2.074$ GB and $1.304$ GB.

\textbf{Storage costs.} The total storage costs of $\mathsf{\Pi}_\mathrm{sen}$ on simulated datasets are already given in Section \ref{subsec:Performance evaluation on sensitive schemes}. When we increase the data density, the party $\mathsf{P_1}$ will generates more PIR queries. However, the processing of each query on the party $\mathsf{P_0}$ requires exactly the same storage complexity $O(l\times m)^{1/d}$, where $d$ is the dimension of the database index. Besides, the other storage costs for the encrypted database on party $\mathsf{P_0}$, the remasked ciphertexts, and the secret shares remain unchanged. Thus, the total cost of $\mathsf{\Pi}_\mathrm{sen}$ varies slightly along the data density variation. As demonstrated by Fig. \ref{fig.exp-storage on sparsity}, when the data density is set as $20\%$, $\mathsf{\Pi}_\mathrm{sen}$ needs roughly half of the full-dataset case, yet $\mathsf{Nospa}$ remains the same storage costs as the communication volumes.

\textbf{Remark.} The comparison scheme $\mathsf{S^{3}Rec}_{\mathrm{sen}}$ has the same asymptotic computation, communication, and storage complexity as $\mathsf{\Pi}_\mathrm{sen}$ when varying the data density. In addition, the overall performance is comprehensively evaluated in Section \ref{subsec:Performance evaluation on sensitive schemes}. Thus, we omit it here due to space limitations.

\subsection{Accuracy Evaluation}
\label{subsec:ACCURACY evaluation}
%From those Tables, we find that: (1) the use of social information can indeed improve the recommendation performance of the rating platform, e.g., 1.193 vs. 1.062 and 0.927 vs. 0.098 in terms of RMSE on Epinions and Lthing, respectively. This result is consistent with existing work from [18, 5];

%Metrics. We will evaluate both accuracy and efficiency of our proposed model. For accuracy, we choose Root Mean Square Error (RMSE) as the evaluation metric, since ratings range in [0, 5].

\renewcommand\tablename{TABLE}
\renewcommand \thetable{\Roman{table}}
\setcounter{table}{1}
\setcounter{figure}{5}
\begin{table}[htb]
\centering
\caption{Accuracy Comparison}
\label{table:accuracy}\
\setlength{\tabcolsep}{4mm}{
\begin{tabular}{ccccc}
\hline
   &MF &$\mathsf{S^{3}Rec}_{\mathrm{sen}}$  &$\mathsf{\Pi}_\mathrm{ins}$  &$\mathsf{\Pi}_\mathrm{sen}$  \\
\hline
Epinions  & 1.197  &   1.063  &   1.064    &   1.062      \\
\hline
LiThing  & 0.925  &  0.907  &  0.909    &   0.907   \\
\hline
\end{tabular}}
\end{table}

In this part, we review the impacts on the accuracy of our proposed privacy-preserving schemes $\mathsf{\Pi}_\mathrm{ins}$, $\mathsf{\Pi}_\mathrm{sen}$ and the comparison scheme $\mathsf{S^{3}Rec}_{\mathrm{sen}}$. The mainstream accuracy measurement Root Mean Square Error (RMSE) \cite{cui2021exploiting} is adopted. To demonstrate the advantage of incorporating the social data for the recommendation, we use the classical matrix factorization (MF) model \cite{bobadilla2013recommender} as the baseline. MF takes only the rating matrix as the input. As shown in Table \ref{table:accuracy}, $\mathsf{\Pi}_\mathrm{ins}$, $\mathsf{\Pi}_\mathrm{sen}$ and $\mathsf{S^{3}Rec}_{\mathrm{sen}}$ achieve higher accuracy than the baseline MF. This demonstrates that the input social data can indeed improve the recommending accuracy. As the used HE and PIR primitives in $\mathsf{\Pi}_\mathrm{ins}$, $\mathsf{\Pi}_\mathrm{sen}$ and $\mathsf{S^{3}Rec}_{\mathrm{sen}}$ preserve the same calculation precision, these three schemes offer roughly the same accuracy.

\section{Conclusion and Future Work}
\label{sec:conlusion and future work}
In this paper, we started with the motivation of boosting the efficiency of privacy-preserving cross-platform recommender systems. Through an in-depth analysis on the target problem, we proposed two lean and fast privacy-preserving schemes. One was designed for the sparse location insensitive setting and the other was designed for the sparse location sensitive setting. We fused versatile advanced message  packing, HE, and PIR primitives into our protocols to guarantee provable security and to fully exploit the input data sparsity. Without compromising the accuracy, our proposed schemes have significantly promoted the overall performance compared with the state-of-the-art work. In the future, we will continuously investigate the sparsity and privacy issues in social data incorporated recommender systems. In addition, we will focus on enabling federated or multiparty recommender systems with attractive features such as model ownership protection.

%\section*{Acknowledgments}
%Thanks to Prof. Zhang and my buddies.

\ifCLASSOPTIONcaptionsoff
\newpage \fi
\bibliographystyle{IEEEtran}
\bibliography{main}
%\vfill
\vspace{-2.16 cm}
\begin{IEEEbiography}[{\includegraphics[width=1in,height=1.25in, clip,keepaspectratio]{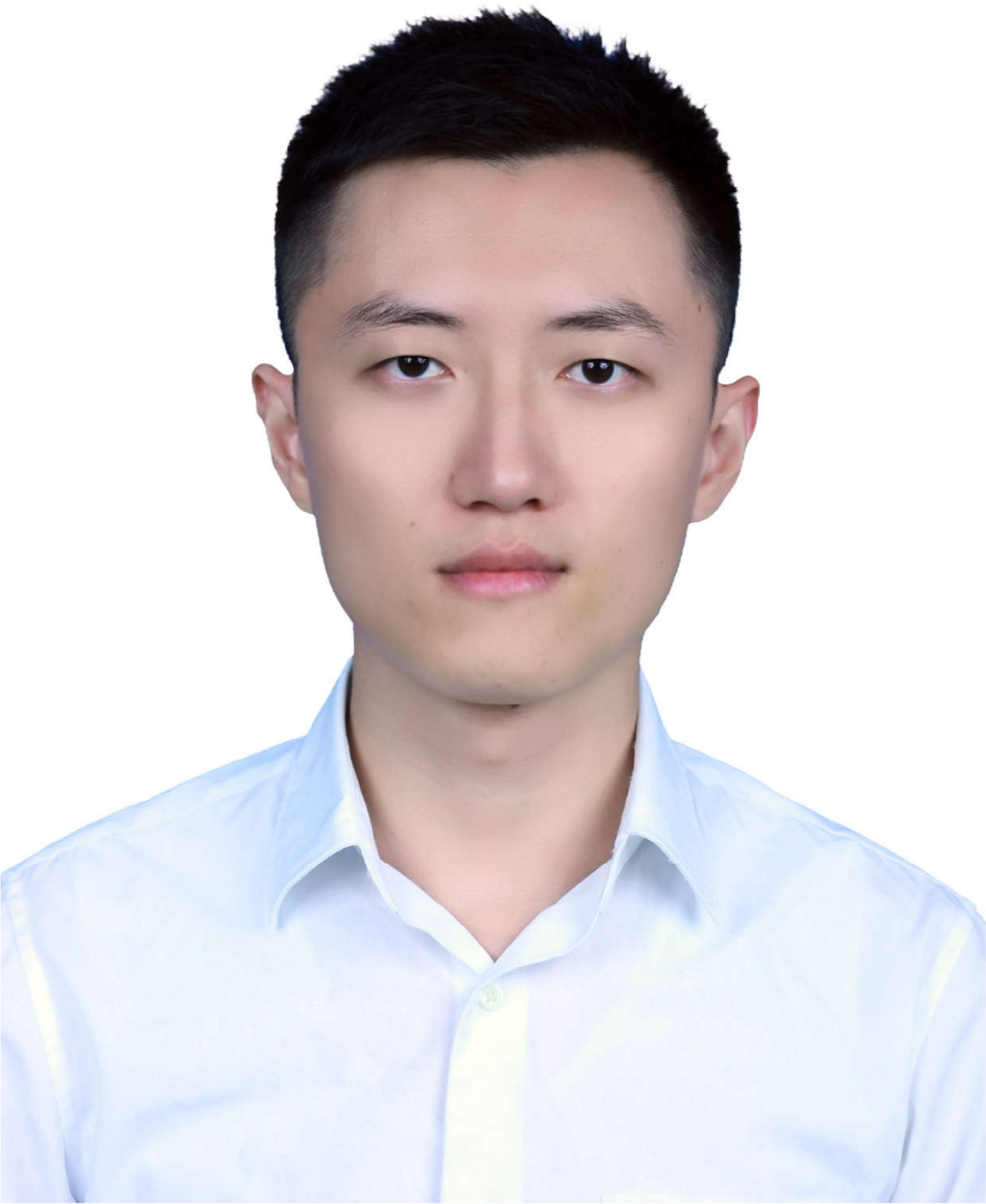}}]{Hao Ren} is currently a Research Fellow at Nanyang Technological University, Singapore. He received his Ph.D. degree in 2020 from the University of Electronic Science and Technology of China. He has published papers in major conferences/journals, including ACM ASIACCS, ACSAC, IEEE TCC, and IEEE Network. His research interests include applied cryptography and  privacy-preserving machine learning.
\end{IEEEbiography}

\vspace{-1.8 cm}

\begin{IEEEbiography}[{\includegraphics[width=1in,height=1.25in, clip,keepaspectratio]{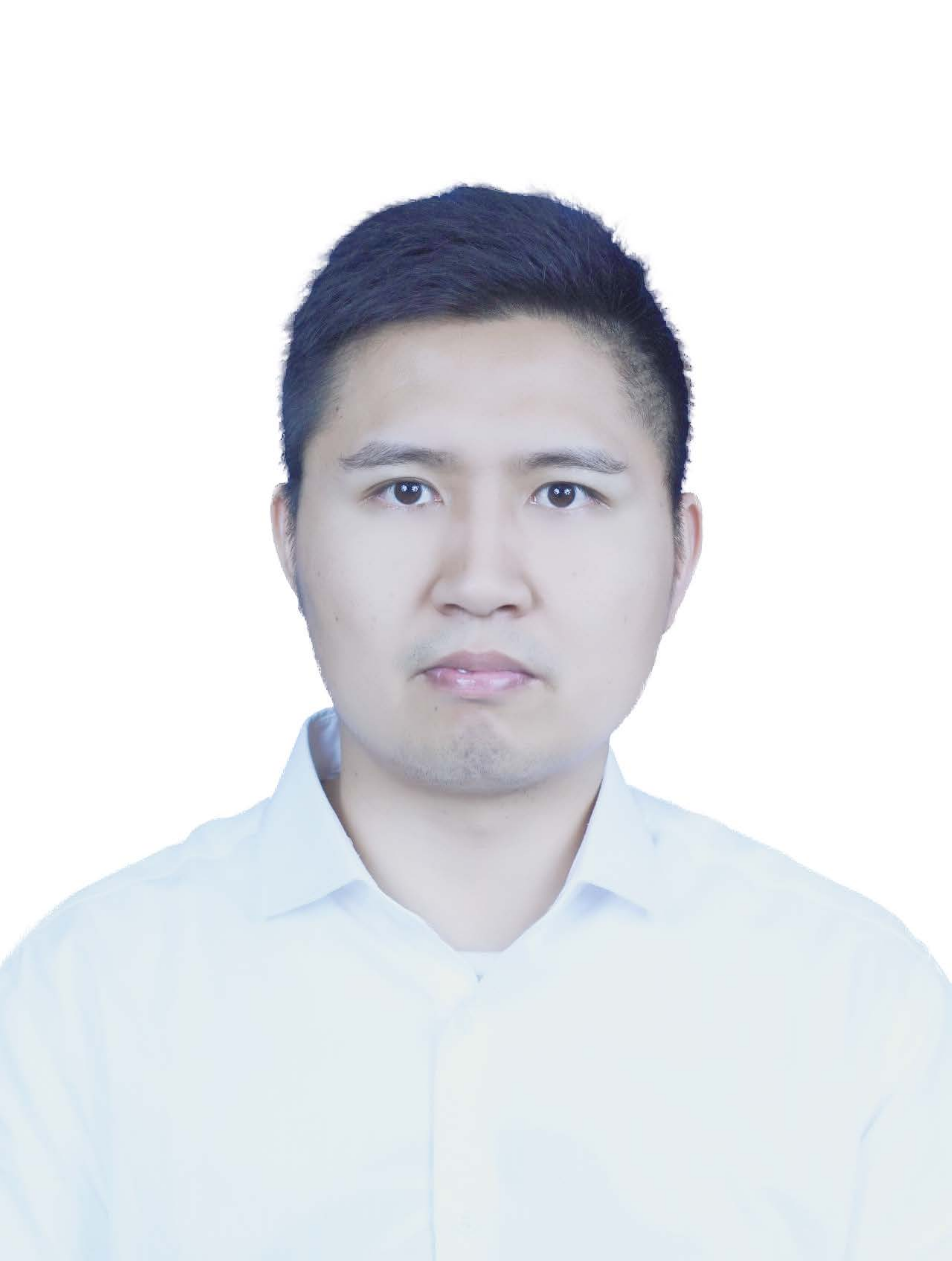}}]{Guowen Xu} is currently a Research Fellow at Nanyang Technological University, Singapore. He received his Ph.D. degree in 2020 from the University of Electronic Science and Technology of China. He has published papers in reputable conferences/journals, including ACM CCS, NeurIPS, ASIACCS, ACSAC, ESORICS, IEEE TIFS, and IEEE TDSC. His research interests include applied cryptography and  privacy-preserving deep learning.
\end{IEEEbiography}

\vspace{-1.8 cm}

\begin{IEEEbiography}[{\includegraphics[width=1in,height=1.25in]{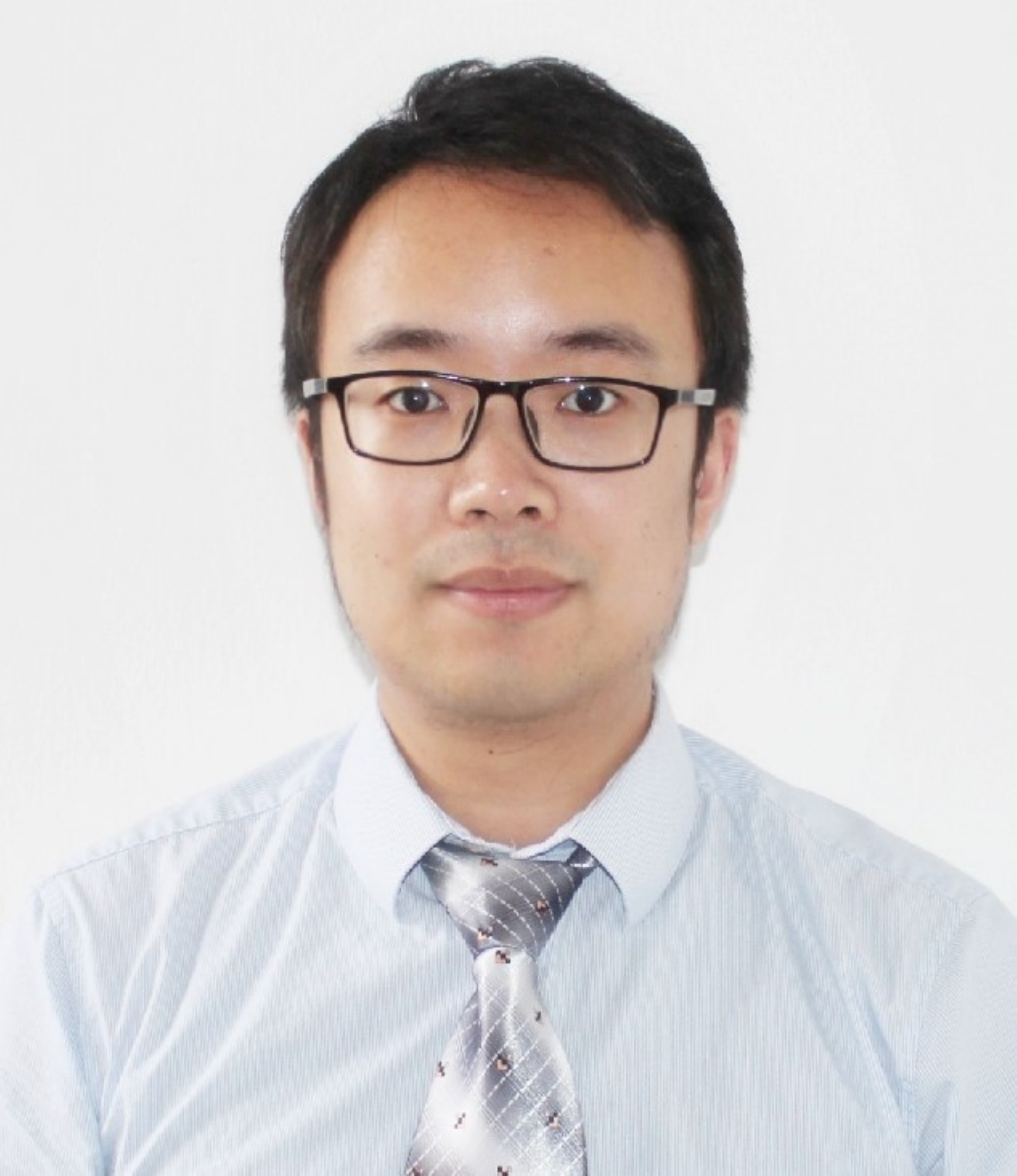}}] {Tianwei Zhang}is an assistant professor at the School of Computer Science and Engineering, at Nanyang Technological University. His research focuses on computer system security. He is particularly interested in security threats and defenses in machine learning systems, autonomous systems, computer architecture, and distributed systems. He received his Bachelor's degree at Peking University in 2011, and his Ph.D. degree at Princeton University in 2017.
\end{IEEEbiography}

\vspace{-1.8 cm}

\begin{IEEEbiography}[{\includegraphics[width=1in,height=1.25in, clip,keepaspectratio]{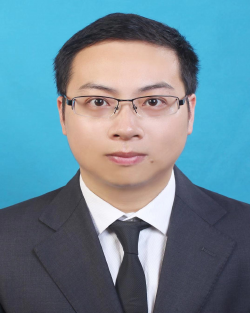}}]{Jianting Ning} is currently a Professor with the Fujian Provincial Key Laboratory of Network Security and Cryptology, College of Computer and Cyber Security, Fujian Normal University, China.  He has published papers in major conferences/journals, such as ACM CCS, NDSS, ASIACRYPT, ESORICS, ACSAC, IEEE Transactions on Information Security and Forensics, and IEEE Transactions on Dependable and Secure Computing. His research interests include applied cryptography and information security.
\end{IEEEbiography}

\vspace{-1.8 cm}

\begin{IEEEbiography}[{\includegraphics[width=1in,height=1.25in]{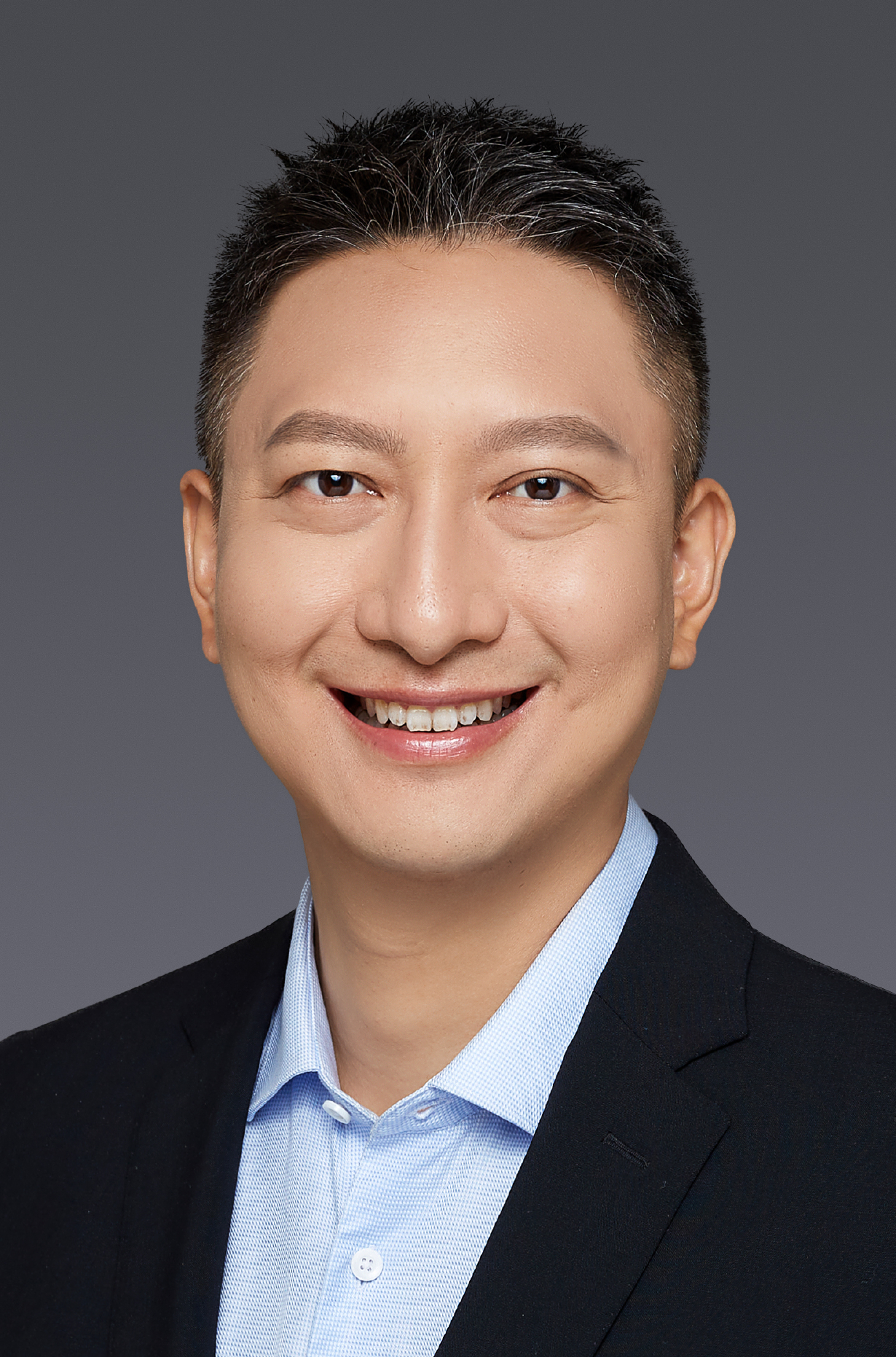}}] {Xinyi Huang} is currently an Associate Professor at the Thrust of Artificial Intelligence, Information Hub, Hong Kong University of Science and Technology (Guangzhou), China. His research interests include cryptography and information security. He is in the Editorial
Board of International Journal of Information Security and SCIENCE CHINA Information Sciences. He has served as the program/general chair or program committee member in over 120 international conferences.
\end{IEEEbiography}

\vspace{-1.8 cm}

\begin{IEEEbiography}[{\includegraphics[width=1in,height=1.25in, clip,keepaspectratio]{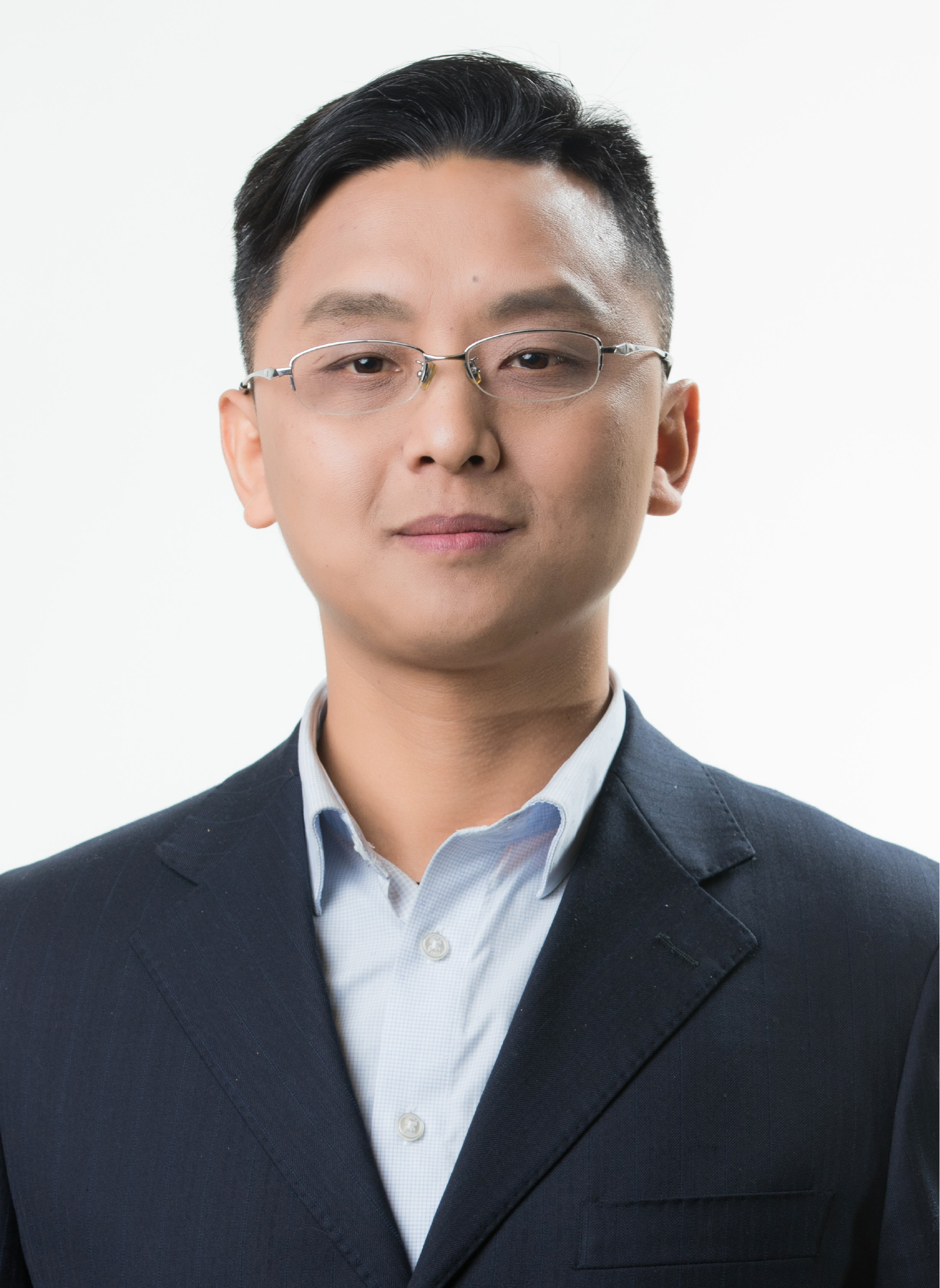}}]{Hongwei Li} is currently the Head and a Professor at the Department of Information Security, School of Computer Science and Engineering, University of Electronic Science and Technology of China.  His research interests include network security and applied cryptography. He is the Senior Member of IEEE and the Distinguished Lecturer of the IEEE Vehicular Technology Society.
\end{IEEEbiography}

\vspace{-1.8 cm}

\begin{IEEEbiography}[{\includegraphics[width=1in,height=1.25in, clip,keepaspectratio]{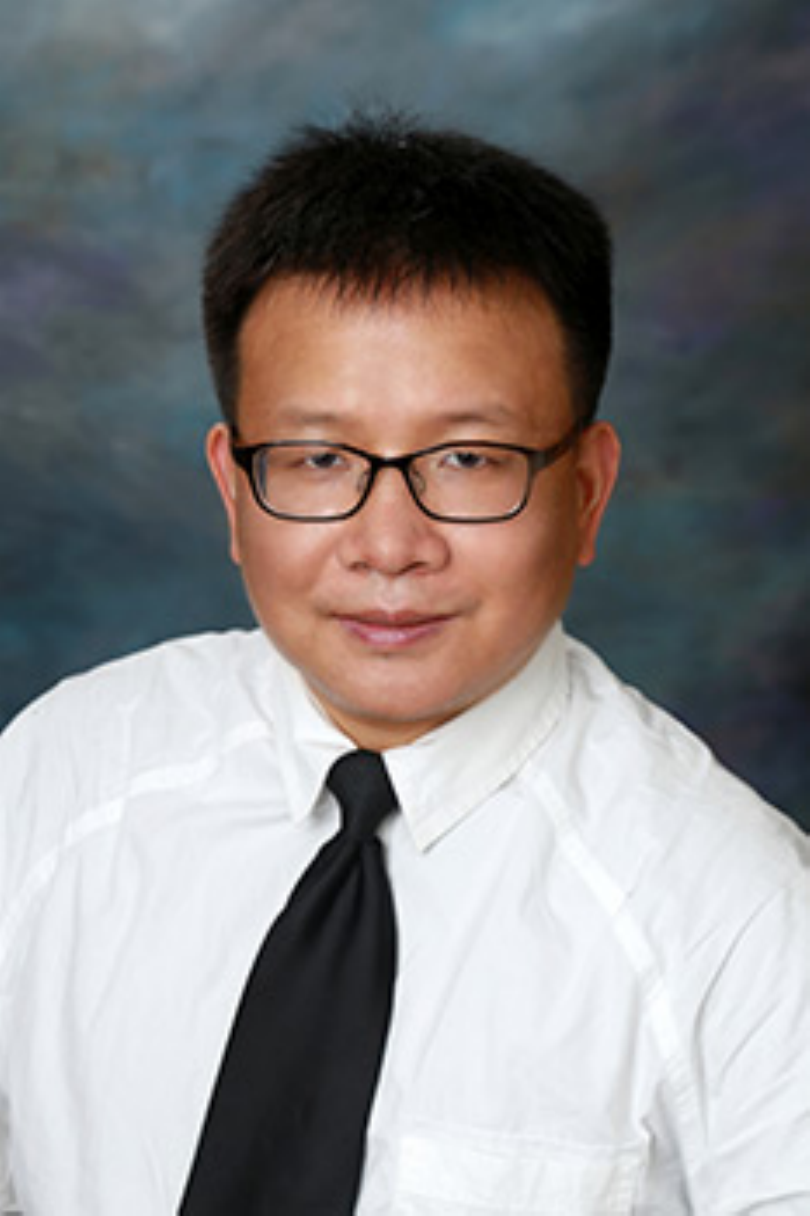}}]{Rongxing Lu} is currently an associate professor at the Faculty of Computer Science (FCS), University of New Brunswick (UNB), Canada. He received his Ph.D. degree from the Department of Electrical \& Computer Engineering, University of Waterloo, Canada, in 2012; and won the 8th IEEE Communications Society (ComSoc) Asia Pacific (AP) Outstanding Young Researcher Award, in 2013. He is presently an IEEE Fellow. Dr. Lu currently serves as the Vice-Chair (Publication) of IEEE ComSoc CIS-TC. Dr. Lu is the Winner of the 2016-17 Excellence in Teaching Award, FCS, UNB.
\end{IEEEbiography}

\end{document}